\documentclass[aps,prb,superscriptaddress,floatfix,reprint,longbibliography]{revtex4-2}

\usepackage{graphicx} 
\usepackage[english]{babel}
\usepackage[utf8]{inputenc}
\usepackage[T1]{fontenc}

\usepackage{amsmath}
\usepackage{amstext}
\usepackage{amssymb}
\usepackage{amsfonts}
\usepackage{amsbsy}
\usepackage{braket}
\usepackage[dvipsnames]{xcolor}
\usepackage{gensymb}
\usepackage{times}

\usepackage[colorlinks=true, urlcolor=teal, linkcolor=teal, citecolor=teal, pdftex]{hyperref}

\usepackage{graphicx}
\usepackage{mathtools}
\usepackage{multirow}
\usepackage{bbold}
\usepackage{csquotes}

\usepackage{tikz}
\usetikzlibrary{calc}
\usetikzlibrary{shapes.geometric}
\usepackage{pgfplots}
\usepackage{pgfplotstable}
\pgfplotsset{width=\columnwidth,compat=1.9}
\usepgfplotslibrary{external}
\usepgfplotslibrary{colormaps}
\newcommand{\equal}{=}

\definecolor{plot1}{RGB}{31, 119, 180}
\definecolor{plot2}{RGB}{255, 127, 14}
\definecolor{plot3}{RGB}{44, 160, 44}
\definecolor{plot4}{RGB}{214, 39, 40}
\definecolor{plot5}{RGB}{148, 103, 189}
\definecolor{plot6}{RGB}{140, 86, 75}
\definecolor{plot7}{RGB}{227, 119, 194}
\definecolor{plot8}{RGB}{127, 127, 127}
\definecolor{plot9}{RGB}{188, 189, 34}
\definecolor{plot10}{RGB}{23, 190, 207}

\definecolor{mathematicaplot1}{rgb}{0.368417, 0.506779, 0.709798}
\definecolor{mathematicaplot2}{rgb}{0.880722, 0.611041, 0.142051}
\definecolor{mathematicaplot3}{rgb}{0.560181, 0.691569, 0.194885}
\definecolor{mathematicaplot4}{rgb}{0.922526, 0.385626, 0.209179}
\definecolor{mathematicaplot5}{rgb}{0.528488, 0.470624, 0.701351}
\definecolor{mathematicaplot6}{rgb}{0.772079, 0.431554, 0.102387}
\definecolor{mathematicaplot7}{rgb}{0.363898, 0.618501, 0.782349}
\definecolor{mathematicaplot8}{rgb}{1, 0.75, 0}
\definecolor{mathematicaplot9}{rgb}{0.647624, 0.37816, 0.614037}
\definecolor{mathematicaplot10}{rgb}{0.571589, 0.586483, 0.}
\definecolor{mathematicaplot11}{rgb}{0.915, 0.3325, 0.2125}
\definecolor{mathematicaplot12}{rgb}{0.400822, 0.522007, 0.85}
\definecolor{mathematicaplot13}{rgb}{0.972829, 0.621644, 0.073362}
\definecolor{mathematicaplot14}{rgb}{0.736783, 0.358, 0.503027}
\definecolor{mathematicaplot15}{rgb}{0.280264, 0.715, 0.429209}

\definecolor{googleB}{HTML}{4285F4}
\definecolor{googleG}{HTML}{34A853}
\definecolor{googleY}{HTML}{FBBC05}
\definecolor{googleR}{HTML}{EA4335}
\definecolor{googleBG}{HTML}{3B96A4}

\begin{document}

\title{Variationally optimizing infinite projected entangled-pair states at large bond dimensions:\\A split corner transfer matrix renormalization group approach}

\newcommand{\FUB}{Freie Universität Berlin, Dahlem Center for Complex Quantum Systems and Institut für Theoretische Physik, Arnimallee 14, 14195 Berlin, Germany}
\newcommand{\UK}{University of Cologne, Institute for Theoretical Physics, 50937 Köln, Germany}

\author{Jan Naumann}
\thanks{Both first authors have contributed equally.}
\affiliation{\FUB}

\author{Erik L. Weerda}
\thanks{Both first authors have contributed equally.}
\affiliation{\UK}

\author{Jens Eisert}
\affiliation{\FUB}
\affiliation{Helmholtz-Zentrum Berlin für Materialien und Energie, Hahn-Meitner-Platz 1, 14109 Berlin, Germany}

\author{Matteo Rizzi}
\affiliation{\UK}
\affiliation{Forschungszentrum Jülich, Institute of Quantum Control, Peter Grünberg Institut (PGI-8), 52425 Jülich, Germany}

\author{Philipp Schmoll}
\email[Correspondence contact: ]{j.naumann@fu-berlin.de}
\email{weerda@thp.uni-koeln.de}
\email{philipp.schmoll@fu-berlin.de}
\affiliation{\FUB}

\date{\today}

\begin{abstract}

Projected entangled-pair states (PEPS) have become a powerful tool for studying quantum many-body systems in the condensed matter and quantum materials context, particularly with advances in variational energy optimization methods. A key challenge within this framework is the computational cost associated with the contraction of the two-dimensional lattice, crucial for calculating state vector norms and expectation values. The conventional approach, using the corner transfer matrix renormalization group  (CTMRG), involves combining two tensor network layers, resulting in significant time and memory demands. In this work, we introduce an alternative ``split-CTMRG'' algorithm, which maintains separate PEPS layers and leverages new environment tensors, reducing computational complexity while preserving accuracy. Benchmarks on quantum lattice models demonstrate substantial speedups for variational energy optimization, rendering this method valuable for large-scale PEPS simulations.
\end{abstract}

\maketitle

\section{Introduction}

One of the main challenges in condensed matter physics is developing tools to predict the properties of interacting quantum matter classically. 
A versatile and important family of classical simulation methods is provided by tensor networks~\cite{Orus-AnnPhys-2014,RevModPhys.93.045003,AreaReview,Handwaving} -- a set of techniques that has originated from the early development of the \emph{density matrix renormalization 
group} approach~\cite{DMRGWhite92,MPSRev} and
that has matured into a large body of powerful methods. 
Of those methods, two-dimensional tensor networks in the form of infinite \emph{projected entangled-pair states} (PEPS)~\cite{verstraete2004renormalizationalgorithmsquantummanybody} have substantially matured in recent years due to the development of variational energy optimization techniques, paving the way for them to become one of the most important tools in the study of quantum many-body systems and two-dimensional quantum materials. 
While the gradient of the energy expectation value used in the optimization was constructed manually by summing up a large number of relevant tensor network diagrams in the pioneering works~\cite{Corboz2016,Vanderstraeten2016}, the more modern approach performs this task in an automated way -- typically by utilizing \emph{automatic differentiation} (AD)~\cite{Liao219,Naumann2024,Francuz2025}. 
These developments have boosted both the accuracy as well as the applicability of PEPS in the study of intricate frustrated quantum magnetism~\cite{Hasik2021,Schmoll2023,Lukin2023,Francesco2023,Zhang2023,Schmoll2024}, topologically ordered phases~\cite{Hasik2022,Xu2023,Weerda2024}, or itinerant electron systems~\cite{Ponsioen2022}. 
However, it has already been pointed out that the high computational cost of the variational optimization can be a limitation for variational PEPS studies of particularly complex models~\cite{Ponsioen2023b}.

One of the key steps in any numerical PEPS simulation is the contraction of the two-dimensional tensor network, required to compute the norm of the state vector as well as expectation values. 
Various methods, such as boundary MPS~\cite{Jordan2008,ZaunerStauber2018,Nietner2020} and tensor coarse-graining techniques~\cite{Levin2007,Xie2009,Zhao2010,Xie2012} have been developed to address this challenge. 
However, the \emph{corner transfer matrix renormalization group} (CTMRG) algorithm~\cite{Nishino1996,Nishino1997,Orus2009, Corboz2010, Corboz2014, Fishman2018, vanAlphen2024} has emerged as the most widely used approach, particularly when combined with automatic differentiation and energy optimization. 
Despite that, it is precisely this CTMRG part that is still the primary bottleneck in the calculations, both in time and memory. 
This is because $(i)$ it implements a computationally expensive iterative renormalization procedure to find the fixed-point tensors representing the (approximate) contraction of the infinite lattice, and $(ii)$ it has to be performed in each evaluation of the energy, which happens at least once -- but likely multiple times -- during every optimization step.

In the original formulation of CTMRG, the norm of the PEPS state vector is computed by combining the two TN layers for
dual vector $\bra{\psi}$ and state
vectors $\ket{\psi}$ into an effective double-layer network. 
While this results in an empirically stable and robust algorithm, it leads to a high computational cost due to the effective squaring of the PEPS bond dimension. 
Here, we instead propose an alternative split-CTMRG scheme that gains an advantage in terms of computational complexity by keeping the two layers of the PEPS separate, and defining individual fixed-point environment tensors for them.

In the past, several proposals were made to improve the efficiency of the CTMRG scheme. 
Refs.~\cite{Xie2017, Haghshenas2019} 
have proposed a substantially more computationally efficient contraction algorithm by flattening the double-layer tensor network for the norm into a single-layer structure, at the cost of expanding the elementary unit cell. 
More recently, Ref.~\cite{Lan2023} presented a variant of the CTMRG algorithm, in which the PEPS tensors and their conjugates are sequentially absorbed into conventional CTMRG environment tensors. 
Both of these approaches were proposed and tested only in the context of ground state search with imaginary-time evolution, i.e.,  the simple or full update~\cite{Jordan2008,Phien2015}.

However, when moving to variational PEPS optimization we are confronted with specific requirements. 
This is because the variational PEPS framework not only has the most to gain from reducing the computational cost of computing environment tensors but also imposes the strictest accuracy requirements on expectation value calculations. 
These strict requirements are due to the fact that variational PEPS optimization relies on accurate calculations of the energy gradient. Without an accurate energy approximation, the optimization will be ineffective, especially near the targeted minimum. 
In order to achieve the optimal accuracy, the projectors used in the CTMRG scheme are of crucial importance. In this work, we therefore employ the philosophy of the best established projectors from the conventional CTMRG~\cite{Corboz2010,Corboz2014,Fishman2018} to define the new projectors for the split-CTMRG algorithm. 
We show that this yields very precise expectation values, while substantially reducing the cost of the algorithm compared to the conventional CTMRG, allowing us to push the limits and applicability of the infinite PEPS ansatz.

This work is structured as follows: In Sec.~\ref{sect:method}, we briefly outline the conventional CTMRG algorithm and introduce the new split-CTMRG version, highlighting important differences and sources of efficiency improvements.  
In Sec.~\ref{sect:Numerical analysis}, we then provide numerical benchmarks on a quantum lattice model, demonstrating state-of-the-art accuracy and showcasing improved results due to significantly reduced computational costs. 
Finally, we conclude this work in Sec.~\ref{sec:conclusionsAndOutlook}.

\section{Method}
\label{sect:method}
Before presenting our alternative version of the CTMRG algorithm in Sec.~\ref{method:splitCTMRG_description}, we briefly revise the context of infinite projected entangled-pair states and the general task of computing their norms in Sec.~\ref{sub:infinitePEPSandCTMRG}.

\subsection{Infinite PEPS and the CTMRG}
\label{sub:infinitePEPSandCTMRG}

Infinite projected entangled-pair states (PEPS) are an ansatz class of variational many-body state vectors of quantum lattice models
\begin{equation}
    \ket{\psi} = \sum_{\{s_i\}} C^{\{s_i\}} \ket{\{s_i\}},
\end{equation}
in which the the coefficient tensor $C^{\{s_i\}}$ is presumed to have a substructure of a tensor network with two dimensional planar connectivity.
On a square lattice, this tensor network consists of rank-five tensors, which are periodically repeated
\begin{align}
    \begin{split}
        \includegraphics[scale = 1.0]{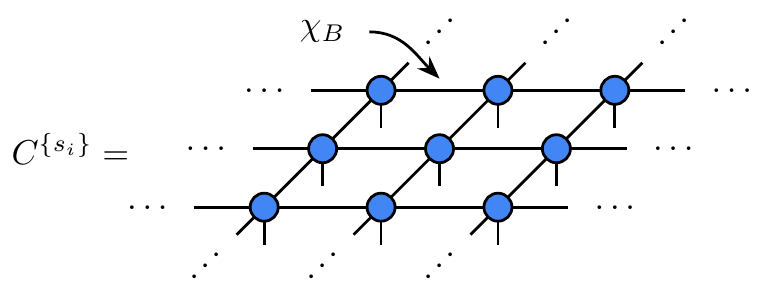}
    \end{split}.
    \label{eq:coefficientTensorNetwork}
\end{align}
While we show here a fully translational invariant network, it is also possible to define a non-trivial unit cell of tensors that is periodically repeated. 
The open legs of the tensors in Eq.~\eqref{eq:coefficientTensorNetwork} correspond to the physical indices of the coefficient tensor, while the virtual horizontal and vertical indices connect the tensors and mediate quantum correlations in the system. 
The dimension of those virtual indices, denoted by the integer \emph{bulk bond dimension} $\chi_B$, is the primary refinement parameter in the system, controlling how accurately the infinite PEPS ansatz can represent a given target state.

When employing the infinite PEPS state vectors as a variational ansatz for the numerical investigation of two dimensional quantum lattice models, a central mathematical object is a double-layer network -- built from the local tensors -- that upon contraction yields the norm
\begin{equation}
    \begin{split}
        \includegraphics[]{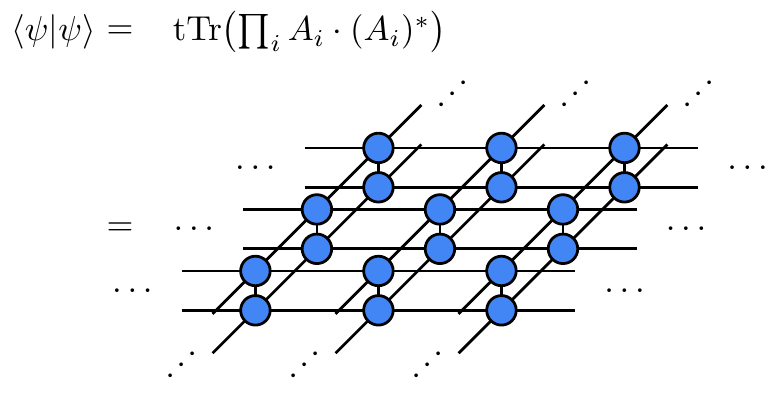}\hspace{-0.02\columnwidth}
    \end{split}
\end{equation}
of the state vector expressed as a tensor network.
\begin{figure}[b!]
    \centering
        \begin{tikzpicture}

        \node at (-3.5, 0) {$\braket{\psi\vert\psi} \approx $};
    
		\def\tensorSep{1}
	    \def\tensorSize{0.2}

		\begin{scope}[shift = {(0.0, 0.0)}]
            \coordinate (ket) at ({-0.5*\tensorSep}, {+0.5*\tensorSep});
            \coordinate (bra) at ({+0.5*\tensorSep}, {-0.5*\tensorSep});
  
		\coordinate (C1) at ({-1.5*\tensorSep}, {+1.5*\tensorSep});
            \coordinate (T1) at ({-0.0*\tensorSep}, {+1.5*\tensorSep});
            \coordinate (C2) at ({+1.5*\tensorSep}, {+1.5*\tensorSep});
            \coordinate (T2) at ({+1.5*\tensorSep}, {+0.0*\tensorSep});
            \coordinate (C3) at ({+1.5*\tensorSep}, {-1.5*\tensorSep});
            \coordinate (T3) at ({-0.0*\tensorSep}, {-1.5*\tensorSep});
            \coordinate (C4) at ({-1.5*\tensorSep}, {-1.5*\tensorSep});
            \coordinate (T4) at ({-1.5*\tensorSep}, {+0.0*\tensorSep});

            \draw[thick, decorate, decoration={coil, aspect=0}] (ket) to node[midway, above right=-1]{$p$} (bra);

            \draw[thick] (bra) to [out = 90, in = 270] (T1);
            \draw[thick] (bra) to [out = 0, in = 180] (T2);
            \draw[thick] (bra) to [out = 270, in = 90] (T3);
            \draw[thick] (bra) to [out = 180, in = 0] (T4);

            \draw[thick] (ket) to [out = 90, in = 270] (T1);
            \draw[white, line width=3pt] (ket) to [out = 0, in = 180] (T2);
            \draw[thick] (ket) to [out = 0, in = 180] (T2);
            \draw[white, line width=3pt] (ket) to [out = 270, in = 90] (T3);
            \draw[thick] (ket) to [out = 270, in = 90] node[midway, below left =1]{$\chi_B$} (T3);
            \draw[thick] (ket) to [out = 180, in = 0] (T4);

            \draw[line width=0.5mm] (C1) to node[midway, above=0]{$\chi_E$} (T1);
            \draw[line width=0.5mm] (T1) to (C2);
            \draw[line width=0.5mm] (C2) to (T2);
            \draw[line width=0.5mm] (T2) to (C3);
            \draw[line width=0.5mm] (C3) to (T3);
            \draw[line width=0.5mm] (T3) to (C4);
            \draw[line width=0.5mm] (C4) to (T4);
            \draw[line width=0.5mm] (T4) to (C1);

            \node[above left=1] at (ket) {$\psi_{x,y}$};
            \draw[thick, fill=googleB] (ket) circle (\tensorSize);
            \node[below right=1] at (bra) {$\psi^{*}_{x,y}$};
            \draw[thick, fill=googleB] (bra) circle (\tensorSize);

		\node[above left=5] at (C1) {$C_{1}$};
		\draw[thick, fill = gray] (C1) circle (\tensorSize);

            \node[above=6] at (T1) {$T_{1}$};
		\draw[thick, fill = gray] (T1) circle (\tensorSize);

            \node[above right=5] at (C2) {$C_{2}$};
		\draw[thick, fill = gray] (C2) circle (\tensorSize);

            \node[right=6] at (T2) {$T_{2}$};
		\draw[thick, fill = gray] (T2) circle (\tensorSize);

            \node[below right=5] at (C3) {$C_{3}$};
		\draw[thick, fill = gray] (C3) circle (\tensorSize);

            \node[below=6] at (T3) {$T_{3}$};
            \draw[thick, fill = gray] (T3) circle (\tensorSize);

            \node[below left=5] at (C4) {$C_{4}$};
		\draw[thick, fill = gray] (C4) circle (\tensorSize);

            \node[left=6] at (T4) {$T_{4}$};
		\draw[thick, fill = gray] (T4) circle (\tensorSize);
        
	\end{scope}
    \end{tikzpicture}
    \caption{Norm of the PEPS state vector computed from effective fixed-point environment tensors in the conventional double-layer CTMRG. The environment bond dimension $\chi_E$ controls the degree of approximations in the contractions of the infinite lattice.}
    \label{fig:CTMRG_Method_2}
\end{figure}
In this expression, the product runs over the sites in the infinite lattice, and the \emph{tensor trace} ($\operatorname{tTr}$) implements the contraction of all virtual indices, which is represented graphically in the second line.

Performing this contraction is computationally hard already in the case of finite PEPS, but generically impossible in the case of an \textit{infinite}, translational invariant PEPS ansatz. 
There are several approaches of approximately contracting this network, that can be broadly categorized as scale-transformation or coarse-graining methods~\cite{Levin07TRG,Evenbly15TNR} and power-type-methods like VUMPS~\cite{ZaunerStauber2018} or the \emph{corner transfer matrix renormalization group} (CTMRG)~\cite{Nishino1996,Nishino1997,Orus2009, Corboz2010, Corboz2014, Fishman2018, vanAlphen2024}. 
The CTMRG method is used to compute effective fixed-point 
environment tensors, cf.~Fig.~\ref{fig:CTMRG_Method_2}, that describe semi-infinite parts of the contracted network and which are generated in an iterative 
absorption scheme that for natural models converges to a fixed point. 
Employing those environment tensors, the norm of the state vector can then be evaluated according to 
Fig.~\ref{fig:CTMRG_Method_2}.
Detailed descriptions of the double-layer CTMRG method can be found in Refs.~\cite{Orus2009,Corboz2014,Bruognolo2021,Naumann2024}.
The unavoidable approximations in the contraction of the infinite PEPS network are controlled by a second refinement parameter, the \emph{environment bond dimension} $\chi_E$. 
In this paper, we exclusively focus on the directional CTMRG formulation, which does not assume any lattice symmetry and is thus capable of treating non-symmetric Hamiltonians and capturing symmetry-broken states.

\subsection{Increasing numerical efficiency: split-CTMRG algorithm}
\label{method:splitCTMRG_description}

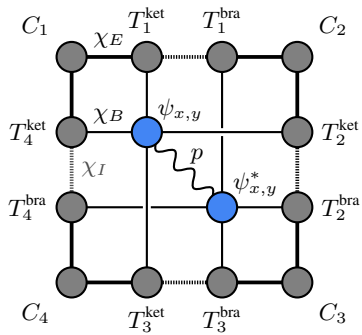
\begin{figure}[tbh]
    \centering
    \begin{tikzpicture}
    \def\tensorSep{1}
    \def\tensorSize{0.2}

    \begin{scope}[shift = {(0.0, 0.0)}]
        \coordinate (ket) at ({-0.5*\tensorSep}, {+0.5*\tensorSep});
        \coordinate (bra) at ({+0.5*\tensorSep}, {-0.5*\tensorSep});

        \coordinate (C1) at ({-1.5*\tensorSep}, {+1.5*\tensorSep});
        \coordinate (T1ket) at ({-0.5*\tensorSep}, {+1.5*\tensorSep});
        \coordinate (T1bra) at ({+0.5*\tensorSep}, {+1.5*\tensorSep});
        \coordinate (C2) at ({+1.5*\tensorSep}, {+1.5*\tensorSep});
        \coordinate (T2ket) at ({+1.5*\tensorSep}, {+0.5*\tensorSep});
        \coordinate (T2bra) at ({+1.5*\tensorSep}, {-0.5*\tensorSep});
        \coordinate (C3) at ({+1.5*\tensorSep}, {-1.5*\tensorSep});
        \coordinate (T3bra) at ({+0.5*\tensorSep}, {-1.5*\tensorSep});
        \coordinate (T3ket) at ({-0.5*\tensorSep}, {-1.5*\tensorSep});
        \coordinate (C4) at ({-1.5*\tensorSep}, {-1.5*\tensorSep});
        \coordinate (T4bra) at ({-1.5*\tensorSep}, {-0.5*\tensorSep});
        \coordinate (T4ket) at ({-1.5*\tensorSep}, {+0.5*\tensorSep});

        \draw[thick, decorate, decoration={coil, aspect=0}] (ket) to node[midway, above right=-1]{$p$} (bra);

        \draw[thick] (bra) to (T1bra);
        \draw[thick] (bra) to (T2bra);
        \draw[thick] (bra) to (T3bra);
        \draw[thick] (bra) to (T4bra);

        \draw[thick] (ket) to (T1ket);
        \draw[white, line width=3pt] (ket) to (T2ket);
        \draw[thick] (ket) to (T2ket);
        \draw[white, line width=3pt] (ket) to (T3ket);
        \draw[thick] (ket) to (T3ket);
        \draw[thick] (ket) to node[midway, above=0]{$\chi_B$} (T4ket);

        \draw[line width=0.5mm] (C1) to node[midway, above=0]{$\chi_E$} (T1ket);
        \draw[line width=0.5mm, dashed, dash pattern=on 0.5pt off 0.5pt] (T1ket) to (T1bra);
        \draw[line width=0.5mm] (T1bra) to (C2);
        \draw[line width=0.5mm] (C2) to (T2ket);
        \draw[line width=0.5mm, dashed, dash pattern=on 0.5pt off 0.5pt] (T2ket) to (T2bra);
        \draw[line width=0.5mm] (T2bra) to (C3);
        \draw[line width=0.5mm] (C3) to (T3bra);
        \draw[line width=0.5mm, dashed, dash pattern=on 0.5pt off 0.5pt] (T3bra) to (T3ket);
        \draw[line width=0.5mm] (T3ket) to (C4);
        \draw[line width=0.5mm] (C4) to (T4bra);
        \draw[line width=0.5mm, dashed, dash pattern=on 0.5pt off 0.5pt, black!70] (T4bra) to node[midway, right=0]{$\chi_I$} (T4ket);
        \draw[line width=0.5mm] (T4ket) to (C1);

        \node[above right=1] at (ket) {$\psi_{x,y}$};
        \draw[thick, fill=googleB] (ket) circle (\tensorSize);
        \node[above right=1] at (bra) {$\psi^{*}_{x,y}$};
        \draw[thick, fill=googleB] (bra) circle (\tensorSize);

        \node[above left=5] at (C1) {$C_{1}$};
        \draw[thick, fill = gray] (C1) circle (\tensorSize);

        \node[above=6] at (T1ket) {$T_{1}^\text{ket}$};
        \draw[thick, fill = gray] (T1ket) circle (\tensorSize);

        \node[above=6] at (T1bra) {$T_{1}^\text{bra}$};
        \draw[thick, fill = gray] (T1bra) circle (\tensorSize);

        \node[above right=5] at (C2) {$C_{2}$};
        \draw[thick, fill = gray] (C2) circle (\tensorSize);

        \node[right=6] at (T2ket) {$T_{2}^\text{ket}$};
        \draw[thick, fill = gray] (T2ket) circle (\tensorSize);

        \node[right=6] at (T2bra) {$T_{2}^\text{bra}$};
        \draw[thick, fill = gray] (T2bra) circle (\tensorSize);

        \node[below right=5] at (C3) {$C_{3}$};
        \draw[thick, fill = gray] (C3) circle (\tensorSize);

        \node[below=6] at (T3bra) {$T_{3}^\text{bra}$};
        \draw[thick, fill = gray] (T3bra) circle (\tensorSize);

        \node[below=6] at (T3ket) {$T_{3}^\text{ket}$};
        \draw[thick, fill = gray] (T3ket) circle (\tensorSize);

        \node[below left=5] at (C4) {$C_{4}$};
        \draw[thick, fill = gray] (C4) circle (\tensorSize);

        \node[left=6] at (T4bra) {$T_{4}^\text{bra}$};
        \draw[thick, fill = gray] (T4bra) circle (\tensorSize);

        \node[left=6] at (T4ket) {$T_{4}^\text{ket}$};
        \draw[thick, fill = gray] (T4ket) circle (\tensorSize);
    \end{scope}
\end{tikzpicture}
  
    \caption{Definition of the fixed-point environment tensors for a PEPS site $\psi_{x,y}$ at position $[x, y]$ in the split-CTMRG algorithm. Here $\chi_B$ denotes the PEPS bulk bond dimension, $\chi_E$ the CTMRG environment refinement parameter and $p$ the physical dimension. The additional link introduced in this work between the split transfer tensors $T^\text{bra}$ and $T^\text{ket}$ is indicated as dashed lines with the new refinement parameter $\chi_I$. We note that the splitting also allows for a gauge transformation to be inserted on these contracted legs.}
    \label{fig:split-transfer-ctmrg}
\end{figure}

In this section, we introduce a variation of the well-established CTMRG scheme that utilizes a different, yet related, set of effective environment tensors, which can be created in an algorithm with a lower computational complexity. 
The new set of environment tensors is illustrated in Fig.~\ref{fig:split-transfer-ctmrg}.
It differs from the conventional setup, outlined in Sec.~\ref{sub:infinitePEPSandCTMRG}, in that we 
define two different transfer tensors $T$ for the \textit{bra}- and \textit{ket}-layer of the double-layer network, so for the dual and state vectors.
As usual, these are defined for every local PEPS tensor and for each direction (top, right, bottom, left). 
The two transfer tensors for the \textit{bra}- and \textit{ket}-layers are 
connected by a new virtual bond of dimension $\chi_I$, the \textit{interlayer} environment bond dimension. 
Graphically, we represent this virtual interlayer bond as a dashed line, as opposed to the regular, solid environment links with bond dimension $\chi_E$. 
The physical index is illustrated with a wiggly line. 
A contraction of the \textit{bra}- and \textit{ket}-layer transfer tensors over this new virtual interlayer bond yields the conventional environment tensors, as defined in the conventional CTMRG.

The heuristic advantage of the definition of the split effective environment tensors is of numerical nature. 
In line with the general mindset of tensor networks, the new method goes further with decomposing larger tensors into structured smaller ones. 
Following this line of thought, the four-index tensors $T$ of the conventional CTMRG are decomposed into products of two three-index tensors, which will generically result in more efficient contractions. 
Furthermore, the separation into \textit{bra}- and \textit{ket}-layer also naturally suggests a sequential absorption and renormalization of the two layers, which can additionally reduce the computational cost.

We now proceed with defining the full split-CTMRG algorithm. 
It produces environment tensors that can be used to very accurately approximate local observables, while significantly reducing the algorithmic complexity of generating them compared to the conventional CTMRG algorithm.
Here, we present the steps of the algorithm corresponding to a \textit{left move} only. A complete (directional) split-CTMRG step, however, includes analogous moves in the other three directions as well. These full split-CTMRG steps are then iterated until the environment tensors converge to a fixed-point.

The left move of the split-CTMRG algorithm consists of the absorption of tensors into the left environment tensors, and the subsequent use of projectors to reduce the connectivity and bond dimension. 
In contrast to the conventional CTMRG algorithm however, the absorption and projection is done separately for the \textit{bra}- and \textit{ket}-layer. 
In the following we show this procedure of absorption and projection in detail first for the corner- and then for the transfer tensors. 
A full, diagrammatic representation of a complete left move can be found in App.~\ref{app:construction of projectors}. 
We discuss how to obtain the optimal choice of projectors in the next section.

\paragraph*{Corner tensors.} When updating the corner tensors $C_1$ and $C_4$ of the split-CTMRG environment (cf.\ Fig.~\ref{fig:split-transfer-ctmrg}) during the left move, we sequentially absorb the transfer tensors of the \textit{ket}- and \textit{bra}-layer into them, and project after each absorption. 
Concretely, the procedure for $C_1$ in a left move is given by
\begin{equation}
    \begin{split}
    \scalebox{0.92}{        \begin{tikzpicture}
		\def\tensorSep{1.29}
	    \def\tensorSize{0.18}

		\begin{scope}[shift = {(-3, 0)}]
            \coordinate (ket) at ({-0.5*\tensorSep}, {0*\tensorSep});
            \coordinate (bra) at ({+0.5*\tensorSep}, {-1.5*\tensorSep});
  
		      \coordinate (C1) at ({-1.5*\tensorSep}, {+1.5*\tensorSep});
            \coordinate (T1ket) at ({-0.5*\tensorSep}, {+1.5*\tensorSep});
            \coordinate (T1bra) at ({+0.5*\tensorSep}, {+1.5*\tensorSep});

            \coordinate (ProjT4ket1Top) at ({-1.0*\tensorSep}, {+1.2*\tensorSep});
            \coordinate (ProjT4ket2Top) at ({-0.25*\tensorSep}, {+0.9*\tensorSep});
            \coordinate (ProjT4ket2Bottom) at ({-0.25*\tensorSep}, {+0.6*\tensorSep});
            \coordinate (ProjT4ket1Bottom) at ({-1.0*\tensorSep}, {+0.3*\tensorSep});

            \draw[line width=0.5mm] (C1) to (T1ket);
            \draw[line width=0.5mm, dashed, dash pattern=on 0.5pt off 0.5pt] (T1ket) to (T1bra);
            \draw[line width=0.5mm,] (T1bra) to ($(T1bra) + ({0.5*\tensorSep}, {0})$);

            \draw[line width=0.5mm,] (C1) to ($(C1) + (0, -0.25*\tensorSep)$);
            \draw[thick] (T1ket) to ($(T1ket) + (0, -0.25*\tensorSep)$);
            \draw[line width=0.5mm,] (ProjT4ket1Top) to ($(ProjT4ket1Top) + (0, -0.225*\tensorSep)$);
            \draw[thick] (T1bra) to ($(T1bra) + (0, -0.525*\tensorSep)$);
            \draw[line width=0.5mm,] (ProjT4ket2Top) to (ProjT4ket2Bottom);
            
		\node[above left=5] at (C1) {$C_{1_{[x, y]}}$};
		\draw[thick, fill = gray] (C1) circle (\tensorSize);

            \node[above=6] at (T1ket) {$T_{1_{[x, y]}}^\text{ket}$};
			\draw[thick, fill = gray] (T1ket) circle (\tensorSize);

            \node[above=6] at (T1bra) {$T_{1_{[x, y]}}^\text{bra}$};
			\draw[thick, fill = gray] (T1bra) circle (\tensorSize);

            \node[isosceles triangle, isosceles triangle apex angle=145, minimum size=8, draw, rotate=-90, fill=googleG] at (ProjT4ket1Top) {};
            
            \node[isosceles triangle, isosceles triangle apex angle=155, minimum size=8, draw, rotate=-90, fill=googleBG] at (ProjT4ket2Top) {};
		\end{scope}

        \begin{scope}[shift = {(-0.75, +1.5*\tensorSep)}]
            \node at (0, 0) {$=$};
        \end{scope}

        \begin{scope}[shift = {(+2.25, 0)}]
		      \coordinate (C1) at ({-1.5*\tensorSep}, {+1.5*\tensorSep});
            \coordinate (C4) at ({-1.5*\tensorSep}, {-3*\tensorSep});
            \coordinate (T4bra) at ({-1.5*\tensorSep}, {-1.5*\tensorSep});
            \coordinate (T4ket) at ({-1.5*\tensorSep}, {0*\tensorSep});

            \draw[line width=0.5mm,] (C1) to ($(C1) + ({0.5*\tensorSep}, {0})$);
            \draw[line width=0.5mm,] (C1) to ($(C1) + ({0}, {-0.5*\tensorSep})$);

			\node[above right=4] at (C1) {$C'_{1_{[x, y+1]}}$};
			\draw[thick, fill = gray] (C1) circle (\tensorSize);

		\end{scope}
    \end{tikzpicture}}
    \end{split}.
    \label{eq:splitCTMRG_C1}
\end{equation}
Here, we first absorb the transfer tensors from the \textit{ket}-layer, $T_1^{\text{ket}}$, into the corner tensors $C_1$ and project the enlarged index from $\chi_E \cdot \chi_B$ back to $\chi_E$ with the help of the green projector illustrated in Eq.~\eqref{eq:splitCTMRG_C1}. 
After that, we proceed with the absorption of the transfer tensors of the \textit{bra}-layer, $T_1^{\text{bra}}$, followed by a truncation using the teal colored projectors in Eq.~\eqref{eq:splitCTMRG_C1} to project the enlarged index of the corner tensor $C_1$ back to their original size $\chi_E$. 
The procedure for the left move for $C_4$ works completely analogously and is given by
\begin{equation}
    \begin{split}
    \scalebox{0.92}{    \begin{tikzpicture}
		\def\tensorSep{1.29}
	    \def\tensorSize{0.18}

		\begin{scope}[shift = {(-3, 0)}]
            \coordinate (ket) at ({-0.5*\tensorSep}, {0*\tensorSep});
            \coordinate (bra) at ({+0.5*\tensorSep}, {-1.5*\tensorSep});
  
		      \coordinate (C1) at ({-1.5*\tensorSep}, {+1.5*\tensorSep});
            \coordinate (T1ket) at ({-0.5*\tensorSep}, {+1.5*\tensorSep});
            \coordinate (T1bra) at ({+0.5*\tensorSep}, {+1.5*\tensorSep});
            \coordinate (T3bra) at ({+0.5*\tensorSep}, {-3*\tensorSep});
            \coordinate (T3ket) at ({-0.5*\tensorSep}, {-3*\tensorSep});
            \coordinate (C4) at ({-1.5*\tensorSep}, {-3*\tensorSep});
            \coordinate (T4bra) at ({-1.5*\tensorSep}, {-1.5*\tensorSep});
            \coordinate (T4ket) at ({-1.5*\tensorSep}, {0*\tensorSep});

            \coordinate (ProjT4ket1Top) at ({-1.0*\tensorSep}, {+1.2*\tensorSep});
            \coordinate (ProjT4ket2Top) at ({-0.25*\tensorSep}, {+0.9*\tensorSep});
            \coordinate (ProjT4ket2Bottom) at ({-0.25*\tensorSep}, {+0.6*\tensorSep});
            \coordinate (ProjT4ket1Bottom) at ({-1.0*\tensorSep}, {+0.3*\tensorSep});

            \coordinate (ProjKetBra1Top) at ({-1.0*\tensorSep}, {-0.3*\tensorSep});
            \coordinate (ProjKetBra2Top) at ({-0.25*\tensorSep}, {-0.6*\tensorSep});
            \coordinate (ProjKetBra2Bottom) at ({-0.25*\tensorSep}, {-0.9*\tensorSep});
            \coordinate (ProjKetBra1Bottom) at ({-1.0*\tensorSep}, {-1.2*\tensorSep});

            \coordinate (ProjT4bra1Top) at ({-1.0*\tensorSep}, {-1.8*\tensorSep});
            \coordinate (ProjT4bra2Top) at ({-0.25*\tensorSep}, {-2.1*\tensorSep});
            \coordinate (ProjT4bra2Bottom) at ({-0.25*\tensorSep}, {-2.4*\tensorSep});
            \coordinate (ProjT4bra1Bottom) at ({-1.0*\tensorSep}, {-2.7*\tensorSep});

            \draw[line width=0.5mm,] (T3bra) to ($(T3bra) + ({0.5*\tensorSep}, {0})$);
            \draw[thick, dashed, dash pattern=on 0.5pt off 0.5pt] (T3bra) to (T3ket);
            \draw[line width=0.5mm,] (T3ket) to (C4);

            \draw[line width=0.5mm,] (ProjT4bra2Top) to (ProjT4bra2Bottom);
            \draw[line width=0.5mm,] (ProjT4bra1Bottom) to ($(ProjT4bra1Bottom) + (0, 0.225*\tensorSep)$);
            \draw[line width=0.5mm,] (C4) to ($(C4) + (0, 0.25*\tensorSep)$);
            \draw[thick] (T3ket) to ($(T3ket) + (0, 0.25*\tensorSep)$);
            \draw[thick] (T3bra) to ($(T3bra) + (0, 0.525*\tensorSep)$);


            \node[below=6] at (T3bra) {$T_{3_{[x, y]}}^\text{bra}$};
			\draw[thick, fill = gray] (T3bra) circle (\tensorSize);

            \node[below=6] at (T3ket) {$T_{3_{[x, y]}}^\text{ket}$};
			\draw[thick, fill = gray] (T3ket) circle (\tensorSize);

            \node[below left=5] at (C4) {$C_{4_{[x, y]}}$};
			\draw[thick, fill = gray] (C4) circle (\tensorSize);

            \node[isosceles triangle, isosceles triangle apex angle=155, minimum size=8, draw, rotate=90, fill=googleBG] at (ProjT4bra2Bottom) {};
            \node[isosceles triangle, isosceles triangle apex angle=145, minimum size=8, draw, rotate=90, fill=googleG] at (ProjT4bra1Bottom) {};
		\end{scope}

        \begin{scope}[shift = {(-0.75, -3*\tensorSep)}]
            \node at (0, 0) {$=$};
        \end{scope}

        \begin{scope}[shift = {(+2.25, 0)}]
		      \coordinate (C1) at ({-1.5*\tensorSep}, {+1.5*\tensorSep});
            \coordinate (C4) at ({-1.5*\tensorSep}, {-3*\tensorSep});
            \coordinate (T4bra) at ({-1.5*\tensorSep}, {-1.5*\tensorSep});
            \coordinate (T4ket) at ({-1.5*\tensorSep}, {0*\tensorSep});
        
            \draw[line width=0.5mm,] (C4) to ($(C4) + ({0.5*\tensorSep}, {0})$);
            \draw[line width=0.5mm,] (C4) to ($(C4) + ({0}, {0.5*\tensorSep})$);

            \node[below right=4] at (C4) {$C'_{4_{[x, y+1]}}$};
			\draw[thick, fill = gray] (C4) circle (\tensorSize);
		\end{scope}
    \end{tikzpicture}}
    \end{split}.
    \label{eq:splitCTMRG_C4}
\end{equation}

\paragraph*{Transfer tensors.} 
In order to update the transfer tensors $T_4^{\text{ket}}$ and $T_4^{\text{bra}}$, shown in Fig.~\ref{fig:split-transfer-ctmrg}, we absorb the local tensor from the \textit{ket}-layer into $T_4^{\text{ket}}$ and only afterwards the local tensor from the \textit{bra}-layer into $T_4^{\text{bra}}$. 
Unlike in the conventional CTMRG algorithm, the physical index now takes active part in the renormalization, i.e.,  the projection step, as it connects the \textit{bra}- and \textit{ket}-layer PEPS tensors.
Furthermore, it is necessary to keep consistent basis choices for the environment tensors such that their contraction for the calculation of local observables is well defined and meaningful.

Let us first examine the expression for the 
updated $T_4^{\text{ket}}$ in a left move, given by
\begin{equation}
    \begin{split}
    \scalebox{0.92}{ \begin{tikzpicture}
		\def\tensorSep{1.29}
	    \def\tensorSize{0.18}

		\begin{scope}[shift = {(-3, 0)}]
            \coordinate (ket) at ({-0.5*\tensorSep}, {0*\tensorSep});
            \coordinate (bra) at ({+0.5*\tensorSep}, {-1.5*\tensorSep});
  
		      \coordinate (C1) at ({-1.5*\tensorSep}, {+1.5*\tensorSep});
            \coordinate (T1ket) at ({-0.5*\tensorSep}, {+1.5*\tensorSep});
            \coordinate (T1bra) at ({+0.5*\tensorSep}, {+1.5*\tensorSep});
            \coordinate (T3bra) at ({+0.5*\tensorSep}, {-3*\tensorSep});
            \coordinate (T3ket) at ({-0.5*\tensorSep}, {-3*\tensorSep});
            \coordinate (C4) at ({-1.5*\tensorSep}, {-3*\tensorSep});
            \coordinate (T4bra) at ({-1.5*\tensorSep}, {-1.5*\tensorSep});
            \coordinate (T4ket) at ({-1.5*\tensorSep}, {0*\tensorSep});

            \coordinate (ProjT4ket1Top) at ({-1.0*\tensorSep}, {+1.2*\tensorSep});
            \coordinate (ProjT4ket2Top) at ({-0.25*\tensorSep}, {+0.9*\tensorSep});
            \coordinate (ProjT4ket2Bottom) at ({-0.25*\tensorSep}, {+0.6*\tensorSep});
            \coordinate (ProjT4ket1Bottom) at ({-1.0*\tensorSep}, {+0.3*\tensorSep});

            \coordinate (ProjKetBra1Top) at ({-1.0*\tensorSep}, {-0.3*\tensorSep});
            \coordinate (ProjKetBra2Top) at ({-0.25*\tensorSep}, {-0.6*\tensorSep});
            \coordinate (ProjKetBra2Bottom) at ({-0.25*\tensorSep}, {-0.9*\tensorSep});
            \coordinate (ProjKetBra1Bottom) at ({-1.0*\tensorSep}, {-1.2*\tensorSep});

            \coordinate (ProjT4bra1Top) at ({-1.0*\tensorSep}, {-1.8*\tensorSep});
            \coordinate (ProjT4bra2Top) at ({-0.25*\tensorSep}, {-2.1*\tensorSep});
            \coordinate (ProjT4bra2Bottom) at ({-0.25*\tensorSep}, {-2.4*\tensorSep});
            \coordinate (ProjT4bra1Bottom) at ({-1.0*\tensorSep}, {-2.7*\tensorSep});

            \draw[thick] (ket) to (T4ket);

            \draw[line width=0.5mm,] (ProjT4ket2Top) to (ProjT4ket2Bottom);
            \draw[line width=0.5mm,] (ProjT4ket1Bottom) to ($(ProjT4ket1Bottom) + (0, 0.225*\tensorSep)$);
            \draw[thick] (ket) to ($(ket) + (0, 0.25*\tensorSep)$);
            \draw[line width=0.5mm,] (T4ket) to ($(T4ket) + (0, 0.25*\tensorSep)$);

            \draw[thick] (ket) to ($(ket) + (0, -0.25*\tensorSep)$);
            \draw[line width=0.5mm, dashed, dash pattern=on 0.5pt off 0.5pt] (T4ket) to ($(T4ket) + (0, -0.25*\tensorSep)$);

            \draw[line width=0.5mm, dashed, dash pattern=on 0.5pt off 0.5pt] (ProjKetBra1Top) to ($(ProjKetBra1Top) + (0, -0.225*\tensorSep)$);
            \draw[thick] ($(ProjT4ket2Bottom) + (0.75*\tensorSep, -0.075*\tensorSep)$) to ($(ProjKetBra2Top) + (0.75*\tensorSep, 0.075*\tensorSep)$);
            
            \draw[thick, decorate, decoration={coil, aspect=0}] ($(ket) + (0.4*\tensorSep, -0.6*\tensorSep)$) to (ket);

            \draw[line width=0.5mm, dashed, dash pattern=on 0.5pt off 0.5pt] (ProjKetBra2Top) to (ProjKetBra2Bottom);

            \draw[white, line width=3pt] (ket) to ($(ket) + ({1.5*\tensorSep}, {0})$);
            \draw[thick] (ket) to ($(ket) + ({1.5*\tensorSep}, {0})$);

            \node[above right=3] at (ket) {$\psi_{x,y}$};
            \draw[thick, fill = googleB] (ket) circle (\tensorSize);
            
            \node[left=6] at (T4ket) {$T_{4_{[x, y]}}^\text{ket}$};
			\draw[thick, fill = gray] (T4ket) circle (\tensorSize);

            \node[isosceles triangle, isosceles triangle apex angle=155, minimum size=8, draw, rotate=90, fill=googleBG] at (ProjT4ket2Bottom) {};
            \node[isosceles triangle, isosceles triangle apex angle=145, minimum size=8, draw, rotate=90, fill=googleG] at (ProjT4ket1Bottom) {};

            \node[isosceles triangle, isosceles triangle apex angle=145, minimum size=8, draw, rotate=-90, fill=googleY] at (ProjKetBra1Top) {};
            \node[isosceles triangle, isosceles triangle apex angle=155, minimum size=8, draw, rotate=-90, fill=googleR] at (ProjKetBra2Top) {};
            
		\end{scope}

        \begin{scope}[shift = {(-0.75, 0*\tensorSep)}]
            \node at (0, 0) {\large $=$};
        \end{scope}

        \begin{scope}[shift = {(+2.25, 0)}]
		\coordinate (C1) at ({-1.5*\tensorSep}, {+1.5*\tensorSep});
            \coordinate (C4) at ({-1.5*\tensorSep}, {-3*\tensorSep});
            \coordinate (T4bra) at ({-1.5*\tensorSep}, {-1.5*\tensorSep});
            \coordinate (T4ket) at ({-1.5*\tensorSep}, {0*\tensorSep});
        
            \draw[line width=0.5mm, dashed, dash pattern=on 0.5pt off 0.5pt] (T4ket) to ($(T4ket) + ({0}, {-0.5*\tensorSep})$);            
            \draw[line width=0.5mm,] (T4ket) to ($(T4ket) + ({0}, {0.5*\tensorSep})$);

            \draw[thick] (T4ket) to ($(T4ket) + ({0.5*\tensorSep}, {0})$);


            \node[above right=4] at (T4ket) {$T_{4_{[x, y+1]}}^{'\text{ket}}$};
		\draw[thick, fill = gray] (T4ket) circle (\tensorSize);

		\end{scope}
    \end{tikzpicture}}
    \end{split}.
    \label{eq:splitCTMRG_Tket}
\end{equation}
Here, the absorption of the local \textit{ket}-layer PEPS tensor is followed by a first renormalization step, in which we use the yellow (green) projectors to truncate the vertical legs from bond dimension $\chi_I \cdot \chi_B$ ($\chi_E \cdot \chi_B$) back to $\chi_I$ ($\chi_E$). 
After this first projection, the physical leg that connects to the \textit{bra}-layer sectionsremains open and is truncated in the next step. 
During the absorption of the \textit{bra}-layer no tensors are explicitly absorbed into $T_4^{\text{ket}}$, but due to the remaining physical leg and the necessity to keep a consistent basis for the resulting new environment tensors, we apply the second layer of red (teal) projectors to truncate to the final desired bond dimensions $\chi_I$ ($\chi_E$). 
This concludes a left move for the tensor $T_4^{\text{ket}}$. 
We highlight the unusual fact, that the red projector in Eq.~\eqref{eq:splitCTMRG_Tket} truncates the physical space together with the virtual spaces from the local and environment tensors into a new virtual index of dimension $\chi_I$. 
The physical index is never explicitly projected in the conventional CTMRG algorithm, as it is contracted while absorbing the \textit{bra}- and \textit{ket}-layer PEPS tensors simultaneously.

The left move of the split-CTMRG algorithm for the transfer tensor from the \textit{bra}-layer $T_4^{\text{bra}}$ works along very similar lines,
\begin{equation}
    \begin{split}
    \scalebox{0.92}{    \begin{tikzpicture}
		\def\tensorSep{1.29}
	    \def\tensorSize{0.18}

		\begin{scope}[shift = {(-3, 0)}]
            \coordinate (ket) at ({-0.5*\tensorSep}, {0*\tensorSep});
            \coordinate (bra) at ({+0.5*\tensorSep}, {-1.5*\tensorSep});
  
		      \coordinate (C1) at ({-1.5*\tensorSep}, {+1.5*\tensorSep});
            \coordinate (T1ket) at ({-0.5*\tensorSep}, {+1.5*\tensorSep});
            \coordinate (T1bra) at ({+0.5*\tensorSep}, {+1.5*\tensorSep});
            \coordinate (T3bra) at ({+0.5*\tensorSep}, {-3*\tensorSep});
            \coordinate (T3ket) at ({-0.5*\tensorSep}, {-3*\tensorSep});
            \coordinate (C4) at ({-1.5*\tensorSep}, {-3*\tensorSep});
            \coordinate (T4bra) at ({-1.5*\tensorSep}, {-1.5*\tensorSep});
            \coordinate (T4ket) at ({-1.5*\tensorSep}, {0*\tensorSep});

            \coordinate (ProjT4ket1Top) at ({-1.0*\tensorSep}, {+1.2*\tensorSep});
            \coordinate (ProjT4ket2Top) at ({-0.25*\tensorSep}, {+0.9*\tensorSep});
            \coordinate (ProjT4ket2Bottom) at ({-0.25*\tensorSep}, {+0.6*\tensorSep});
            \coordinate (ProjT4ket1Bottom) at ({-1.0*\tensorSep}, {+0.3*\tensorSep});

            \coordinate (ProjKetBra1Top) at ({-1.0*\tensorSep}, {-0.3*\tensorSep});
            \coordinate (ProjKetBra2Top) at ({-0.25*\tensorSep}, {-0.6*\tensorSep});
            \coordinate (ProjKetBra2Bottom) at ({-0.25*\tensorSep}, {-0.9*\tensorSep});
            \coordinate (ProjKetBra1Bottom) at ({-1.0*\tensorSep}, {-1.2*\tensorSep});

            \coordinate (ProjT4bra1Top) at ({-1.0*\tensorSep}, {-1.8*\tensorSep});
            \coordinate (ProjT4bra2Top) at ({-0.25*\tensorSep}, {-2.1*\tensorSep});
            \coordinate (ProjT4bra2Bottom) at ({-0.25*\tensorSep}, {-2.4*\tensorSep});
            \coordinate (ProjT4bra1Bottom) at ({-1.0*\tensorSep}, {-2.7*\tensorSep});


            \draw[thick] (bra) to ($(bra) + ({0.5*\tensorSep}, {0})$);
            \draw[thick] (bra) to (T4bra);

            \draw[line width=0.5mm, dashed, dash pattern=on 0.5pt off 0.5pt] (ProjKetBra2Top) to (ProjKetBra2Bottom);
            \draw[thick, decorate, decoration={coil, aspect=0}] ($(bra) + (-0.4*\tensorSep, 0.6*\tensorSep)$) to (bra);
            \draw[thick] (bra) to ($(ProjKetBra2Bottom) + (0.75*\tensorSep, -0.075*\tensorSep)$);
            \draw[line width=0.5mm, dashed, dash pattern=on 0.5pt off 0.5pt] (ProjKetBra1Bottom) to ($(ProjKetBra1Bottom) + (0, 0.225*\tensorSep)$);
            \draw[line width=0.5mm, dashed, dash pattern=on 0.5pt off 0.5pt] (T4bra) to ($(T4bra) + (0, 0.25*\tensorSep)$);

            \draw[line width=0.5mm,] (T4bra) to ($(T4bra) + (0, -0.25*\tensorSep)$);
            
            \draw[white, line width=3pt] ($(ProjKetBra1Bottom) + (0.5*\tensorSep, -0.075*\tensorSep)$) to ($(ProjT4bra1Top) + (0.5*\tensorSep, 0.075*\tensorSep)$);
            \draw[thick] ($(ProjKetBra1Bottom) + (0.5*\tensorSep, -0.075*\tensorSep)$) to ($(ProjT4bra1Top) + (0.5*\tensorSep, 0.075*\tensorSep)$);

            \draw[line width=0.5mm,] (ProjT4bra1Top) to ($(ProjT4bra1Top) + (0, -0.225*\tensorSep)$);
            \draw[thick] (bra) to ($(ProjT4bra2Top) + (0.75*\tensorSep, 0.075*\tensorSep)$);
            
            \draw[line width=0.5mm,] (ProjT4bra2Top) to (ProjT4bra2Bottom);

            
            \node[above right=3] at (bra) {$\psi^{*}_{x,y}$};
            \draw[thick, fill = googleB] (bra) circle (\tensorSize);

            \node[left=6] at (T4bra) {$T_{4_{[x, y]}}^\text{bra}$};
			\draw[thick, fill = gray] (T4bra) circle (\tensorSize);
            
            \node[isosceles triangle, isosceles triangle apex angle=155, minimum size=8, draw, rotate=90, fill=googleR] at (ProjKetBra2Bottom) {};
            \node[isosceles triangle, isosceles triangle apex angle=145, minimum size=8, draw, rotate=90, fill=googleY] at (ProjKetBra1Bottom) {};

            \node[isosceles triangle, isosceles triangle apex angle=145, minimum size=8, draw, rotate=-90, fill=googleG] at (ProjT4bra1Top) {};
            \node[isosceles triangle, isosceles triangle apex angle=155, minimum size=8, draw, rotate=-90, fill=googleBG] at (ProjT4bra2Top) {};
            
		\end{scope}

        \begin{scope}[shift = {(-0.75, -1.5*\tensorSep)}]
            \node at (0, 0) {\large $=$};
        \end{scope}

        \begin{scope}[shift = {(+2.25, 0)}]
		  \coordinate (C1) at ({-1.5*\tensorSep}, {+1.5*\tensorSep});
            \coordinate (C4) at ({-1.5*\tensorSep}, {-3*\tensorSep});
            \coordinate (T4bra) at ({-1.5*\tensorSep}, {-1.5*\tensorSep});
            \coordinate (T4ket) at ({-1.5*\tensorSep}, {0*\tensorSep});

            \draw[thick] (T4bra) to ($(T4bra) + ({0.5*\tensorSep}, {0})$);
            \draw[line width=0.5mm, dashed, dash pattern=on 0.5pt off 0.5pt] (T4bra) to ($(T4bra) + ({0}, {0.5*\tensorSep})$);
            \draw[line width=0.5mm,] (T4bra) to ($(T4bra) + ({0}, {-0.5*\tensorSep})$);

            \node[below right=4] at (T4bra) {$T_{4_{[x, y+1]}}^{'\text{bra}}$};
			\draw[thick, fill = gray] (T4bra) circle (\tensorSize);

		\end{scope}
    \end{tikzpicture}}
    \end{split}.
    \label{eq:splitCTMRG_Tbra}
\end{equation}
This time, there is no tensor explicitly absorbed into $T_4^{\text{bra}}$ during the absorption of the \textit{ket}-layer, so that we apply the green (yellow) projectors in Eq.~\eqref{eq:splitCTMRG_Tbra} only to keep a consistent choice of basis during the left move. 
When we subsequently absorb the \textit{bra}-layer, we use the red (teal) projectors to truncate to the final desired environment bond dimensions. 
Again we truncate the physical leg together with virtual legs in the red projector, to the new virtual index of the interlayer bond dimension $\chi_I$.

Having explained the absorption step of the left move above, we now present the calculations of the corresponding projectors below.
Constructing the projectors in a way that is both computationally efficient and accurate is central to the success of the CTMRG algorithm, both in the conventional formulation as well as in the new split-CTMRG variation proposed here. 
The construction we propose here follows the general philosophy of defining projectors in the conventional CTMRG, that has proven to be very accurate in many practical applications.

\begin{figure*}[bth]
    \centering
    \begin{minipage}[b]{.98\columnwidth}
        \centering
            \begin{tikzpicture}

        \node at (-4.0, +3.5) {(a)};

		\def\tensorSep{1.1}
	    \def\tensorSize{0.2}

		\begin{scope}[shift = {(0.0, 1.125*\tensorSep)}]
            \coordinate (ket) at ({-0.5*\tensorSep}, {+0.5*\tensorSep});
            \coordinate (bra) at ({+0.5*\tensorSep}, {-0.5*\tensorSep});
  
			\coordinate (C1) at ({-1.5*\tensorSep}, {+1.5*\tensorSep});
            \coordinate (T1ket) at ({-0.5*\tensorSep}, {+1.5*\tensorSep});
            \coordinate (T1bra) at ({+0.5*\tensorSep}, {+1.5*\tensorSep});
            \coordinate (C2) at ({+1.5*\tensorSep}, {+1.5*\tensorSep});
            \coordinate (T2ket) at ({+1.5*\tensorSep}, {+0.5*\tensorSep});
            \coordinate (T2bra) at ({+1.5*\tensorSep}, {-0.5*\tensorSep});
            \coordinate (C3) at ({+1.5*\tensorSep}, {-1.5*\tensorSep});
            \coordinate (T3bra) at ({+0.5*\tensorSep}, {-1.5*\tensorSep});
            \coordinate (T3ket) at ({-0.5*\tensorSep}, {-1.5*\tensorSep});
            \coordinate (C4) at ({-1.5*\tensorSep}, {-1.5*\tensorSep});
            \coordinate (T4bra) at ({-1.5*\tensorSep}, {-0.5*\tensorSep});
            \coordinate (T4ket) at ({-1.5*\tensorSep}, {+0.5*\tensorSep});

            \coordinate (text) at ({-3.5*\tensorSep}, {+0.5*\tensorSep});

            \draw[rounded corners, gray] ({-3*\tensorSep}, {+2.25*\tensorSep}) --  ({3*\tensorSep}, {+2.25*\tensorSep}) -- ({3*\tensorSep}, {-1.05*\tensorSep}) -- ({-3*\tensorSep}, {-1.05*\tensorSep}) -- cycle;

            \draw[thick, decorate, decoration={coil, aspect=0}] (ket) to (bra);

            \draw[thick] (bra) to (T1bra);
            \draw[thick] (bra) to (T2bra);
            \draw[thick] (bra) to (T4bra);

            \draw[thick] (ket) to (T1ket);
            \draw[white, line width=3pt] (ket) to (T2ket);
            \draw[thick] (ket) to (T2ket);
            \draw[thick] (ket) to (T4ket);

            \draw[line width=0.5mm,] (C1) to (T1ket);
            \draw[line width=0.5mm, dashed, dash pattern=on 0.5pt off 0.5pt] (T1ket) to (T1bra);
            \draw[line width=0.5mm,] (T1bra) to (C2);
            \draw[line width=0.5mm,] (C2) to (T2ket);
            \draw[line width=0.5mm, dashed, dash pattern=on 0.5pt off 0.5pt] (T2ket) to (T2bra);
            \draw[line width=0.5mm, dashed, dash pattern=on 0.5pt off 0.5pt] (T4bra) to (T4ket);
            \draw[line width=0.5mm,] (T4ket) to (C1);
            
            \draw[white, line width=3pt] (ket) to ($(ket) + (0, -1.5*\tensorSep)$);
            \draw[thick] (ket) to ($(ket) + (0, -1.5*\tensorSep)$);
            \draw[thick] (bra) to ($(bra) + (0, -0.5*\tensorSep)$);
            \draw[line width=0.5mm,] (T2bra) to ($(T2bra) + (0, -0.5*\tensorSep)$);
            \draw[line width=0.5mm,] (T4bra) to ($(T4bra) + (0, -0.5*\tensorSep)$);

            \node[above right=1] at (ket) {$\psi_{x,y}$};
            \draw[thick, fill=googleB] (ket) circle (\tensorSize);
            \node[below right=1] at (bra) {$\psi^{*}_{x,y}$};
            \draw[thick, fill=googleB] (bra) circle (\tensorSize);

			\node[above left=5] at (C1) {$C_{1_{[x, y]}}$};
			\draw[thick, fill = gray] (C1) circle (\tensorSize);

            \node[above=6] at (T1ket) {$T_{1_{[x, y]}}^\text{ket}$};
			\draw[thick, fill = gray] (T1ket) circle (\tensorSize);

            \node[above=6] at (T1bra) {$T_{1_{[x, y]}}^\text{bra}$};
			\draw[thick, fill = gray] (T1bra) circle (\tensorSize);

            \node[above right=5] at (C2) {$C_{2_{[x, y]}}$};
			\draw[thick, fill = gray] (C2) circle (\tensorSize);

            \node[right=6] at (T2ket) {$T_{2_{[x, y]}}^\text{ket}$};
			\draw[thick, fill = gray] (T2ket) circle (\tensorSize);

            \node[right=6] at (T2bra) {$T_{2_{[x, y]}}^\text{bra}$};
			\draw[thick, fill = gray] (T2bra) circle (\tensorSize);





            \node[left=6] at (T4bra) {$T_{4_{[x, y]}}^\text{bra}$};
			\draw[thick, fill = gray] (T4bra) circle (\tensorSize);

            \node[left=6] at (T4ket) {$T_{4_{[x, y]}}^\text{ket}$};
			\draw[thick, fill = gray] (T4ket) circle (\tensorSize);

            \node at (text) {$\rho_\text{green}^T =$};
		\end{scope}

        \begin{scope}[shift = {(0.0, -1.125*\tensorSep)}]
            \coordinate (ket) at ({-0.5*\tensorSep}, {+0.5*\tensorSep});
            \coordinate (bra) at ({+0.5*\tensorSep}, {-0.5*\tensorSep});
  
			\coordinate (C1) at ({-1.5*\tensorSep}, {+1.5*\tensorSep});
            \coordinate (T1ket) at ({-0.5*\tensorSep}, {+1.5*\tensorSep});
            \coordinate (T1bra) at ({+0.5*\tensorSep}, {+1.5*\tensorSep});
            \coordinate (C2) at ({+1.5*\tensorSep}, {+1.5*\tensorSep});
            \coordinate (T2ket) at ({+1.5*\tensorSep}, {+0.5*\tensorSep});
            \coordinate (T2bra) at ({+1.5*\tensorSep}, {-0.5*\tensorSep});
            \coordinate (C3) at ({+1.5*\tensorSep}, {-1.5*\tensorSep});
            \coordinate (T3bra) at ({+0.5*\tensorSep}, {-1.5*\tensorSep});
            \coordinate (T3ket) at ({-0.5*\tensorSep}, {-1.5*\tensorSep});
            \coordinate (C4) at ({-1.5*\tensorSep}, {-1.5*\tensorSep});
            \coordinate (T4bra) at ({-1.5*\tensorSep}, {-0.5*\tensorSep});
            \coordinate (T4ket) at ({-1.5*\tensorSep}, {+0.5*\tensorSep});

            \coordinate (text) at ({-3.5*\tensorSep}, {-0.5*\tensorSep});

            \draw[rounded corners, gray] ({-3*\tensorSep}, {+1.05*\tensorSep}) --  ({3*\tensorSep}, {+1.05*\tensorSep}) -- ({3*\tensorSep}, {-2.25*\tensorSep}) -- ({-3*\tensorSep}, {-2.25*\tensorSep}) -- cycle;

            \draw[thick, decorate, decoration={coil, aspect=0}] (ket) to (bra);

            \draw[thick] (bra) to (T2bra);
            \draw[thick] (bra) to (T3bra);
            \draw[thick] (bra) to (T4bra);
            \draw[thick] (bra) to ($(bra) + (0, 1.5*\tensorSep)$);

            \draw[white, line width=3pt] (ket) to (T2ket);
            \draw[thick] (ket) to (T2ket);
            \draw[white, line width=3pt] (ket) to (T3ket);
            \draw[thick] (ket) to (T3ket);
            \draw[thick] (ket) to (T4ket);

            \draw[line width=0.5mm, dashed, dash pattern=on 0.5pt off 0.5pt] (T2ket) to (T2bra);
            \draw[line width=0.5mm,] (T2bra) to (C3);
            \draw[line width=0.5mm,] (C3) to (T3bra);
            \draw[line width=0.5mm, dashed, dash pattern=on 0.5pt off 0.5pt] (T3bra) to (T3ket);
            \draw[line width=0.5mm,] (T3ket) to (C4);
            \draw[line width=0.5mm,] (C4) to (T4bra);
            \draw[line width=0.5mm, dashed, dash pattern=on 0.5pt off 0.5pt] (T4bra) to (T4ket);

            \draw[thick] (ket) to ($(ket) + (0, 0.5*\tensorSep)$);
            \draw[line width=0.5mm,] (T2ket) to ($(T2ket) + (0, 0.5*\tensorSep)$);
            \draw[line width=0.5mm,] (T4ket) to ($(T4ket) + (0, 0.5*\tensorSep)$);

            \node[above right=1] at (ket) {$\psi_{x+1,y}$};
            \draw[thick, fill=googleB] (ket) circle (\tensorSize);
            \node[below right=1] at (bra) {$\psi^{*}_{x+1,y}$};
            \draw[thick, fill=googleB] (bra) circle (\tensorSize);





            \node[right=6] at (T2ket) {$T_{2_{[x+1, y]}}^\text{ket}$};
			\draw[thick, fill = gray] (T2ket) circle (\tensorSize);

            \node[right=6] at (T2bra) {$T_{2_{[x+1, y]}}^\text{bra}$};
			\draw[thick, fill = gray] (T2bra) circle (\tensorSize);

            \node[below right=5] at (C3) {$C_{3_{[x+1, y]}}$};
			\draw[thick, fill = gray] (C3) circle (\tensorSize);

            \node[below=6] at (T3bra) {$T_{3_{[x+1, y]}}^\text{bra}$};
			\draw[thick, fill = gray] (T3bra) circle (\tensorSize);

            \node[below=6] at (T3ket) {$T_{3_{[x+1, y]}}^\text{ket}$};
			\draw[thick, fill = gray] (T3ket) circle (\tensorSize);

            \node[below left=5] at (C4) {$C_{4_{[x+1, y]}}$};
			\draw[thick, fill = gray] (C4) circle (\tensorSize);

            \node[left=6] at (T4bra) {$T_{4_{[x+1, y]}}^\text{bra}$};
			\draw[thick, fill = gray] (T4bra) circle (\tensorSize);

            \node[left=6] at (T4ket) {$T_{4_{[x+1, y]}}^\text{ket}$};
			\draw[thick, fill = gray] (T4ket) circle (\tensorSize);

            \node at (text) {$\rho_\text{green}^B =$};
		\end{scope}
    \end{tikzpicture}
    \end{minipage}
    \begin{minipage}[b]{.98\columnwidth}
      \centering
      \begin{minipage}[b]{.98\columnwidth}
        \centering
        \begin{align*}
                \begin{tikzpicture}[baseline=-0.65ex]%

        \node at (-3.4, +1.5) {(b)};
        
		\def\tensorSep{1.1}%
	    \def\tensorSize{0.2}%
		\begin{scope}[shift={(-0.5, 0.75*\tensorSep)}]%
            \coordinate (rhoTop) at ({0*\tensorSep}, {0*\tensorSep});%
            \coordinate (rhoBottom) at ({0*\tensorSep}, {-1.5*\tensorSep});%
            \node at (-2, -0.75) {$\mathcal{M}_\text{green} =$};
            \draw[line width=0.5mm,] ($(rhoTop) + (-0.8*\tensorSep, 0)$) -- ($(rhoTop) + (-0.8*\tensorSep, -0.75*\tensorSep)$);%
            \draw[thick] ($(rhoTop) + (-0.4*\tensorSep, 0)$) -- ($(rhoTop) + (-0.4*\tensorSep, -0.75*\tensorSep)$);%
            \draw[line width=0.5mm,] ($(rhoTop) + (0.8*\tensorSep, 0)$) -- ($(rhoTop) + (0.8*\tensorSep, -0.6*\tensorSep)$);%
            \draw[thick] ($(rhoTop) + (0.4*\tensorSep, 0)$) -- ($(rhoTop) + (0.4*\tensorSep, -0.6*\tensorSep)$);%
            \draw[line width=0.5mm,] ($(rhoBottom) + (-0.8*\tensorSep, 0)$) -- ($(rhoBottom) + (-0.8*\tensorSep, 0.75*\tensorSep)$);%
            \draw[thick] ($(rhoBottom) + (-0.4*\tensorSep, 0)$) -- ($(rhoBottom) + (-0.4*\tensorSep, 0.75*\tensorSep)$);%
            \draw[line width=0.5mm,] ($(rhoBottom) + (0.8*\tensorSep, 0)$) -- ($(rhoBottom) + (0.8*\tensorSep, 0.6*\tensorSep)$);%
            \draw[thick] ($(rhoBottom) + (0.4*\tensorSep, 0)$) -- ($(rhoBottom) + (0.4*\tensorSep, 0.6*\tensorSep)$);%
            \draw[draw,fill=gray!30] ($(rhoTop) - (1.0*\tensorSep, 0.25*\tensorSep)$) rectangle ++(2.0*\tensorSep, 0.5*\tensorSep);%
            \node at (rhoTop) {$\rho_\text{green}^T$};%
            \draw[draw,fill=gray!30] ($(rhoBottom) - (1.0*\tensorSep, 0.25*\tensorSep)$) rectangle ++(2.0*\tensorSep, 0.5*\tensorSep);%
            \node at (rhoBottom) {$\rho_\text{green}^B$};%
		\end{scope}%
      \end{tikzpicture}%
          ~=~ %
                \begin{tikzpicture}[baseline=-0.65ex]%
		\def\tensorSep{1.1}%
	    \def\tensorSize{0.2}%
		\begin{scope}[shift={(-0.5, 0.75*\tensorSep)}]%
            \coordinate (Vh) at ({0*\tensorSep}, {0*\tensorSep});%
            \coordinate (S) at ({-0.6*\tensorSep}, {-0.75*\tensorSep});%
            \coordinate (U) at ({0*\tensorSep}, {-1.5*\tensorSep});%
            \draw[line width=0.5mm,] ($(Vh) + (-0.8*\tensorSep, 0)$) -- ($(Vh) + (-0.8*\tensorSep, -0.75*\tensorSep)$);%
            \draw[thick] ($(Vh) + (-0.4*\tensorSep, 0)$) -- ($(Vh) + (-0.4*\tensorSep, -0.75*\tensorSep)$);%
            \draw[line width=0.5mm,] ($(Vh) + (0.8*\tensorSep, 0)$) -- ($(Vh) + (0.8*\tensorSep, -0.6*\tensorSep)$);%
            \draw[thick] ($(Vh) + (0.4*\tensorSep, 0)$) -- ($(Vh) + (0.4*\tensorSep, -0.6*\tensorSep)$);%
            \draw[line width=0.5mm,] ($(U) + (-0.8*\tensorSep, 0)$) -- ($(U) + (-0.8*\tensorSep, 0.75*\tensorSep)$);%
            \draw[thick] ($(U) + (-0.4*\tensorSep, 0)$) -- ($(U) + (-0.4*\tensorSep, 0.75*\tensorSep)$);%
            \draw[line width=0.5mm,] ($(U) + (0.8*\tensorSep, 0)$) -- ($(U) + (0.8*\tensorSep, 0.6*\tensorSep)$);%
            \draw[thick] ($(U) + (0.4*\tensorSep, 0)$) -- ($(U) + (0.4*\tensorSep, 0.6*\tensorSep)$);%
            \draw[draw,fill=gray!30] ($(U) - (1.0*\tensorSep, 0.25*\tensorSep)$) rectangle ++(2.0*\tensorSep, 0.5*\tensorSep);%
            \node at (U) {$U_\text{green}$};%
            \draw[draw,fill=gray!30] ($(S) - (0.4*\tensorSep, 0.25*\tensorSep)$) rectangle ++(0.8*\tensorSep, 0.5*\tensorSep);%
            \node at (S) {$S_\text{green}$};%
            \draw[draw,fill=gray!30] ($(Vh) - (1.0*\tensorSep, 0.25*\tensorSep)$) rectangle ++(2.0*\tensorSep, 0.5*\tensorSep);%
            \node at (Vh) {$V_\text{green}^\dagger$};%
		\end{scope}%
      \end{tikzpicture}%
        \end{align*}
      \end{minipage}\\[\baselineskip]
      \begin{minipage}[b]{.98\columnwidth}
        \centering
        \begin{align*}
                \begin{tikzpicture}[baseline=-0.65ex]%

        \node at (-2.0, +2.0) {(c)};%
      
		\def\tensorSep{1.1}%
	    \def\tensorSize{0.2}%
		\begin{scope}[shift={(0, 0.75*\tensorSep)}]%
            \coordinate (projTop) at ({0*\tensorSep}, {0*\tensorSep});%
            \coordinate (projBottom) at ({0*\tensorSep}, {-1.5*\tensorSep});%
            \node at (-1.75, {-0.75*\tensorSep}) {$\mathbb{1} \approx$};%
            \draw[line width=0.5mm,] (projTop) to (projBottom);
            \draw[line width=0.5mm,] ($(projTop) + (-0.5*\tensorSep, 0)$) -- ($(projTop) + (-0.5*\tensorSep, 0.5*\tensorSep)$);%
            \draw[thick] ($(projTop) + (0.5*\tensorSep, 0)$) -- ($(projTop) + (0.5*\tensorSep, 0.5*\tensorSep)$);%
            \draw[line width=0.5mm,] ($(projBottom) + (-0.5*\tensorSep, 0)$) -- ($(projBottom) + (-0.5*\tensorSep, -0.5*\tensorSep)$);%
            \draw[thick] ($(projBottom) + (0.5*\tensorSep, 0)$) -- ($(projBottom) + (0.5*\tensorSep, -0.5*\tensorSep)$);%
            \node[isosceles triangle, isosceles triangle apex angle=100, draw, shape border rotate=-90, fill=googleG] at (projTop) {$P_T$};%
            \node[isosceles triangle, isosceles triangle apex angle=100, minimum size=1, draw, shape border rotate=90, fill=googleG] at (projBottom) {$P_B$};%
		\end{scope}%
      \end{tikzpicture}%
          ~=~~ %
                \begin{tikzpicture}[baseline=-0.65ex]%
		\def\tensorSep{1.1}%
	    \def\tensorSize{0.2}%
		\begin{scope}%
            \coordinate (Uh) at ({0*\tensorSep}, {1.25*\tensorSep});%
            \coordinate (Stop) at ({0.6*\tensorSep}, {0.5*\tensorSep});%
            \coordinate (Sbottom) at ({0.6*\tensorSep}, {-0.5*\tensorSep});%
            \coordinate (V) at ({0*\tensorSep}, {-1.25*\tensorSep});%
            \coordinate (rhoBottom) at ({-1*\tensorSep}, {0.5*\tensorSep});%
            \coordinate (rhoTop) at ({-1*\tensorSep}, {-0.5*\tensorSep});%
            \draw[line width=0.5mm,] ($(rhoBottom) + (0.65*\tensorSep, 0)$) -- ($(rhoBottom) + (0.65*\tensorSep, 0.75*\tensorSep)$);%
            \draw[thick] ($(rhoBottom) + (0.25*\tensorSep, 0)$) -- ($(rhoBottom) + (0.25*\tensorSep, 0.75*\tensorSep)$);%
            \draw[line width=0.5mm,] ($(rhoBottom) + (-0.65*\tensorSep, 0)$) -- ($(rhoBottom) + (-0.65*\tensorSep, 0.75*\tensorSep)$);%
            \draw[thick] ($(rhoBottom) + (-0.25*\tensorSep, 0)$) -- ($(rhoBottom) + (-0.25*\tensorSep, 0.75*\tensorSep)$);%
            \draw[line width=0.5mm,] ($(rhoTop) + (0.65*\tensorSep, 0)$) -- ($(rhoTop) + (0.65*\tensorSep, -0.75*\tensorSep)$);%
            \draw[thick] ($(rhoTop) + (0.25*\tensorSep, 0)$) -- ($(rhoTop) + (0.25*\tensorSep, -0.75*\tensorSep)$);%
            \draw[line width=0.5mm,] ($(rhoTop) + (-0.65*\tensorSep, 0)$) -- ($(rhoTop) + (-0.65*\tensorSep, -0.75*\tensorSep)$);%
            \draw[thick] ($(rhoTop) + (-0.25*\tensorSep, 0)$) -- ($(rhoTop) + (-0.25*\tensorSep, -0.75*\tensorSep)$);%
            \draw[line width=0.5mm,] ($(Stop) + (0, 0.1*\tensorSep)$) to ($(Sbottom) + (0, -0.1*\tensorSep)$);%
            \draw[line width=0.5mm,out=90,in=0] ($(Stop) + (0, 0.25*\tensorSep)$) to ($(Uh) + (0.25*\tensorSep, 0)$);%
            \draw[line width=0.5mm,out=270,in=0] ($(Sbottom) + (0, -0.25*\tensorSep)$) to ($(V) + (0.25*\tensorSep, 0)$);
            %
            \draw[draw,fill=gray!30] ($(V) - (1.0*\tensorSep, 0.25*\tensorSep)$) -- ($(V) + (-1.0*\tensorSep, 0.25*\tensorSep)$) -- ($(V) + (0*\tensorSep, 0.25*\tensorSep)$) -- ($(V) + (0.26*\tensorSep, 0*\tensorSep)$) -- ($(V) + (0*\tensorSep, -0.25*\tensorSep)$) -- cycle;%
            \node at ($(V) + (-0.375*\tensorSep, 0)$) {$\widetilde{V}_\text{green}$};%
            \draw[draw,fill=gray!30] ($(Stop) - (0.4*\tensorSep, 0.25*\tensorSep)$) rectangle ++(0.8*\tensorSep, 0.5*\tensorSep);%
            \node at (Stop) {$\widetilde{S}_\text{green}^{-1/2}$};%
            \draw[draw,fill=gray!30] ($(Sbottom) - (0.4*\tensorSep, 0.25*\tensorSep)$) rectangle ++(0.8*\tensorSep, 0.5*\tensorSep);%
            \node at (Sbottom) {$\widetilde{S}_\text{green}^{-1/2}$};%
            \draw[draw,fill=gray!30] ($(Uh) - (1.0*\tensorSep, 0.25*\tensorSep)$) --  ($(Uh) + (-1.0*\tensorSep, 0.25*\tensorSep)$) -- ($(Uh) + (0.*\tensorSep, 0.25*\tensorSep)$) -- ($(Uh) + (0.26*\tensorSep, 0.*\tensorSep)$) -- ($(Uh) + (0.*\tensorSep, -0.25*\tensorSep)$) -- cycle;
            \node at ($(Uh) - (0.375*\tensorSep, 0)$) {$\widetilde{U}_\text{green}^\dagger$};%
            \draw[draw,fill=gray!30] ($(rhoBottom) - (0.85*\tensorSep, 0.25*\tensorSep)$) rectangle ++(1.7*\tensorSep, 0.5*\tensorSep);%
            \node at (rhoBottom) {$\rho^B_\text{green}$};%
            \draw[draw,fill=gray!30] ($(rhoTop) - (0.85*\tensorSep, 0.25*\tensorSep)$) rectangle ++(1.7*\tensorSep, 0.5*\tensorSep);%
            \node at (rhoTop) {$\rho^T_\text{green}$};%
		\end{scope}%
      \end{tikzpicture}%
        \end{align*}
      \end{minipage}
    \end{minipage}
    \caption{Calculation of the green projectors in a left absorption step. (a) Initial networks $\rho_\text{green}^B$ and $\rho_\text{green}^T$ used as a starting point for the calculation. (b) The product of $\rho_\text{green}^B$ and $\rho_\text{green}^T$ over the bonds we wish to renormalize is decomposed using a singular value decomposition. (c) Finally, the green projectors can be defined from the tensors in (a) and (b) as an approximate resolution of the identity. The singular values are truncated to the desired bond dimension $\chi_E$. The calculation of the other three projectors (teal, yellow, red) is performed along similar schemes and is described in App.~\ref{app:construction of projectors}.}
    \label{fig:proj_plots_green}
\end{figure*}
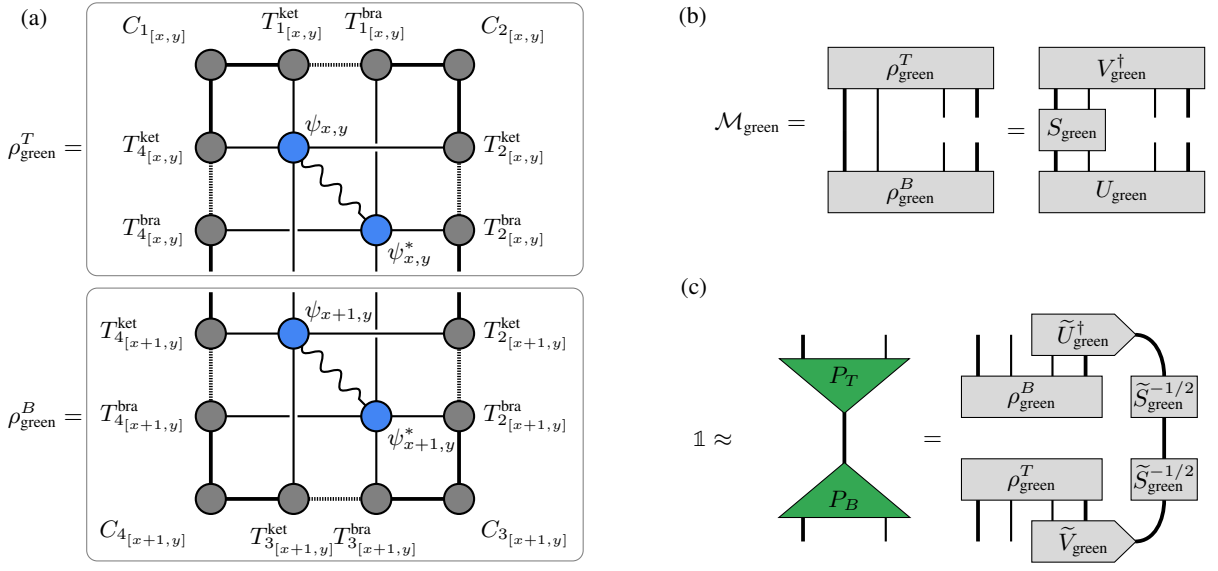

We begin by constructing the green projectors shown in the equations above, which are applied during the projection step following the absorption of the \textit{ket}-layer. To do so, we define the objects $\rho_\text{green}^B$ and $\rho_\text{green}^T$ as shown in Fig.~\ref{fig:proj_plots_green}(a), which are built from a patch of local PEPS tensors with their respective environment tensors, such that their contraction approximates a large part of the double-layer network. 
Importantly, those networks are chosen such that their open indices include those indices that we wish to renormalize by applying the projector to them.

We then use the two initial networks to construct a matrix
\begin{equation}
    \mathcal{M}_\text{green} \coloneq \rho_\text{green}^B \rho_\text{green}^T = U_\text{green} S_\text{green} V_\text{green}^\dagger,
\end{equation}
by contracting over the indices of $\rho_\text{green}^B$ and $\rho_\text{green}^T$ that correspond to the vector spaces we aim to truncate with the projectors we are defining. 
We further perform a \emph{singular value decomposition} (SVD) of $\mathcal M_\text{green}$ as shown in Fig.~\ref{fig:proj_plots_green}(b).
The goal is to obtain a pair of projectors $P_B$ and $P_T$, computed such that their insertion into the bonds to be truncated only changes $\mathcal{M}_\text{green}$ minimally, i.e.,
\begin{equation}
    \widetilde{\mathcal{M}}_\text{green} = \rho_\text{green}^B P_B P_T \rho_\text{green}^T \approx \mathcal{M}_\text{green}.
\end{equation}
To this end use the singular value decomposition of $\mathcal{M}_\text{green}$ to approximate its inverse with a matrix of the rank we want to truncate to (typically $\chi_I$ or $\chi_E$). With this at hand, we can construct the projectors by creating an approximate expression for the identity
\begin{equation}
    \begin{split}
        \mathcal{M}_\text{green}^{-1}& = (\rho_\text{green}^T)^{-1} (\rho_\text{green}^B)^{-1} \approx \widetilde{V}_\text{green} \widetilde{S}_\text{green}^{-1} \widetilde{U}_\text{green}^\dagger\\
        &\implies \mathbb{1} \approx \rho_\text{green}^T \widetilde{V}_\text{green} \widetilde{S}_\text{green}^{-1} \widetilde{U}_\text{green}^\dagger \rho_\text{green}^B,
    \end{split}
\end{equation}
where $\widetilde{S}_\text{green}$ represents the matrix with the truncated singular values and $\widetilde{V}_\text{green}$ and $\widetilde{U}_\text{green}$ indicate we truncated the corresponding singular vectors. 
The resulting projectors are illustrated in Fig.~\ref{fig:proj_plots_green}(c).
We stress again that this type of construction of the projectors has been established to be very accurate in many practical state-of-the-art infinite PEPS calculations.

The remaining projectors shown in Sec.~\ref{method:splitCTMRG_description} are built using the same philosophy, with adjustments to the initial networks chosen for $\rho^B$ and $\rho^T$ to ensure that they encompass the specific vector spaces targeted for truncation by these projectors. 
We note that this process requires the projectors for the \textit{bra}-layer to incorporate those of the \textit{ket}-layer into the definitions of $\rho^B$ and $\rho^T$, as detailed in App.~\ref{app:construction of projectors}.

As in the conventional CTMRG, the computational bottleneck in the split-CTMRG algorithm remains the construction of the projectors, as it involves computationally expensive tensor contractions and decompositions. 
Assuming a small physical dimension (i.e.,  $p < \chi_B$), the conventional CTMRG has a scaling of
\begin{equation}
    \mathcal{O}(\chi_E^3 \chi_B^6) \xrightarrow{\chi_E \, \approx \, \chi_B^2} \mathcal{O}(\chi_{B}^{12})
\end{equation}
in terms of bulk bond dimension $\chi_B$ and environment bond dimension $\chi_E$. 
In contrast, the scaling of the split-CTMRG scheme introduced here is 
\begin{equation}
    \mathcal{O}(\chi_E^3\chi_B^4) \xrightarrow{\chi_E \,   \approx \, \chi_B^2} \mathcal{O}(\chi_B^{10}),
\end{equation}
which is the cost of constructing the teal colored projectors. 
We note that this leading cost can be further reduced by modifying the projectors, which we discuss in App.~\ref{app:complexity and alternative projectors}.\\\\

\section{Numerical analysis and benchmarking}
\label{sect:Numerical analysis}

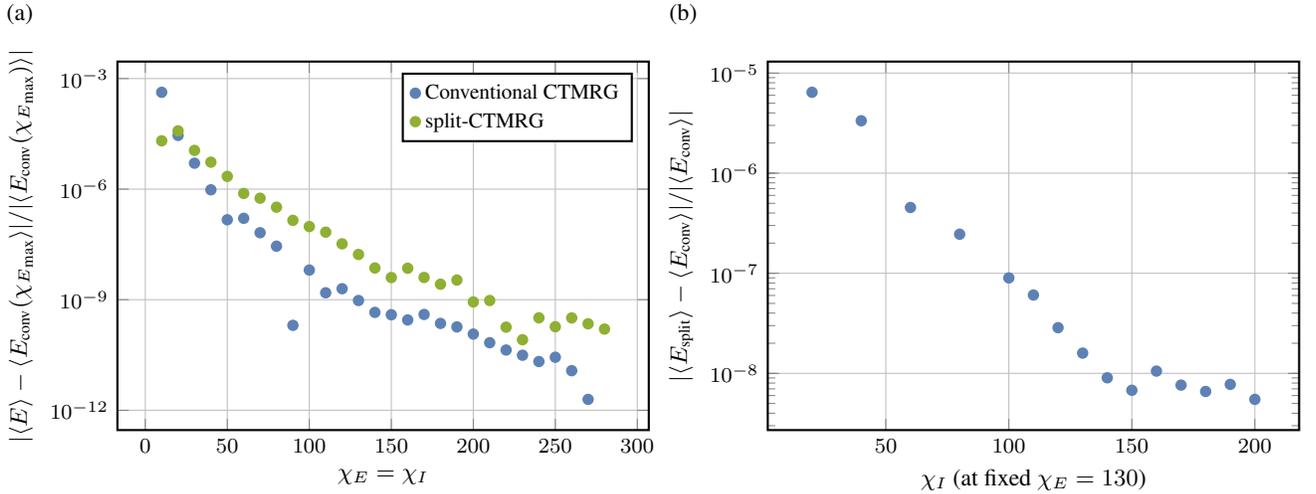
\begin{figure*}[tb]
    \centering
        \tikzsetnextfilename{energy_comparison_classic_split}
    \begin{tikzpicture}

    \node at (-0.15\columnwidth, 0.64\columnwidth) {(a)};
    \node at (0.87\columnwidth, 0.64\columnwidth) {(b)};

    \begin{semilogyaxis}[
        xlabel=$\chi_E \equal \chi_I$,
        ylabel=$|\langle E \rangle - \langle E_\text{conv}({\chi_E}_\text{max}\rangle| / |\langle E_\text{conv}({\chi_E}_\text{max})\rangle|$,
        width=\columnwidth,
        height=0.75\columnwidth,
        legend style={font=\footnotesize},
        legend cell align={left},
        label style={font=\small},
        tick label style={font=\footnotesize},
        every axis/.append style={thick},
        /tikz/mark size=1.75pt,
        legend pos=north east,
        xmajorgrids=true,
        ymajorgrids=true,
    ]
    
    \addplot[mathematicaplot1, mark=*, only marks] table [x=chi,y=EdiffClassical] {data/energy_runtime_honeycomb_D8.txt};
    \addlegendentry{Conventional CTMRG}

    \addplot[mathematicaplot3, mark=*, only marks] table [x=chi,y=EdiffSplit] {data/energy_runtime_honeycomb_D8.txt};
    \addlegendentry{split-CTMRG}
    
    \end{semilogyaxis}

    \begin{semilogyaxis}[
        xlabel=$\chi_I$ (at fixed $\chi_E \equal 130$),
        ylabel=$|\langle E_\text{split} \rangle - \langle E_\text{conv} \rangle| / |\langle E_\text{conv} \rangle|$,
        width=\columnwidth,
        height=0.75\columnwidth,
        legend style={font=\small},
        label style={font=\small},
        tick label style={font=\footnotesize},
        every axis/.append style={thick},
        /tikz/mark size=1.75pt,
        legend pos=north west,
        xmajorgrids=true,
        ymajorgrids=true,
        xshift=\columnwidth
    ]
    
    \addplot[mathematicaplot1, mark=*, only marks] table [x=chiI,y=Ediff] {data/interlayer_D_8.txt};
    
    \end{semilogyaxis}
    
    \end{tikzpicture}
    \caption{Accuracy benchmarks for the split-CTMRG algorithm. We use a low-energy honeycomb Heisenberg state at bulk bond dimension $\chi_B = 8$ for the analysis. (a) Relative energy difference between the split-CTMRG and the conventional CTMRG, taking $E_\text{conv}$ at ${\chi_E}_\text{max} = 280$ as the reference value. In the split-CTMRG, the interlayer bond dimension is equal to the environment bond dimension, i.e., $\chi_E = \chi_I$. (b) Dependence of the energy expectation value on the interlayer bond dimension $\chi_I$ in the split-CTMRG algorithm, fixing $\chi_E = 130$. The relative difference drops to $10^{-8}$ when the two approach each other, which justifies the choice of $\chi_I$ in panel (a) and other numerical tests.}
    \label{fig:benchmark_justification}
    \label{fig:Energy_comp}
    \label{fig:interlayer_test}
\end{figure*}

\begin{figure*}[t]
        \tikzsetnextfilename{optimizer_energy_over_runtime}
    \begin{tikzpicture}

    \node at (-0.105\columnwidth, 0.63\columnwidth) {(a)};
    \node at (0.855\columnwidth, 0.63\columnwidth) {(b)};

    \begin{loglogaxis}[
        xlabel=$\chi_B$ (at fixed $\chi_E\equal\chi_I\equal170$),
        ylabel=Time (s) of single absorption step,
        width=\columnwidth,
        height=0.75\columnwidth,
        legend style={font=\footnotesize},
        legend cell align={left},
        label style={font=\small},
        tick label style={font=\footnotesize},
        every axis/.append style={thick},
        /tikz/mark size=1.75pt,
        legend pos=north west,
        xmajorgrids=true,
        ymajorgrids=true,
        xtick={3,4,5,6,7,8,9,10},
        xticklabels={3,4,5,6,7,8,9,10},
    ]
    
    \addplot[mathematicaplot1, mark=*, only marks] table [x=chiB,y=T] {data/scaling_classic_chiE_170.txt};
    \addlegendentry{Conventional CTMRG}

    \addplot[mathematicaplot3, mark=*, only marks] table [x=chiB,y=T] {data/scaling_split_chiE_170.txt};
    \addlegendentry{split-CTMRG}

    \addplot[densely dashed, mathematicaplot1!70!black, domain=2.9:10.2] {exp(6.010285506178981*ln(x) -4.952236501792971)};
    \addlegendentry{Linear fit: ${\chi_B}^{6.0\pm0.2}$}

    \addplot[densely dashed, mathematicaplot3!70!black, domain=2.9:10.2] {exp(2.9784689527428907*ln(x) -0.49722018178460115)};
    \addlegendentry{Linear fit: ${\chi_B}^{3.0\pm0.2}$}
    \end{loglogaxis}

    \begin{axis}[
        xlabel=Time (h),
        ylabel=$\langle E \rangle$,
        width=\columnwidth,
        height=0.75\columnwidth,
        legend style={font=\footnotesize},
        yticklabel style={
            /pgf/number format/fixed,
            /pgf/number format/precision=4
        },
        scaled x ticks=false,
        scaled y ticks=false,
        legend cell align={left},
        label style={font=\small},
        tick label style={font=\footnotesize},
        every axis/.append style={thick},
        /tikz/mark size=1.75pt,
        legend pos=north east,
        xmajorgrids=true,
        ymajorgrids=true,
        ylabel shift=-8,
        xshift=\columnwidth,
        xmin=-10,
        xmax=260
    ]
    
    \addplot[mathematicaplot1, mark=*, only marks] table [x expr=\thisrow{T}/60,y=E] {data/optimizer_energy_over_runtime_classic.txt};
    \addlegendentry{Conventional CTMRG}

    \addplot[mathematicaplot3, mark=square, only marks] table [x expr=\thisrow{T}/60,y=E] {data/optimizer_energy_over_runtime_split.txt};
    \addlegendentry{split-CTMRG}

    
    \end{axis}
    
    \end{tikzpicture}
    \caption{Efficiency benchmarks for the split-CTMRG algorithm. (a) Time for a single split-CTMRG absorption step for different bulk bond dimensions $\chi_B$ at fixed $\chi_E = \chi_I = 170$. The reduced complexity compared to the conventional double-layer approach results in different scaling laws for large bulk bond dimensions. For the linear fit in the log-log scale the data points for $\chi_B \geq 6$ have been considered. (b) Energy expectation value of the honeycomb Heisenberg model over time for a variational optimization at bulk bond dimension $\chi_B = 10$ and $\chi_E = \chi_I = 100$, started from the same initial preconverged state.}
    \label{fig:scaling}
    \label{fig:timing_at_otpimization}
\end{figure*}
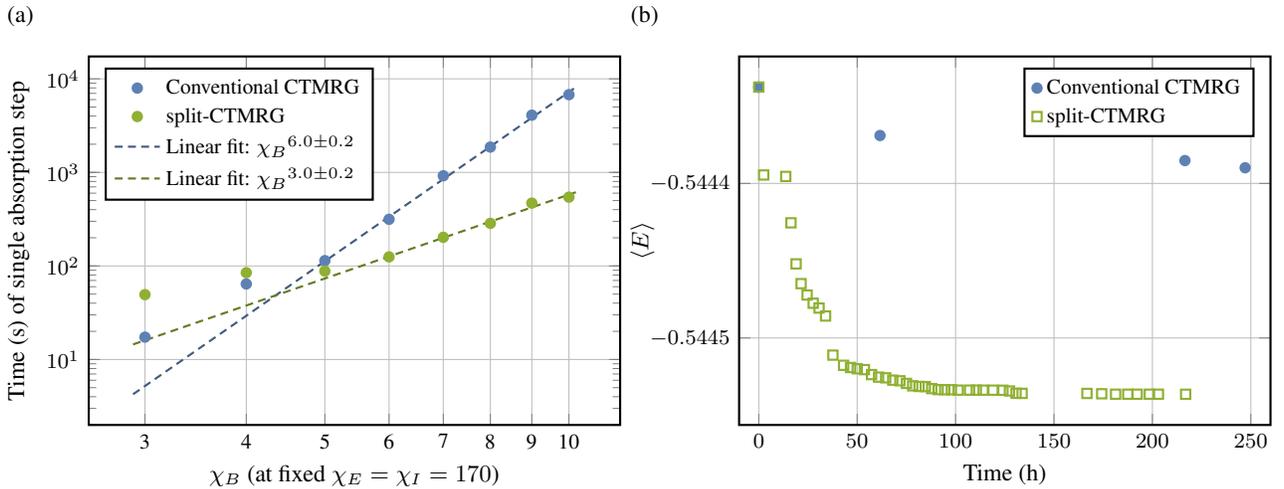

In this section we demonstrate the retained numerical accuracy as well as the advantageous computational complexity of the split-CTMRG algorithm, as described in Sec.~\ref{method:splitCTMRG_description}. 
We further show how this enables variational PEPS optimization at higher bond dimensions, thereby pushing its applicability to largely unexplored regimes. 
Concretely, we benchmark our algorithm on the antiferromagnetic spin-$1/2$ Heisenberg model
\begin{align}
    \mathcal H = \sum_{\langle i,j \rangle} \vec S_i \cdot \vec S_j
\end{align}
on the honeycomb lattice, where $\langle i,j \rangle$ denotes nearest neighboring pairs. 
To simulate this system using a single-site PEPS ansatz, we coarse-grain two physical sites, thereby effectively increasing the local Hilbert space dimension to $p = 4$. 

Setting aside any variational optimization, we first present the calculation of local observables to demonstrate the accuracy of the split-CTMRG in comparison to the established, conventional CTMRG.
For the latter, we use the state-of-the-art projectors introduced in Refs.~\cite{Corboz2010, Corboz2014} as a trustworthy benchmark. 
In Fig.~\ref{fig:Energy_comp}(a), we show the relative difference of the energy expectation value $\langle \hat{H} \rangle(\chi_E, \chi_I)$ for a low-energy state of the honeycomb Heisenberg model at different environment and interlayer bond dimensions. As a reference value for the relative difference, we choose the expectation value obtained from the conventional CTMRG at $\chi_E=280.$
In this comparison, we choose $\chi_I = \chi_E$, while the freedom to choose those two bond dimensions independently will be discussed later. 
Our results show that both the conventional as well as the split-CTMRG algorithm converge to the same expectation value in practice. 
This demonstrates that the split-CTMRG algorithm can compute local observables with state-of-the-art accuracy, a crucial prerequisite for any method that aims to improve the efficiency of gradient-based variational optimization. 
As mentioned above, the definition of the split-CTMRG environment tensors (cf.\ Fig.~\ref{fig:split-transfer-ctmrg}) introduces an additional interlayer bond dimension $\chi_I$. 
In order to find a reasonable choice for this auxiliary tensor index, we again compare the numerical energies obtained via the split-CTMRG at varying $\chi_I$ to those obtained by the conventional CTMRG. 
The results are shown in Fig.~\ref{fig:interlayer_test}(b) and reveal only minimal relative energy differences, which are on the order of $10^{-8}$ if the two bond dimensions coincide. 
We therefore fix $\chi_I = \chi_E$ as a rather conservative choice in the remainder of our benchmarks, and note that smaller values may suffice in practical calculations. 
This choice is particularly appreciable, as the maximal theoretical value would be $\chi_I = \chi_E \cdot \chi_B$ if one were to directly decompose the conventional edge environment tensors to split them (cf.\ Fig.~\ref{fig:CTMRG_Method_2}). 
The significant reduction of $\chi_I$ is only achievable because the projectors inserted between the \textit{bra}- and \textit{ket}-layers effectively renormalize the tensors to the most relevant subspaces. This is a crucial ingredient of the split-CTMRG method.

Having established the numerical accuracy of the split-CTMRG algorithm, we next move towards numerically analyzing its computational cost. 
The reduced complexity, as discussed in Sec.~\ref{method:splitCTMRG_description}, manifests itself in a different scaling of the computational time for the contraction of the infinite PEPS network at different bond dimensions $\chi_B$, for fixed environment bond dimensions ($\chi_E, \chi_I$).
In Fig.~\ref{fig:scaling}(a), we show evidence that the two methods indeed follow different scaling laws for their computational complexity. 
We observe that the computational time scales as $\mathcal{O}(\chi_B^6)$ for the conventional CTMRG, as expected. 
For the split-CTMRG we find a scaling closer to $\chi_B^3$. 
The reason for this is that the dominant contribution to the scaling in the split-CTMRG algorithm is given by the construction of the teal projectors, which only have to be constructed once (per tensor in the unit cell) in each iteration step. 
All other components of the algorithm have a more favorable scaling and dominate for bond dimensions up to $\chi_B = 10$. 
We thus expect that the leading scaling contribution becomes visible only at even larger bulk bond dimensions.

Transitioning to the realm of gradient-based variational energy optimization, the computational advantages reveal the full potential of our algorithm. 
This is because the approximate contraction of the infinite lattice, e.g., by using the (split)-CTMRG, has to be performed at least once (but possibly also multiple times) per optimization step, along with the calculation of the energy gradient~\cite{Corboz2016, Vanderstraeten2016}. 
When employing reverse-mode automatic differentiation for this task~\cite{Liao219, Ponsioen2022,Naumann2024,Francuz2025}, the computational complexity for generating the gradient is identical to the complexity of the CTMRG. 
Thus by using the split-CTMRG algorithm, the gradient calculation benefits from the same improvements. 
Therefore, the contraction of the infinite lattice is \emph{the} central bottleneck that limits the variational energy optimization at larger bond dimensions $\chi_B$.

In practice, simulations are typically constrained to \mbox{$\chi_B \lesssim 8$}, as long as no global symmetries can be exploited in the tensor network. 
In contrast, the split-CTMRG method allows us to push the variational optimization up to bond dimension $\chi_B = 10$ at least. 
To demonstrate this, we show all optimization steps performed over a fixed time interval and the corresponding energy expectation values in Fig.~\ref{fig:timing_at_otpimization}(b). 
For the fixed duration of ten days, the gradient-based optimization procedure using a conventional CTMRG performed only three optimization steps. 
In contrast, the optimization using the split-CTMRG algorithm was able to perform 44 optimization steps and already tends to converge in the energy. 
To eliminate some bias, both simulations were initialized with the same preconverged PEPS state and optimized with $\chi_B = 10$ and $\chi_E = 100$. 
The drastic increase in the number of optimization steps together with the retained accuracy of the evaluation of the energy density reveals the potential of the split-CTMRG for its application in state-of-the-art calculations with the PEPS ansatz. It furthermore has the potential to benefit post-optimization analysis of variational PEPS simulations, such as finite-entanglement scaling~\cite{Rader2016,Corboz2018,Vanhecke2023}, where large environment bond dimensions can be required.

\section{Conclusions and outlook}
\label{sec:conclusionsAndOutlook}

In this work, we have introduced what we call the \emph{split-CTMRG algorithm}, a revised and more efficient approach for contracting infinite PEPS at large bond dimensions. This applies both in principle as a matter of scaling as well as in practical use of the algorithm. 
To achieve this, we have maintained separate environment tensors for each layer in the double-layer PEPS network. 
The method addresses a key bottleneck in the computational task of variational energy optimization by reducing the contraction complexity while preserving established accuracies. 
A numerical benchmark for the honeycomb lattice antiferromagnetic Heisenberg model showcases the algorithm’s enhanced efficiency in optimizing PEPS at large bond dimensions. 
This not only leads to lower achievable variational energies in practice, but also increases the amount of accessible data for post-optimization finite-entanglement scaling analysis. 
Therefore, our approach represents a substantial step 
forward in computational methods for two-dimensional 
strongly correlated quantum many-body systems in the context of
condensed matter physics and quantum materials 
using projected entangled-pair states.

The split-CTMRG framework is readily compatible with the implementation of global symmetries~\cite{Singh2010,Weichselbaum2012,Silvi2019,Schmoll2020}, as well as its combination with the spiral PEPS ansatz~\cite{Hasik2024}. 
Furthermore, the method can potentially extend to other algorithms based on the CTMRG, such as the calculation of excited states~\cite{Vanderstraeten2015,Vanderstraeten2019,Ponsioen2020,Ponsioen2022} and structure factors~\cite{Vanderstraeten2022,Ponsioen2023a,Tu2024}. 
A scheme following the idea of the split-CTMRG algorithm could also improve CTMRG-based schemes in three-dimensional settings~\cite{Vlaar2021,Vlaar2023}. It is the hope that such method development, as the one proposed here, pushes the boundary of what two-dimensional condensed matter and quantum materials systems can be classically reliably simulated.

\paragraph*{Algorithm and open source code.}
An implementation of the proposed algorithm is available in our open source variPEPS library~\cite{variPEPS_GitHub, naumann24_varipeps_python}. 

\paragraph*{Data availability.} 
The data that support the findings of this article are openly available~\cite{raw_data_split_ctmrg}.

\begin{acknowledgments}

E.~L.~W.\ thanks the Studienstiftung des deutschen Volkes for support.  
This work has been funded by the Deutsche Forschungsgemeinschaft (DFG, German Research Foundation) under the project number 277101999 -- CRC 183 (projects A04 and B01), for which this constitutes an inter-node publication involving both Cologne and Berlin, and the BMBF (MUNIQC-Atoms, FermiQP).
This work was supported by Horizon Europe programme HORIZON-CL4-2022-QUANTUM-02-SGA via the project 101113690 (PASQuanS2.1).
It has also received funding from the Cluster of Excellence MATH+, and Berlin Quantum.
We would like to thank the ZEDV (IT support) of the physics department, Freie Universität Berlin, for computing time and their technical support, particularly we thank Jörg Behrmann, Cornelius 
Hoffmann and Jens Dreger. 
We also acknowledge the computing time provided by the HPC Service of FUB-IT, Freie Universität Berlin~\cite{Bennett2020}.

The authors gratefully acknowledge the Gauss Centre for Supercomputing e.V.\ (www.gauss-centre.eu) for funding this project by providing computing time through the \emph{John von Neumann Institute for Computing}  (NIC) on the GCS Supercomputer JUWELS~\cite{JUWELS} (Grant NeTeNeSyQuMa) and the FZ J\"ulich for computing time on JURECA~\cite{JURECA2021}  (institute project PGI-8) at \emph{J\"ulich Supercomputing Centre}  (JSC).
This work has been initiated during the workshop \enquote{Entanglement in Strongly Correlated Systems (2023)} at the Centro de Ciencias de Benasque Pedro Pascual. 
We thank the organizers and the venue for their hospitality.
\end{acknowledgments}

\appendix

\section{Construction of projectors}
\label{app:construction of projectors}

\begin{figure}[thb]
    \centering
    \input{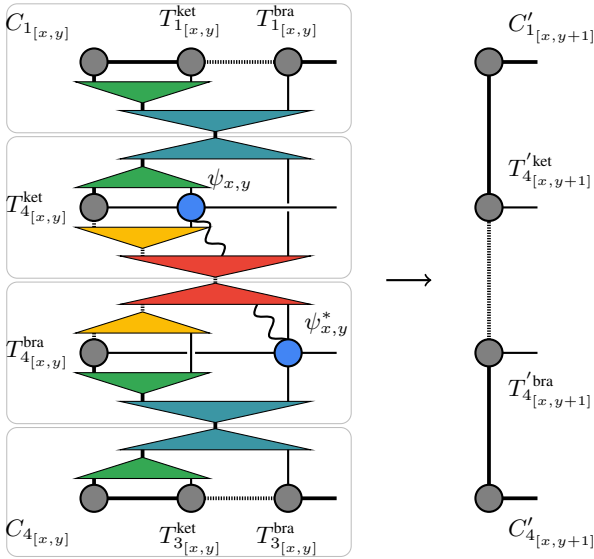}
    \caption{Left absorption step for the split-transfer CTMRG with full renormalization. A column of PEPS and environment tensors is absorbed into the left environment tensors $C_1$, $T_4^\text{ket}$, $T_4^\text{bra}$ and $C_4$ while using the projectors discussed in detail in the text.}
    \label{fig:left_absorption_full}
\end{figure}

In the following, we describe the construction of the projectors that were not explicitly discussed in the main text. 
For reference, Fig.~\ref{fig:left_absorption_full} illustrates the role of the four different projectors in a full left move of the split-CTMRG step. 
The projectors used in the other directional absorption steps (top, right, bottom) are constructed analogously.

\begin{figure*}[t]
    \centering
    \begin{minipage}{1.3\columnwidth}
        \include{figures/rho_T_and_B_teal_proj}
    \end{minipage}
    \begin{minipage}{.7\columnwidth}
      \begin{minipage}{\textwidth}
        \centering
        \begin{align*}
          \mathcal{M}_{\text{teal}} &=~ %
              \begin{tikzpicture}[baseline=-0.65ex]%
    \def\tensorSep{1.1}%
    \def\tensorSize{0.2}%
    \begin{scope}[shift={(0, 0.75*\tensorSep)}]%
        \coordinate (rhoTop) at ({0*\tensorSep}, {0*\tensorSep});%
        \coordinate (rhoBottom) at ({0*\tensorSep}, {-1.5*\tensorSep});%
        \draw[line width=0.5mm,] ($(rhoTop) + (-0.8*\tensorSep, 0)$) -- ($(rhoTop) + (-0.8*\tensorSep, -0.75*\tensorSep)$);%
        \draw[thick] ($(rhoTop) + (-0.4*\tensorSep, 0)$) -- ($(rhoTop) + (-0.4*\tensorSep, -0.75*\tensorSep)$);%
        \draw[line width=0.5mm,] ($(rhoTop) + (0.8*\tensorSep, 0)$) -- ($(rhoTop) + (0.8*\tensorSep, -0.6*\tensorSep)$);%
        \draw[thick] ($(rhoTop) + (0.4*\tensorSep, 0)$) -- ($(rhoTop) + (0.4*\tensorSep, -0.6*\tensorSep)$);%
        \draw[line width=0.5mm,] ($(rhoBottom) + (-0.8*\tensorSep, 0)$) -- ($(rhoBottom) + (-0.8*\tensorSep, 0.75*\tensorSep)$);%
        \draw[thick] ($(rhoBottom) + (-0.4*\tensorSep, 0)$) -- ($(rhoBottom) + (-0.4*\tensorSep, 0.75*\tensorSep)$);%
        \draw[line width=0.5mm,] ($(rhoBottom) + (0.8*\tensorSep, 0)$) -- ($(rhoBottom) + (0.8*\tensorSep, 0.6*\tensorSep)$);%
        \draw[thick] ($(rhoBottom) + (0.4*\tensorSep, 0)$) -- ($(rhoBottom) + (0.4*\tensorSep, 0.6*\tensorSep)$);%
        \draw[draw,fill=gray!30] ($(rhoTop) - (1.0*\tensorSep, 0.25*\tensorSep)$) rectangle ++(2.0*\tensorSep, 0.5*\tensorSep);%
        \node at (rhoTop) {$\rho_{\text{teal}}^T$};%
        \draw[draw,fill=gray!30] ($(rhoBottom) - (1.0*\tensorSep, 0.25*\tensorSep)$) rectangle ++(2.0*\tensorSep, 0.5*\tensorSep);%
        \node at (rhoBottom) {$\rho_{\text{teal}}^B$};%
    \end{scope}%
  \end{tikzpicture}%
            ~=~ 
            \begin{tikzpicture}[baseline=-0.65ex]%
\def\tensorSep{1.1}%
\def\tensorSize{0.2}%
\begin{scope}[shift={(0, 0.75*\tensorSep)}]%
    \coordinate (Vh) at ({0*\tensorSep}, {0*\tensorSep});%
    \coordinate (S) at ({-0.6*\tensorSep}, {-0.75*\tensorSep});%
    \coordinate (U) at ({0*\tensorSep}, {-1.5*\tensorSep});%
    \draw[line width=0.5mm,] ($(Vh) + (-0.8*\tensorSep, 0)$) -- ($(Vh) + (-0.8*\tensorSep, -0.75*\tensorSep)$);%
    \draw[thick] ($(Vh) + (-0.4*\tensorSep, 0)$) -- ($(Vh) + (-0.4*\tensorSep, -0.75*\tensorSep)$);%
    
    \draw[line width=0.5mm,] ($(Vh) + (0.8*\tensorSep, 0)$) -- ($(Vh) + (0.8*\tensorSep, -0.6*\tensorSep)$);%
    \draw[thick] ($(Vh) + (0.4*\tensorSep, 0)$) -- ($(Vh) + (0.4*\tensorSep, -0.6*\tensorSep)$);%
    \draw[line width=0.5mm,] ($(U) + (-0.8*\tensorSep, 0)$) -- ($(U) + (-0.8*\tensorSep, 0.75*\tensorSep)$);%
    \draw[thick] ($(U) + (-0.4*\tensorSep, 0)$) -- ($(U) + (-0.4*\tensorSep, 0.75*\tensorSep)$);%
    
    \draw[line width=0.5mm,] ($(U) + (0.8*\tensorSep, 0)$) -- ($(U) + (0.8*\tensorSep, 0.6*\tensorSep)$);%
    \draw[thick] ($(U) + (0.4*\tensorSep, 0)$) -- ($(U) + (0.4*\tensorSep, 0.6*\tensorSep)$);%
    \draw[draw,fill=gray!30] ($(U) - (1.0*\tensorSep, 0.25*\tensorSep)$) rectangle ++(2.0*\tensorSep, 0.5*\tensorSep);%
    \node at (U) {$U_{\text{teal}}$};%
    \draw[draw,fill=gray!30] ($(S) - (0.4*\tensorSep, 0.25*\tensorSep)$) rectangle ++(0.8*\tensorSep, 0.5*\tensorSep);%
    \node at (S) {$S_{\text{teal}}$};%
    \draw[draw,fill=gray!30] ($(Vh) - (1.0*\tensorSep, 0.25*\tensorSep)$) rectangle ++(2.0*\tensorSep, 0.5*\tensorSep);%
    \node at (Vh) {$V_{\text{teal}}^\dagger$};%
\end{scope}%
\end{tikzpicture}%
        \end{align*}
      \end{minipage}\\[3\baselineskip]
      \begin{minipage}{\textwidth}
        \centering
        \begin{align*}
          \mathbb{1} \approx~ %
          \begin{tikzpicture}[baseline=-0.65ex]%
		\def\tensorSep{1.1}%
	    \def\tensorSize{0.2}%
		\begin{scope}[shift={(0, 0.75*\tensorSep)}]%
            \coordinate (projTop) at ({0*\tensorSep}, {0*\tensorSep});%
            \coordinate (projBottom) at ({0*\tensorSep}, {-1.5*\tensorSep});%
            \draw[line width=0.5mm,] (projTop) to (projBottom);
            \draw[line width=0.5mm,] ($(projTop) + (-0.5*\tensorSep, 0)$) -- ($(projTop) + (-0.5*\tensorSep, 0.5*\tensorSep)$);%
            \draw[thick] ($(projTop) + (0.5*\tensorSep, 0)$) -- ($(projTop) + (0.5*\tensorSep, 0.5*\tensorSep)$);%
            \draw[line width=0.5mm,] ($(projBottom) + (-0.5*\tensorSep, 0)$) -- ($(projBottom) + (-0.5*\tensorSep, -0.5*\tensorSep)$);%
            \draw[thick] ($(projBottom) + (0.5*\tensorSep, 0)$) -- ($(projBottom) + (0.5*\tensorSep, -0.5*\tensorSep)$);%
            \node[isosceles triangle, isosceles triangle apex angle=100, draw, shape border rotate=-90, fill=googleBG] at (projTop) {$P_T$};%
            \node[isosceles triangle, isosceles triangle apex angle=100, minimum size=1, draw, shape border rotate=90, fill=googleBG] at (projBottom) {$P_B$};%
		\end{scope}%
      \end{tikzpicture}%
          ~=~~ %
          \begin{tikzpicture}[baseline=-0.65ex]%
		\def\tensorSep{1.1}%
	    \def\tensorSize{0.2}%
		\begin{scope}%
            \coordinate (Uh) at ({0*\tensorSep}, {1.25*\tensorSep});%
            \coordinate (Stop) at ({0.6*\tensorSep}, {0.5*\tensorSep});%
            \coordinate (Sbottom) at ({0.6*\tensorSep}, {-0.5*\tensorSep});%
            \coordinate (V) at ({0*\tensorSep}, {-1.25*\tensorSep});%
            \coordinate (rhoBottom) at ({-1*\tensorSep}, {0.5*\tensorSep});%
            \coordinate (rhoTop) at ({-1*\tensorSep}, {-0.5*\tensorSep});%
            \draw[line width=0.5mm,] ($(rhoBottom) + (0.65*\tensorSep, 0)$) -- ($(rhoBottom) + (0.65*\tensorSep, 0.75*\tensorSep)$);%
            \draw[thick] ($(rhoBottom) + (0.25*\tensorSep, 0)$) -- ($(rhoBottom) + (0.25*\tensorSep, 0.75*\tensorSep)$);%
            \draw[line width=0.5mm,] ($(rhoBottom) + (-0.65*\tensorSep, 0)$) -- ($(rhoBottom) + (-0.65*\tensorSep, 0.75*\tensorSep)$);%
            \draw[thick] ($(rhoBottom) + (-0.25*\tensorSep, 0)$) -- ($(rhoBottom) + (-0.25*\tensorSep, 0.75*\tensorSep)$);%
            \draw[line width=0.5mm,] ($(rhoTop) + (0.65*\tensorSep, 0)$) -- ($(rhoTop) + (0.65*\tensorSep, -0.75*\tensorSep)$);%
            \draw[thick] ($(rhoTop) + (0.25*\tensorSep, 0)$) -- ($(rhoTop) + (0.25*\tensorSep, -0.75*\tensorSep)$);%
            \draw[line width=0.5mm,] ($(rhoTop) + (-0.65*\tensorSep, 0)$) -- ($(rhoTop) + (-0.65*\tensorSep, -0.75*\tensorSep)$);%
            \draw[thick] ($(rhoTop) + (-0.25*\tensorSep, 0)$) -- ($(rhoTop) + (-0.25*\tensorSep, -0.75*\tensorSep)$);%
            \draw[line width=0.5mm,] ($(Stop) + (0, 0.1*\tensorSep)$) to ($(Sbottom) + (0, -0.1*\tensorSep)$);%
            \draw[line width=0.5mm,out=90,in=0] ($(Stop) + (0, 0.25*\tensorSep)$) to ($(Uh) + (0.25*\tensorSep, 0)$);%
            \draw[line width=0.5mm,out=-90,in=0] ($(Sbottom) + (0, -0.25*\tensorSep)$) to ($(V) + (0.25*\tensorSep, 0)$);%
            %
            \draw[draw,fill=gray!30] ($(V) - (1.0*\tensorSep, 0.25*\tensorSep)$) -- ($(V) + (-1.0*\tensorSep, 0.25*\tensorSep)$) -- ($(V) + (0*\tensorSep, 0.25*\tensorSep)$) -- ($(V) + (0.26*\tensorSep, 0*\tensorSep)$) -- ($(V) + (0*\tensorSep, -0.25*\tensorSep)$) -- cycle;
            \node at ($(V) + (-0.375*\tensorSep, 0)$) {$\widetilde{V}_{\text{teal}}$};%
            \draw[draw,fill=gray!30] ($(Stop) - (0.4*\tensorSep, 0.25*\tensorSep)$) rectangle ++(0.8*\tensorSep, 0.5*\tensorSep);%
            \node at (Stop) {$\widetilde{S}_{\text{teal}}^{-1/2}$};%
            \draw[draw,fill=gray!30] ($(Sbottom) - (0.4*\tensorSep, 0.25*\tensorSep)$) rectangle ++(0.8*\tensorSep, 0.5*\tensorSep);%
            \node at (Sbottom) {$\widetilde{S}_{\text{teal}}^{-1/2}$};%
            %
            \draw[draw,fill=gray!30] ($(Uh) - (1.0*\tensorSep, 0.25*\tensorSep)$) -- ($(Uh) + (-1.0*\tensorSep, 0.25*\tensorSep)$) -- ($(Uh) + (0*\tensorSep, 0.25*\tensorSep)$) -- ($(Uh) + (0.26*\tensorSep, 0*\tensorSep)$) -- ($(Uh) + (0*\tensorSep, -0.25*\tensorSep)$) -- cycle;
            \node at ($(Uh) + (-0.375*\tensorSep, 0)$) {$\widetilde{U}_{\text{teal}}^\dagger$};%
            \draw[draw,fill=gray!30] ($(rhoBottom) - (0.85*\tensorSep, 0.25*\tensorSep)$) rectangle ++(1.7*\tensorSep, 0.5*\tensorSep);%
            \node at (rhoBottom) {$\rho^B_{\text{teal}}$};%
            \draw[draw,fill=gray!30] ($(rhoTop) - (0.85*\tensorSep, 0.25*\tensorSep)$) rectangle ++(1.7*\tensorSep, 0.5*\tensorSep);%
            \node at (rhoTop) {$\rho^T_{\text{teal}}$};%
		\end{scope}%
      \end{tikzpicture}%
        \end{align*}
      \end{minipage}
    \end{minipage}
    \caption{Construction of the teal projectors used in the left absorption step in Fig.~\ref{fig:left_absorption_full}. The initial networks for $\rho^B_{\text{teal}}$ and $\rho^T_{\text{teal}}$ already contain the green projectors from the preceding absorption of the \textit{ket}-layer. A truncated SVD of $\mathcal M_\text{teal}$ is used to generate an approximate resolution of the identity, from which the final projectors are defined.}
    \label{fig:constructionTealProjectors}
\end{figure*}
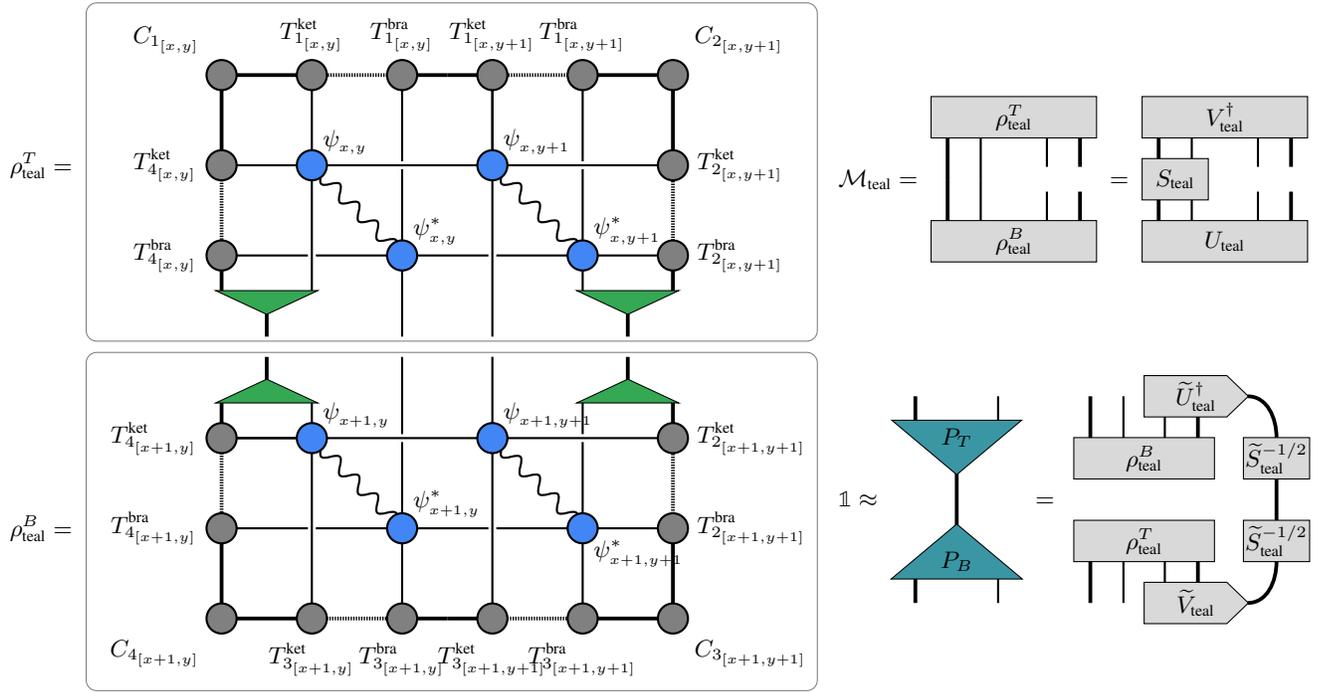

The first projectors we focus on are the teal projectors in Fig.~\ref{fig:left_absorption_full}, which are applied after absorbing the \textit{bra}-layer to truncate vector spaces that have already been projected using the green projectors. 
Consequently, a suitable tensor network for defining $\rho^B_{\text{teal}}$ and $\rho^T_{\text{teal}}$, already incorporating these green projectors, is shown in Fig.~\ref{fig:constructionTealProjectors}. 
Following the same approach as for the projectors discussed so far, the teal projectors are constructed via a singular value decomposition of the product of $\rho^B_{\text{teal}}$ and $\rho^T_{\text{teal}}$, as illustrated in Fig.~\ref{fig:constructionTealProjectors}. 
Here, the singular values are truncated to the environment bond dimension $\chi_E$.

\begin{figure*}[t]
    \centering
    \begin{minipage}{1.1\columnwidth}
        \centering
            \begin{tikzpicture}
		\def\tensorSep{1.1}
	    \def\tensorSize{0.2}

		\begin{scope}[shift = {(0.0, 0.25*\tensorSep)}]
            \coordinate (ket) at ({-0.5*\tensorSep}, {+0.5*\tensorSep});
            \coordinate (bra) at ({+0.5*\tensorSep}, {-0.5*\tensorSep});
  
			\coordinate (C1) at ({-1.5*\tensorSep}, {+1.5*\tensorSep});
            \coordinate (T1ket) at ({-0.5*\tensorSep}, {+1.5*\tensorSep});
            \coordinate (T1bra) at ({+0.5*\tensorSep}, {+1.5*\tensorSep});
            \coordinate (C2) at ({+1.5*\tensorSep}, {+1.5*\tensorSep});
            \coordinate (T2ket) at ({+1.5*\tensorSep}, {+0.5*\tensorSep});
            \coordinate (T2bra) at ({+1.5*\tensorSep}, {-0.5*\tensorSep});
            \coordinate (C3) at ({+1.5*\tensorSep}, {-1.5*\tensorSep});
            \coordinate (T3bra) at ({+0.5*\tensorSep}, {-1.5*\tensorSep});
            \coordinate (T3ket) at ({-0.5*\tensorSep}, {-1.5*\tensorSep});
            \coordinate (C4) at ({-1.5*\tensorSep}, {-1.5*\tensorSep});
            \coordinate (T4bra) at ({-1.5*\tensorSep}, {-0.5*\tensorSep});
            \coordinate (T4ket) at ({-1.5*\tensorSep}, {+0.5*\tensorSep});

            \coordinate (text) at ({-3.5*\tensorSep}, {+1*\tensorSep});

            \draw[rounded corners, gray] ({-3*\tensorSep}, {+2.25*\tensorSep}) --  ({3*\tensorSep}, {+2.25*\tensorSep}) -- ({3*\tensorSep}, {-0.05*\tensorSep}) -- ({-3*\tensorSep}, {-0.05*\tensorSep}) -- cycle;

            \draw[thick, decorate, decoration={coil, aspect=0}] ($(ket) + (0.3333333333333333*\tensorSep, -0.5*\tensorSep)$) to (ket);


            \draw[line width=0.5mm,] (C1) to (T1ket);
            \draw[line width=0.5mm, dashed, dash pattern=on 0.5pt off 0.5pt] (T1ket) to (T1bra);
            \draw[line width=0.5mm,] (T1bra) to (C2);
            \draw[line width=0.5mm,] (C2) to (T2ket);
            \draw[line width=0.5mm,] (T4ket) to (C1);

            \draw[thick] (ket) to ($(ket) + (0, -0.5*\tensorSep)$);
            \draw[line width=0.5mm, dashed, dash pattern=on 0.5pt off 0.5pt] (T4ket) to ($(T4ket) + (0, -0.5*\tensorSep)$);
            \draw[line width=0.5mm, dashed, dash pattern=on 0.5pt off 0.5pt] (T2ket) to ($(T2ket) + (0, -0.5*\tensorSep)$);
            \draw[thick] (T1bra) to ($(T1bra) + (0, -1.5*\tensorSep)$);

            \draw[thick] (ket) to (T1ket);
            \draw[white, line width=3pt] (ket) to (T2ket);
            \draw[thick] (ket) to (T2ket);
            \draw[thick] (ket) to (T4ket);

            \node[above right=1] at (ket) {$\psi_{x,y}$};
            \draw[thick, fill=googleB] (ket) circle (\tensorSize);

			\node[above left=5] at (C1) {$C_{1_{[x, y]}}$};
			\draw[thick, fill = gray] (C1) circle (\tensorSize);

            \node[above=6] at (T1ket) {$T_{1_{[x, y]}}^\text{ket}$};
			\draw[thick, fill = gray] (T1ket) circle (\tensorSize);

            \node[above=6] at (T1bra) {$T_{1_{[x, y]}}^\text{bra}$};
			\draw[thick, fill = gray] (T1bra) circle (\tensorSize);

            \node[above right=5] at (C2) {$C_{2_{[x, y]}}$};
			\draw[thick, fill = gray] (C2) circle (\tensorSize);

            \node[right=6] at (T2ket) {$T_{2_{[x, y]}}^\text{ket}$};
			\draw[thick, fill = gray] (T2ket) circle (\tensorSize);







            \node[left=6] at (T4ket) {$T_{4_{[x, y]}}^\text{ket}$};
			\draw[thick, fill = gray] (T4ket) circle (\tensorSize);

            \node at (text) {$\rho_{\text{yellow}}^T =$};
		\end{scope}

        \begin{scope}[shift = {(0.0, 0.0*\tensorSep)}]
            \coordinate (ket) at ({-0.5*\tensorSep}, {+0.5*\tensorSep});
            \coordinate (bra) at ({+0.5*\tensorSep}, {-0.5*\tensorSep});
  
			\coordinate (C1) at ({-1.5*\tensorSep}, {+1.5*\tensorSep});
            \coordinate (T1ket) at ({-0.5*\tensorSep}, {+1.5*\tensorSep});
            \coordinate (T1bra) at ({+0.5*\tensorSep}, {+1.5*\tensorSep});
            \coordinate (C2) at ({+1.5*\tensorSep}, {+1.5*\tensorSep});
            \coordinate (T2ket) at ({+1.5*\tensorSep}, {+0.5*\tensorSep});
            \coordinate (T2bra) at ({+1.5*\tensorSep}, {-0.5*\tensorSep});
            \coordinate (C3) at ({+1.5*\tensorSep}, {-1.5*\tensorSep});
            \coordinate (T3bra) at ({+0.5*\tensorSep}, {-1.5*\tensorSep});
            \coordinate (T3ket) at ({-0.5*\tensorSep}, {-1.5*\tensorSep});
            \coordinate (C4) at ({-1.5*\tensorSep}, {-1.5*\tensorSep});
            \coordinate (T4bra) at ({-1.5*\tensorSep}, {-0.5*\tensorSep});
            \coordinate (T4ket) at ({-1.5*\tensorSep}, {+0.5*\tensorSep});

            \coordinate (text) at ({-3.5*\tensorSep}, {-1.0*\tensorSep});

            \draw[rounded corners, gray] ({-3*\tensorSep}, {+0.05*\tensorSep}) --  ({3*\tensorSep}, {+0.05*\tensorSep}) -- ({3*\tensorSep}, {-2.25*\tensorSep}) -- ({-3*\tensorSep}, {-2.25*\tensorSep}) -- cycle;

            \draw[thick, decorate, decoration={coil, aspect=0}] ($(bra) + (-0.3333333333333333*\tensorSep, 0.5*\tensorSep)$) to (bra);


            \draw[thick] (bra) to (T2bra);
            \draw[thick] (bra) to (T3bra);
            \draw[thick] (bra) to (T4bra);

            \draw[line width=0.5mm,] (T2bra) to (C3);
            \draw[line width=0.5mm,] (C3) to (T3bra);
            \draw[line width=0.5mm, dashed, dash pattern=on 0.5pt off 0.5pt] (T3bra) to (T3ket);
            \draw[line width=0.5mm,] (T3ket) to (C4);
            \draw[line width=0.5mm,] (C4) to (T4bra);

            \draw[thick] (bra) to ($(bra) + (0, 0.5*\tensorSep)$);
            \draw[white, line width=3pt] (T3ket) to ($(T3ket) + (0, 1.5*\tensorSep)$);
            \draw[thick] (T3ket) to ($(T3ket) + (0, 1.5*\tensorSep)$);
            \draw[line width=0.5mm, dashed, dash pattern=on 0.5pt off 0.5pt] (T2bra) to ($(T2bra) + (0, 0.5*\tensorSep)$);
            \draw[line width=0.5mm, dashed, dash pattern=on 0.5pt off 0.5pt] (T4bra) to ($(T4bra) + (0, 0.5*\tensorSep)$);

            \node[above right=1] at (bra) {$\psi^{*}_{x,y}$};
            \draw[thick, fill=googleB] (bra) circle (\tensorSize);






            \node[right=6] at (T2bra) {$T_{2_{[x, y]}}^\text{bra}$};
			\draw[thick, fill = gray] (T2bra) circle (\tensorSize);

            \node[below right=5] at (C3) {$C_{3_{[x, y]}}$};
			\draw[thick, fill = gray] (C3) circle (\tensorSize);

            \node[below=6] at (T3bra) {$T_{3_{[x, y]}}^\text{bra}$};
			\draw[thick, fill = gray] (T3bra) circle (\tensorSize);

            \node[below=6] at (T3ket) {$T_{3_{[x, y]}}^\text{ket}$};
			\draw[thick, fill = gray] (T3ket) circle (\tensorSize);

            \node[below left=5] at (C4) {$C_{4_{[x, y]}}$};
			\draw[thick, fill = gray] (C4) circle (\tensorSize);

            \node[left=6] at (T4bra) {$T_{4_{[x, y]}}^\text{bra}$};
			\draw[thick, fill = gray] (T4bra) circle (\tensorSize);


            \node at (text) {$\rho_{\text{yellow}}^B =$};
		\end{scope}
    \end{tikzpicture}   
    \end{minipage}
    \begin{minipage}{0.9\columnwidth}
      \begin{minipage}{\textwidth}
        \centering
        \begin{align*}
            \mathcal{M}_{\text{yellow}} &=~ %
            \begin{tikzpicture}[baseline=-0.65ex]%
		\def\tensorSep{1.1}%
	    \def\tensorSize{0.2}%
		\begin{scope}[shift={(0, 0.75*\tensorSep)}]%
            \coordinate (rhoTop) at ({0*\tensorSep}, {0*\tensorSep});%
            \coordinate (rhoBottom) at ({0*\tensorSep}, {-1.5*\tensorSep});%
            \draw[line width=0.5mm, dashed, dash pattern=on 0.5pt off 0.5pt] ($(rhoTop) + (-0.8*\tensorSep, 0)$) -- ($(rhoTop) + (-0.8*\tensorSep, -0.75*\tensorSep)$);%
            \draw[thick] ($(rhoTop) + (-0.4*\tensorSep, 0)$) -- ($(rhoTop) + (-0.4*\tensorSep, -0.75*\tensorSep)$);%
            \draw[line width=0.5mm, dashed, dash pattern=on 0.5pt off 0.5pt] ($(rhoTop) + (0.8*\tensorSep, 0)$) -- ($(rhoTop) + (0.8*\tensorSep, -0.6*\tensorSep)$);%
            \draw[thick] ($(rhoTop) + (0.4*\tensorSep, 0)$) -- ($(rhoTop) + (0.4*\tensorSep, -0.6*\tensorSep)$);%
            \draw[thick, decorate, decoration={coil, aspect=0}] ($(rhoTop) + (0.1*\tensorSep, -0.6*\tensorSep)$) -- ($(rhoTop) + (0.1*\tensorSep, 0)$);%
            \draw[line width=0.5mm, dashed, dash pattern=on 0.5pt off 0.5pt] ($(rhoBottom) + (-0.8*\tensorSep, 0)$) -- ($(rhoBottom) + (-0.8*\tensorSep, 0.75*\tensorSep)$);%
            \draw[thick] ($(rhoBottom) + (-0.4*\tensorSep, 0)$) -- ($(rhoBottom) + (-0.4*\tensorSep, 0.75*\tensorSep)$);%
            \draw[line width=0.5mm, dashed, dash pattern=on 0.5pt off 0.5pt] ($(rhoBottom) + (0.8*\tensorSep, 0)$) -- ($(rhoBottom) + (0.8*\tensorSep, 0.6*\tensorSep)$);%
            \draw[thick] ($(rhoBottom) + (0.4*\tensorSep, 0)$) -- ($(rhoBottom) + (0.4*\tensorSep, 0.6*\tensorSep)$);%
            \draw[thick, decorate, decoration={coil, aspect=0}] ($(rhoBottom) + (0.1*\tensorSep, 0.6*\tensorSep)$) -- ($(rhoBottom) + (0.1*\tensorSep, 0)$);%
            \draw[draw,fill=gray!30] ($(rhoTop) - (1.0*\tensorSep, 0.25*\tensorSep)$) rectangle ++(2.0*\tensorSep, 0.5*\tensorSep);%
            \node at (rhoTop) {$\rho_{\text{yellow}}^T$};%
            \draw[draw,fill=gray!30] ($(rhoBottom) - (1.0*\tensorSep, 0.25*\tensorSep)$) rectangle ++(2.0*\tensorSep, 0.5*\tensorSep);%
            \node at (rhoBottom) {$\rho_{\text{yellow}}^B$};%
		\end{scope}%
      \end{tikzpicture}%
            ~=~ %
            \begin{tikzpicture}[baseline=-0.65ex]%
\def\tensorSep{1.1}%
\def\tensorSize{0.2}%
\begin{scope}[shift={(0, 0.75*\tensorSep)}]%
    \coordinate (Vh) at ({0*\tensorSep}, {0*\tensorSep});%
    \coordinate (S) at ({-0.6*\tensorSep}, {-0.75*\tensorSep});%
    \coordinate (U) at ({0*\tensorSep}, {-1.5*\tensorSep});%
    \draw[line width=0.5mm, dashed, dash pattern=on 0.5pt off 0.5pt] ($(Vh) + (-0.8*\tensorSep, 0)$) -- ($(Vh) + (-0.8*\tensorSep, -0.75*\tensorSep)$);%
    \draw[thick] ($(Vh) + (-0.4*\tensorSep, 0)$) -- ($(Vh) + (-0.4*\tensorSep, -0.75*\tensorSep)$);%
    
    \draw[thick, decorate, decoration={coil, aspect=0}] ($(Vh) + (-0.0*\tensorSep, -0.6*\tensorSep)$) -- ($(Vh) + (-0.0*\tensorSep, 0)$);%
    
    \draw[line width=0.5mm, dashed, dash pattern=on 0.5pt off 0.5pt] ($(Vh) + (0.8*\tensorSep, 0)$) -- ($(Vh) + (0.8*\tensorSep, -0.6*\tensorSep)$);%
    \draw[thick] ($(Vh) + (0.4*\tensorSep, 0)$) -- ($(Vh) + (0.4*\tensorSep, -0.6*\tensorSep)$);%
    \draw[line width=0.5mm, dashed, dash pattern=on 0.5pt off 0.5pt] ($(U) + (-0.8*\tensorSep, 0)$) -- ($(U) + (-0.8*\tensorSep, 0.75*\tensorSep)$);%
    \draw[thick] ($(U) + (-0.4*\tensorSep, 0)$) -- ($(U) + (-0.4*\tensorSep, 0.75*\tensorSep)$);%

    \draw[thick, decorate, decoration={coil, aspect=0}] ($(U) + (-0.0*\tensorSep, 0.64*\tensorSep)$) -- ($(U) + (-0.0*\tensorSep, 0)$);%
    
    \draw[line width=0.5mm, dashed, dash pattern=on 0.5pt off 0.5pt] ($(U) + (0.8*\tensorSep, 0)$) -- ($(U) + (0.8*\tensorSep, 0.6*\tensorSep)$);%
    \draw[thick] ($(U) + (0.4*\tensorSep, 0)$) -- ($(U) + (0.4*\tensorSep, 0.6*\tensorSep)$);%
    \draw[draw,fill=gray!30] ($(U) - (1.0*\tensorSep, 0.25*\tensorSep)$) rectangle ++(2.0*\tensorSep, 0.5*\tensorSep);%
    \node at (U) {$U_{\text{yellow}}$};%
    \draw[draw,fill=gray!30] ($(S) - (0.4*\tensorSep, 0.25*\tensorSep)$) rectangle ++(0.8*\tensorSep, 0.5*\tensorSep);%
    \node at (S) {$S_{\text{yellow}}$};%
    \draw[draw,fill=gray!30] ($(Vh) - (1.0*\tensorSep, 0.25*\tensorSep)$) rectangle ++(2.0*\tensorSep, 0.5*\tensorSep);%
    \node at (Vh) {$V_{\text{yellow}}^\dagger$};%
\end{scope}%
\end{tikzpicture}%
        \end{align*}
      \end{minipage}
      \begin{minipage}{\textwidth}
        \centering
        \begin{align*}
            \mathbb{1} \approx~ %
            \begin{tikzpicture}[baseline=-0.65ex]%
		\def\tensorSep{1.1}%
	    \def\tensorSize{0.2}%
		\begin{scope}[shift={(0, 0.75*\tensorSep)}]%
            \coordinate (projTop) at ({0*\tensorSep}, {0*\tensorSep});%
            \coordinate (projBottom) at ({0*\tensorSep}, {-1.5*\tensorSep});%
            \draw[line width=0.5mm, dashed, dash pattern=on 0.5pt off 0.5pt] (projTop) to (projBottom);
            \draw[line width=0.5mm, dashed, dash pattern=on 0.5pt off 0.5pt] ($(projTop) + (-0.5*\tensorSep, 0)$) -- ($(projTop) + (-0.5*\tensorSep, 0.5*\tensorSep)$);%
            \draw[thick] ($(projTop) + (0.5*\tensorSep, 0)$) -- ($(projTop) + (0.5*\tensorSep, 0.5*\tensorSep)$);%
            \draw[line width=0.5mm, dashed, dash pattern=on 0.5pt off 0.5pt] ($(projBottom) + (-0.5*\tensorSep, 0)$) -- ($(projBottom) + (-0.5*\tensorSep, -0.5*\tensorSep)$);%
            \draw[thick] ($(projBottom) + (0.5*\tensorSep, 0)$) -- ($(projBottom) + (0.5*\tensorSep, -0.5*\tensorSep)$);%
            \node[isosceles triangle, isosceles triangle apex angle=100, draw, shape border rotate=-90, fill=googleY] at (projTop) {$P_T$};%
            \node[isosceles triangle, isosceles triangle apex angle=100, minimum size=1, draw, shape border rotate=90, fill=googleY] at (projBottom) {$P_B$};%
		\end{scope}%
      \end{tikzpicture}%
            ~=~~ %
            \begin{tikzpicture}[baseline=-0.65ex]%
		\def\tensorSep{1.1}%
	    \def\tensorSize{0.2}%
		\begin{scope}%
            \coordinate (Uh) at ({0*\tensorSep}, {1.25*\tensorSep});%
            \coordinate (Stop) at ({0.6*\tensorSep}, {0.5*\tensorSep});%
            \coordinate (Sbottom) at ({0.6*\tensorSep}, {-0.5*\tensorSep});%
            \coordinate (V) at ({0*\tensorSep}, {-1.25*\tensorSep});%
            \coordinate (rhoBottom) at ({-1*\tensorSep}, {0.5*\tensorSep});%
            \coordinate (rhoTop) at ({-1*\tensorSep}, {-0.5*\tensorSep});%
            \draw[thick] ($(rhoBottom) + (0.55*\tensorSep, 0)$) -- ($(rhoBottom) + (0.55*\tensorSep, 0.75*\tensorSep)$);%
            \draw[line width=0.5mm, dashed, dash pattern=on 0.5pt off 0.5pt] ($(rhoBottom) + (0.75*\tensorSep, 0)$) -- ($(rhoBottom) + (0.75*\tensorSep, 0.75*\tensorSep)$);%
            \draw[thick, decorate, decoration={coil, aspect=0}] ($(rhoBottom) + (0.15*\tensorSep, 0.75*\tensorSep)$) -- ($(rhoBottom) + (0.15*\tensorSep, 0)$);%
            
            \draw[line width=0.5mm, dashed, dash pattern=on 0.5pt off 0.5pt] ($(rhoBottom) + (-0.65*\tensorSep, 0)$) -- ($(rhoBottom) + (-0.65*\tensorSep, 0.75*\tensorSep)$);%
            
            \draw[thick] ($(rhoBottom) + (-0.25*\tensorSep, 0)$) -- ($(rhoBottom) + (-0.25*\tensorSep, 0.75*\tensorSep)$);%
            \draw[thick] ($(rhoTop) + (0.55*\tensorSep, 0)$) -- ($(rhoTop) + (0.55*\tensorSep, -0.75*\tensorSep)$);%
            
            \draw[line width=0.5mm, dashed, dash pattern=on 0.5pt off 0.5pt] ($(rhoTop) + (0.75*\tensorSep, 0)$) -- ($(rhoTop) + (0.75*\tensorSep, -0.75*\tensorSep)$);%
            
            \draw[thick, decorate, decoration={coil, aspect=0}] ($(rhoTop) + (0.15*\tensorSep, 0)$) -- ($(rhoTop) + (0.15*\tensorSep, -0.75*\tensorSep)$);%
            
            \draw[line width=0.5mm, dashed, dash pattern=on 0.5pt off 0.5pt] ($(rhoTop) + (-0.65*\tensorSep, 0)$) -- ($(rhoTop) + (-0.65*\tensorSep, -0.75*\tensorSep)$);%
            
            \draw[thick] ($(rhoTop) + (-0.25*\tensorSep, 0)$) -- ($(rhoTop) + (-0.25*\tensorSep, -0.75*\tensorSep)$);%
            \draw[line width=0.5mm, dashed, dash pattern=on 0.5pt off 0.5pt] ($(Stop) + (0, 0.1*\tensorSep)$) to ($(Sbottom) + (0, -0.1*\tensorSep)$);%
            
            \draw[line width=0.5mm, dashed, dash pattern=on 0.5pt off 0.5pt, out=90, in=0] (Stop) to ($(Uh) + (0.25*\tensorSep, 0)$);%
            \draw[line width=0.5mm, dashed, dash pattern=on 0.5pt off 0.5pt, out=270, in=0] (Sbottom) to ($(V) + (0.25*\tensorSep, 0)$);%
            %
            \draw[draw,fill=gray!30] ($(V) - (1.0*\tensorSep, 0.25*\tensorSep)$) -- ($(V) + (-1.0*\tensorSep, 0.25*\tensorSep)$) -- ($(V) + (0*\tensorSep, 0.25*\tensorSep)$) -- ($(V) + (0.26*\tensorSep, 0*\tensorSep)$) -- ($(V) + (0*\tensorSep, -0.25*\tensorSep)$) -- cycle;
            \node at ($(V) + (-0.375*\tensorSep, 0)$) {$\widetilde{V}_{\text{yellow}}$};%
            \draw[draw,fill=gray!30] ($(Stop) - (0.4*\tensorSep, 0.25*\tensorSep)$) rectangle ++(0.8*\tensorSep, 0.5*\tensorSep);%
            \node at (Stop) {$\widetilde{S}_{\text{yellow}}^{-1/2}$};%
            \draw[draw,fill=gray!30] ($(Sbottom) - (0.4*\tensorSep, 0.25*\tensorSep)$) rectangle ++(0.8*\tensorSep, 0.5*\tensorSep);%
            \node at (Sbottom) {$\widetilde{S}_{\text{yellow}}^{-1/2}$};%
            %
            \draw[draw,fill=gray!30] ($(Uh) - (1.0*\tensorSep, 0.25*\tensorSep)$) -- ($(Uh) + (-1.0*\tensorSep, 0.25*\tensorSep)$) -- ($(Uh) + (0*\tensorSep, 0.25*\tensorSep)$) -- ($(Uh) + (0.26*\tensorSep, 0*\tensorSep)$) -- ($(Uh) + (0*\tensorSep, -0.25*\tensorSep)$) -- cycle;
            \node at ($(Uh) + (-0.375*\tensorSep, 0)$) {$\widetilde{U}_{\text{yellow}}^\dagger$};%
            \draw[draw,fill=gray!30] ($(rhoBottom) - (0.85*\tensorSep, 0.25*\tensorSep)$) rectangle ++(1.7*\tensorSep, 0.5*\tensorSep);%
            \node at (rhoBottom) {$\rho_{\text{yellow}}^B$};%
            \draw[draw,fill=gray!30] ($(rhoTop) - (0.85*\tensorSep, 0.25*\tensorSep)$) rectangle ++(1.7*\tensorSep, 0.5*\tensorSep);%
            \node at (rhoTop) {$\rho_{\text{yellow}}^T$};%
		\end{scope}%
      \end{tikzpicture}%
        \end{align*}
      \end{minipage}
    \end{minipage}
    \caption{Construction of the yellow projectors used in the left absorption step in Fig.~\ref{fig:left_absorption_full}. A truncated SVD of $\mathcal M_\text{yellow}$ is used to generate an approximate resolution of the identity, from which the final projectors are defined.}
    \label{fig:constructionYellowProjectors}
\end{figure*}
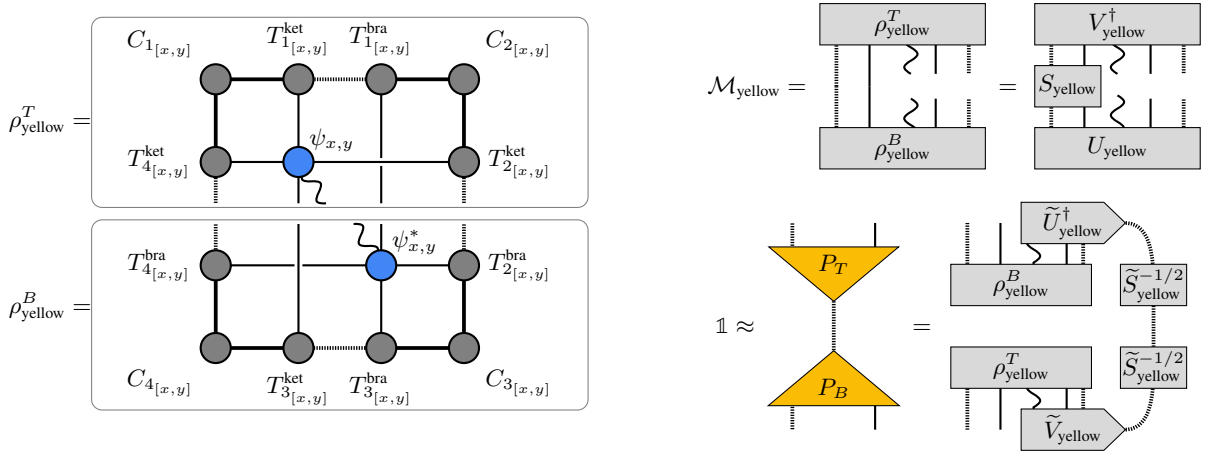

\begin{figure*}[t]
    \centering
    \begin{minipage}{1.1\columnwidth}
        \input{figures/rhoT_rhoB_red_proj}
    \end{minipage}
    \begin{minipage}{.9\columnwidth}
      \begin{minipage}{\textwidth}
        \begin{align*}
          \mathcal{M}_{\text{red}} &=~ %
             \begin{tikzpicture}[baseline=-0.65ex]%
    \def\tensorSep{1.1}%
    \def\tensorSize{0.2}%
    \begin{scope}[shift={(0, 0.75*\tensorSep)}]%
        \coordinate (rhoTop) at ({0*\tensorSep}, {0*\tensorSep});%
        \coordinate (rhoBottom) at ({0*\tensorSep}, {-1.5*\tensorSep});%
        \draw[line width=0.5mm, dashed, dash pattern=on 0.5pt off 0.5pt] ($(rhoTop) + (-0.8*\tensorSep, 0)$) -- ($(rhoTop) + (-0.8*\tensorSep, -0.75*\tensorSep)$);%
        
        \draw[thick] ($(rhoTop) + (-0.4*\tensorSep, 0)$) -- ($(rhoTop) + (-0.4*\tensorSep, -0.75*\tensorSep)$);%
        
        \draw[thick, decorate, decoration={coil, aspect=0}] ($(rhoBottom) + (-0.6*\tensorSep, 1.6*\tensorSep)$) -- ($(rhoBottom) + (-0.6*\tensorSep, 0)$);%

        \draw[line width=0.5mm, dashed, dash pattern=on 0.5pt off 0.5pt] ($(rhoTop) + (0.8*\tensorSep, 0)$) -- ($(rhoTop) + (0.8*\tensorSep, -0.6*\tensorSep)$);%
        \draw[thick] ($(rhoTop) + (0.4*\tensorSep, 0)$) -- ($(rhoTop) + (0.4*\tensorSep, -0.6*\tensorSep)$);%
        \draw[thick, decorate, decoration={coil, aspect=0}] ($(rhoTop) + (0.6*\tensorSep, -0.6*\tensorSep)$) -- ($(rhoTop) + (0.6*\tensorSep, 0)$);%

        \draw[thick, decorate, decoration={coil, aspect=0}] ($(rhoBottom) + (0.6*\tensorSep, 0.6*\tensorSep)$) -- ($(rhoBottom) + (0.6*\tensorSep, 0)$);%
        
        \draw[line width=0.5mm, dashed, dash pattern=on 0.5pt off 0.5pt] ($(rhoBottom) + (-0.8*\tensorSep, 0)$) -- ($(rhoBottom) + (-0.8*\tensorSep, 0.75*\tensorSep)$);%
        \draw[thick] ($(rhoBottom) + (-0.4*\tensorSep, 0)$) -- ($(rhoBottom) + (-0.4*\tensorSep, 0.75*\tensorSep)$);%
        \draw[line width=0.5mm, dashed, dash pattern=on 0.5pt off 0.5pt] ($(rhoBottom) + (0.8*\tensorSep, 0)$) -- ($(rhoBottom) + (0.8*\tensorSep, 0.6*\tensorSep)$);%
        \draw[thick] ($(rhoBottom) + (0.4*\tensorSep, 0)$) -- ($(rhoBottom) + (0.4*\tensorSep, 0.6*\tensorSep)$);%
        \draw[draw,fill=gray!30] ($(rhoTop) - (1.0*\tensorSep, 0.25*\tensorSep)$) rectangle ++(2.0*\tensorSep, 0.5*\tensorSep);%
        \node at (rhoTop) {$\rho_{\text{red}}^T$};%
        \draw[draw,fill=gray!30] ($(rhoBottom) - (1.0*\tensorSep, 0.25*\tensorSep)$) rectangle ++(2.0*\tensorSep, 0.5*\tensorSep);%
        \node at (rhoBottom) {$\rho_{\text{red}}^B$};%
    \end{scope}%
  \end{tikzpicture}%
            ~=~ 
            \begin{tikzpicture}[baseline=-0.65ex]%
\def\tensorSep{1.1}%
\def\tensorSize{0.2}%
\begin{scope}[shift={(0, 0.75*\tensorSep)}]%
    \coordinate (Vh) at ({0*\tensorSep}, {0*\tensorSep});%
    \coordinate (S) at ({-0.6*\tensorSep}, {-0.75*\tensorSep});%
    \coordinate (U) at ({0*\tensorSep}, {-1.5*\tensorSep});%
    \draw[line width=0.5mm, dashed, dash pattern=on 0.5pt off 0.5pt] ($(Vh) + (-0.8*\tensorSep, 0)$) -- ($(Vh) + (-0.8*\tensorSep, -0.75*\tensorSep)$);%
    \draw[thick] ($(Vh) + (-0.4*\tensorSep, 0)$) -- ($(Vh) + (-0.4*\tensorSep, -0.75*\tensorSep)$);%
    
    \draw[line width=0.5mm, dashed, dash pattern=on 0.5pt off 0.5pt] ($(Vh) + (0.8*\tensorSep, 0)$) -- ($(Vh) + (0.8*\tensorSep, -0.6*\tensorSep)$);%
    \draw[thick] ($(Vh) + (0.4*\tensorSep, 0)$) -- ($(Vh) + (0.4*\tensorSep, -0.6*\tensorSep)$);%
    \draw[line width=0.5mm, dashed, dash pattern=on 0.5pt off 0.5pt] ($(U) + (-0.8*\tensorSep, 0)$) -- ($(U) + (-0.8*\tensorSep, 0.75*\tensorSep)$);%
    \draw[thick] ($(U) + (-0.4*\tensorSep, 0)$) -- ($(U) + (-0.4*\tensorSep, 0.75*\tensorSep)$);%

    \draw[thick, decorate, decoration={coil, aspect=0}] ($(rhoTop) + (0.6*\tensorSep, -0.6*\tensorSep)$) -- ($(rhoTop) + (0.6*\tensorSep, 0)$);%

    \draw[thick, decorate, decoration={coil, aspect=0}] ($(rhoBottom) + (-0.6*\tensorSep, 1.6*\tensorSep)$) -- ($(rhoBottom) + (-0.6*\tensorSep, 0)$);%
    \draw[thick, decorate, decoration={coil, aspect=0}] ($(rhoBottom) + (0.6*\tensorSep, 0.6*\tensorSep)$) -- ($(rhoBottom) + (0.6*\tensorSep, 0)$);%

    \draw[line width=0.5mm, dashed, dash pattern=on 0.5pt off 0.5pt] ($(U) + (0.8*\tensorSep, 0)$) -- ($(U) + (0.8*\tensorSep, 0.6*\tensorSep)$);%
    \draw[thick] ($(U) + (0.4*\tensorSep, 0)$) -- ($(U) + (0.4*\tensorSep, 0.6*\tensorSep)$);%
    \draw[draw,fill=gray!30] ($(U) - (1.0*\tensorSep, 0.25*\tensorSep)$) rectangle ++(2.0*\tensorSep, 0.5*\tensorSep);%
    \node at (U) {$U_{\text{teal}}$};%
    \draw[draw,fill=gray!30] ($(S) - (0.4*\tensorSep, 0.25*\tensorSep)$) rectangle ++(0.8*\tensorSep, 0.5*\tensorSep);%
    \node at (S) {$S_{\text{red}}$};%
    \draw[draw,fill=gray!30] ($(Vh) - (1.0*\tensorSep, 0.25*\tensorSep)$) rectangle ++(2.0*\tensorSep, 0.5*\tensorSep);%
    \node at (Vh) {$V_{\text{red}}^\dagger$};%
\end{scope}%
\end{tikzpicture}%
        \end{align*}
      \end{minipage}
      \begin{minipage}{\textwidth}
        \begin{align*}
            \mathbb{1} \approx~ %
            \begin{tikzpicture}[baseline=-0.65ex]%
		\def\tensorSep{1.1}%
	    \def\tensorSize{0.2}%
		\begin{scope}[shift={(0, 0.75*\tensorSep)}]%
            \coordinate (projTop) at ({0*\tensorSep}, {0*\tensorSep});%
            \coordinate (projBottom) at ({0*\tensorSep}, {-1.5*\tensorSep});%
            \draw[line width=0.5mm, dashed, dash pattern=on 0.5pt off 0.5pt] (projTop) to (projBottom);
            \draw[line width=0.5mm, dashed, dash pattern=on 0.5pt off 0.5pt] ($(projTop) + (-0.5*\tensorSep, 0)$) -- ($(projTop) + (-0.5*\tensorSep, 0.5*\tensorSep)$);%
            \draw[thick, decorate, decoration={coil, aspect=0}] ($(projTop) + (-0.0*\tensorSep, 0)$) -- ($(projTop) + (-0.0*\tensorSep, 0.5*\tensorSep)$);
            \draw[thick] ($(projTop) + (0.5*\tensorSep, 0)$) -- ($(projTop) + (0.5*\tensorSep, 0.5*\tensorSep)$);%
            \draw[line width=0.5mm, dashed, dash pattern=on 0.5pt off 0.5pt] ($(projBottom) + (-0.5*\tensorSep, 0)$) -- ($(projBottom) + (-0.5*\tensorSep, -0.5*\tensorSep)$);%
            \draw[thick, decorate, decoration={coil, aspect=0}] ($(projBottom) + (-0.0*\tensorSep, 0)$) -- ($(projBottom) + (-0.0*\tensorSep, -0.5*\tensorSep)$);%
            %
            \draw[thick] ($(projBottom) + (0.5*\tensorSep, 0)$) -- ($(projBottom) + (0.5*\tensorSep, -0.5*\tensorSep)$);%
            \node[isosceles triangle, isosceles triangle apex angle=100, draw, shape border rotate=-90, fill=googleR] at (projTop) {$P_T$};%
            \node[isosceles triangle, isosceles triangle apex angle=100, minimum size=1, draw, shape border rotate=90, fill=googleR] at (projBottom) {$P_B$};%
		\end{scope}%
      \end{tikzpicture}%
            ~=~~ %
                  \begin{tikzpicture}[baseline=-0.65ex]%
		\def\tensorSep{1.1}%
	    \def\tensorSize{0.2}%
		\begin{scope}%
            \coordinate (Uh) at ({0*\tensorSep}, {1.25*\tensorSep});%
            \coordinate (Stop) at ({0.6*\tensorSep}, {0.5*\tensorSep});%
            \coordinate (Sbottom) at ({0.6*\tensorSep}, {-0.5*\tensorSep});%
            \coordinate (V) at ({0*\tensorSep}, {-1.25*\tensorSep});%
            \coordinate (rhoBottom) at ({-1*\tensorSep}, {0.5*\tensorSep});%
            \coordinate (rhoTop) at ({-1*\tensorSep}, {-0.5*\tensorSep});%
            \draw[line width=0.5mm, dashed, dash pattern=on 0.5pt off 0.5pt] ($(rhoBottom) + (0.55*\tensorSep, 0)$) -- ($(rhoBottom) + (0.55*\tensorSep, 0.75*\tensorSep)$);%
            \draw[thick] ($(rhoBottom) + (0.15*\tensorSep, 0)$) -- ($(rhoBottom) + (0.15*\tensorSep, 0.75*\tensorSep)$);%
            \draw[thick, decorate, decoration={coil, aspect=0}] ($(rhoBottom) + (0.35*\tensorSep, 0.75*\tensorSep)$) -- ($(rhoBottom) + (0.35*\tensorSep, 0)$);%
            \draw[line width=0.5mm, dashed, dash pattern=on 0.5pt off 0.5pt] ($(rhoBottom) + (-0.65*\tensorSep, 0)$) -- ($(rhoBottom) + (-0.65*\tensorSep, 0.75*\tensorSep)$);%
            \draw[thick] ($(rhoBottom) + (-0.15*\tensorSep, 0)$) -- ($(rhoBottom) + (-0.15*\tensorSep, 0.75*\tensorSep)$);%
            \draw[thick, decorate, decoration={coil, aspect=0}] ($(rhoBottom) + (-0.4*\tensorSep, 0.75*\tensorSep)$) -- ($(rhoBottom) + (-0.4*\tensorSep, 0)$);%

            \draw[line width=0.5mm, dashed, dash pattern=on 0.5pt off 0.5pt] ($(rhoTop) + (0.55*\tensorSep, 0)$) -- ($(rhoTop) + (0.55*\tensorSep, -0.75*\tensorSep)$);%
            \draw[thick] ($(rhoTop) + (0.15*\tensorSep, 0)$) -- ($(rhoTop) + (0.15*\tensorSep, -0.75*\tensorSep)$);%
            \draw[thick, decorate, decoration={coil, aspect=0}] ($(rhoTop) + (0.35*\tensorSep, -0.75*\tensorSep)$) -- ($(rhoTop) + (0.35*\tensorSep, 0)$);%

            \draw[line width=0.5mm, dashed, dash pattern=on 0.5pt off 0.5pt] ($(rhoTop) + (-0.65*\tensorSep, 0)$) -- ($(rhoTop) + (-0.65*\tensorSep, -0.75*\tensorSep)$);%
            \draw[thick] ($(rhoTop) + (-0.15*\tensorSep, 0)$) -- ($(rhoTop) + (-0.15*\tensorSep, -0.75*\tensorSep)$);%
            \draw[thick, decorate, decoration={coil, aspect=0}] ($(rhoTop) + (-0.4*\tensorSep, -0.75*\tensorSep)$) -- ($(rhoTop) + (-0.4*\tensorSep, 0)$);%
            \draw[line width=0.5mm, dashed, dash pattern=on 0.5pt off 0.5pt] ($(Stop) + (0, 0.1*\tensorSep)$) to ($(Sbottom) + (0, -0.1*\tensorSep)$);%
            \draw[line width=0.5mm, dashed, dash pattern=on 0.5pt off 0.5pt, out=90, in=0] ($(Stop) + (0, 0.25*\tensorSep)$) to ($(Uh) + (0.25*\tensorSep, 0)$);%
            \draw[line width=0.5mm, dashed, dash pattern=on 0.5pt off 0.5pt, out=270, in=0] ($(Sbottom) + (0, -0.25*\tensorSep)$) to ($(V) + (0.25*\tensorSep, 0)$);%
            %
            \draw[draw,fill=gray!30] ($(V) - (1.0*\tensorSep, 0.25*\tensorSep)$) -- ($(V) + (-1.0*\tensorSep, 0.25*\tensorSep)$) -- ($(V) + (0*\tensorSep, 0.25*\tensorSep)$) -- ($(V) + (0.26*\tensorSep, 0*\tensorSep)$) -- ($(V) + (0*\tensorSep, -0.25*\tensorSep)$) -- cycle;
            \node at ($(V) + (-0.375*\tensorSep, 0)$) {$\widetilde{V}_{\text{red}}$};%
            \draw[draw,fill=gray!30] ($(Stop) - (0.4*\tensorSep, 0.25*\tensorSep)$) rectangle ++(0.8*\tensorSep, 0.5*\tensorSep);%
            \node at (Stop) {$\widetilde{S}_{\text{red}}^{-1/2}$};%
            \draw[draw,fill=gray!30] ($(Sbottom) - (0.4*\tensorSep, 0.25*\tensorSep)$) rectangle ++(0.8*\tensorSep, 0.5*\tensorSep);%
            \node at (Sbottom) {$\widetilde{S}_{\text{red}}^{-1/2}$};%
            %
            \draw[draw,fill=gray!30] ($(Uh) - (1.0*\tensorSep, 0.25*\tensorSep)$) -- ($(Uh) + (-1.0*\tensorSep, 0.25*\tensorSep)$) -- ($(Uh) + (0*\tensorSep, 0.25*\tensorSep)$) -- ($(Uh) + (0.26*\tensorSep, 0*\tensorSep)$) -- ($(Uh) + (0*\tensorSep, -0.25*\tensorSep)$) -- cycle;%
            \node at ($(Uh) + (-0.375*\tensorSep, 0)$) {$\widetilde{U}_{\text{red}}^\dagger$};%
            \draw[draw,fill=gray!30] ($(rhoBottom) - (0.85*\tensorSep, 0.25*\tensorSep)$) rectangle ++(1.7*\tensorSep, 0.5*\tensorSep);%
            \node at (rhoBottom) {$\rho_{\text{red}}^B$};%
            \draw[draw,fill=gray!30] ($(rhoTop) - (0.85*\tensorSep, 0.25*\tensorSep)$) rectangle ++(1.7*\tensorSep, 0.5*\tensorSep);%
            \node at (rhoTop) {$\rho_{\text{red}}^T$};%
		\end{scope}%
      \end{tikzpicture}%
        \end{align*}
      \end{minipage}
    \end{minipage}
    \caption{Construction of the red projectors used in the left absorption step in Fig.~\ref{fig:left_absorption_full}. A truncated SVD of $\mathcal M_\text{red}$ is used to generate an approximate resolution of the identity, 
    from which the final projectors are defined.}
    \label{fig:constructionRedProjectors}
\end{figure*}

The two remaining projectors that we need for a full left move, i.e., the yellow and red projectors, have one notable difference to the ones we have discussed so far. 
This is due to the fact that they are inserted between the \textit{bra}- and \textit{ket}-layer transfer tensors to truncate to the newly introduced \textit{interlayer} environment bond dimension $\chi_I$. 
While the yellow projectors do not renormalize the physical Hilbert spaces of the PEPS tensors, they however do appear as open indices in the canonical choice of the initial networks for $\rho_{\text{yellow}}^B$ and $\rho^T_{\text{yellow}}$, as shown in Fig.~\ref{fig:constructionYellowProjectors}. 
The spaces we do want to truncate with those yellow projectors are then contracted followed by a singular value decomposition of the resulting tensor, ultimately resulting in the final projectors. 
Here, the singular values are truncated to the interlayer bond dimension $\chi_I$.
We note that it is computationally advantageous to perform a singular value decomposition on tensors $\rho_{\text{yellow}}^B$ and $\rho_{\text{yellow}}^T$ before multiplying them together. 
While a similar step has been introduced in Ref.~\cite{Fishman2018} to achieve a better precondition for the inversion of the singular values, here it has the additional benefit of reducing the computational cost of a subsequent operation. 
In fact, since $\rho_{\text{yellow}}^B$ and $\rho_{\text{yellow}}^T$ only have rank $\chi_I \cdot \chi_B$, we can truncate to this bond dimension using the two preceding decompositions. 
This leads to a reduced cost in the successive SVD, from which the final projectors are computed (cf.\ Fig.~\ref{fig:constructionYellowProjectors}).

Lastly, the red projectors in the left absorption step of Fig.~\ref{fig:left_absorption_full} act on vector spaces that have already undergone projection with the yellow projectors. 
Hence, analogous to the teal projectors, we define the networks for $\rho^B_{\text{red}}$ and $\rho^T_{\text{red}}$ with these yellow projectors included, as shown in Fig.~\ref{fig:constructionRedProjectors}. 
The red projects then renormalize the enlarged space back to the interlayer bond dimension $\chi_I$.
Contrary to the three previous one however, they are the only ones to include the local physical Hilbert spaces in the truncation.

\section{Complexity and alternative projectors}
\label{app:complexity and alternative projectors}

\begin{figure}[tbh]
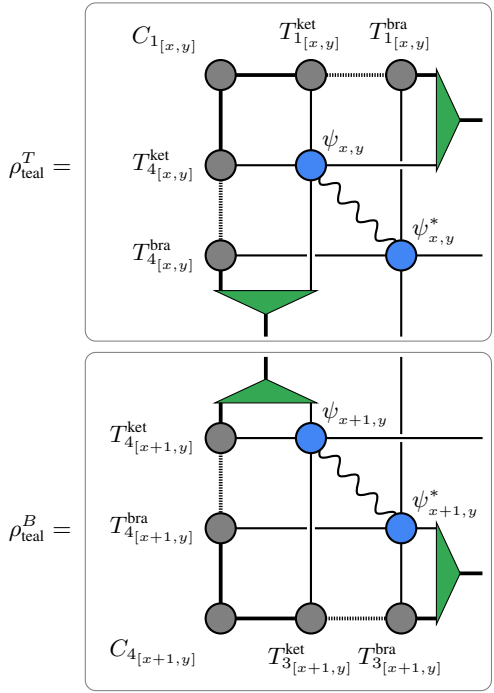

    \centering
    \include{figures/rho_T_and_B_teal_half_proj}
    \caption{Reduced initial networks $\rho^B_{\text{teal}}$ and $\rho^T_{\text{teal}}$ for the construction of teal half projectors.}
    \label{fig:constructionTealHalfProjectors}
\end{figure}

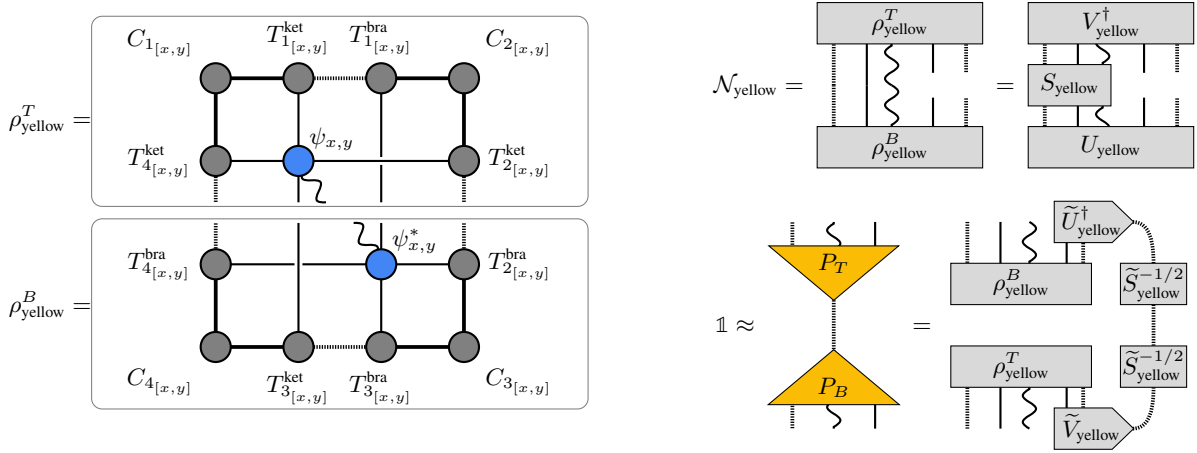
\begin{figure*}[t]
    \centering
    \begin{minipage}{1.1\columnwidth}
            \begin{tikzpicture}
		\def\tensorSep{1.1}
	    \def\tensorSize{0.2}

		\begin{scope}[shift = {(0.0, 0.25*\tensorSep)}]
            \coordinate (ket) at ({-0.5*\tensorSep}, {+0.5*\tensorSep});
            \coordinate (bra) at ({+0.5*\tensorSep}, {-0.5*\tensorSep});
  
			\coordinate (C1) at ({-1.5*\tensorSep}, {+1.5*\tensorSep});
            \coordinate (T1ket) at ({-0.5*\tensorSep}, {+1.5*\tensorSep});
            \coordinate (T1bra) at ({+0.5*\tensorSep}, {+1.5*\tensorSep});
            \coordinate (C2) at ({+1.5*\tensorSep}, {+1.5*\tensorSep});
            \coordinate (T2ket) at ({+1.5*\tensorSep}, {+0.5*\tensorSep});
            \coordinate (T2bra) at ({+1.5*\tensorSep}, {-0.5*\tensorSep});
            \coordinate (C3) at ({+1.5*\tensorSep}, {-1.5*\tensorSep});
            \coordinate (T3bra) at ({+0.5*\tensorSep}, {-1.5*\tensorSep});
            \coordinate (T3ket) at ({-0.5*\tensorSep}, {-1.5*\tensorSep});
            \coordinate (C4) at ({-1.5*\tensorSep}, {-1.5*\tensorSep});
            \coordinate (T4bra) at ({-1.5*\tensorSep}, {-0.5*\tensorSep});
            \coordinate (T4ket) at ({-1.5*\tensorSep}, {+0.5*\tensorSep});

            \coordinate (text) at ({-3.5*\tensorSep}, {+1*\tensorSep});

            \draw[rounded corners, gray] ({-3*\tensorSep}, {+2.25*\tensorSep}) --  ({3*\tensorSep}, {+2.25*\tensorSep}) -- ({3*\tensorSep}, {-0.05*\tensorSep}) -- ({-3*\tensorSep}, {-0.05*\tensorSep}) -- cycle;

            \draw[thick, decorate, decoration={coil, aspect=0}] ($(ket) + (0.3333333333333333*\tensorSep, -0.5*\tensorSep)$) to (ket);


            \draw[line width=0.5mm,] (C1) to (T1ket);
            \draw[line width=0.5mm, dashed, dash pattern=on 0.5pt off 0.5pt] (T1ket) to (T1bra);
            \draw[line width=0.5mm,] (T1bra) to (C2);
            \draw[line width=0.5mm,] (C2) to (T2ket);
            \draw[line width=0.5mm,] (T4ket) to (C1);

            \draw[thick] (ket) to ($(ket) + (0, -0.5*\tensorSep)$);
            \draw[line width=0.5mm, dashed, dash pattern=on 0.5pt off 0.5pt] (T4ket) to ($(T4ket) + (0, -0.5*\tensorSep)$);
            \draw[line width=0.5mm, dashed, dash pattern=on 0.5pt off 0.5pt] (T2ket) to ($(T2ket) + (0, -0.5*\tensorSep)$);
            \draw[thick] (T1bra) to ($(T1bra) + (0, -1.5*\tensorSep)$);

            \draw[thick] (ket) to (T1ket);
            \draw[white, line width=3pt] (ket) to (T2ket);
            \draw[thick] (ket) to (T2ket);
            \draw[thick] (ket) to (T4ket);

            \node[above right=1] at (ket) {$\psi_{x,y}$};
            \draw[thick, fill=googleB] (ket) circle (\tensorSize);

			\node[above left=5] at (C1) {$C_{1_{[x, y]}}$};
			\draw[thick, fill = gray] (C1) circle (\tensorSize);

            \node[above=6] at (T1ket) {$T_{1_{[x, y]}}^\text{ket}$};
			\draw[thick, fill = gray] (T1ket) circle (\tensorSize);

            \node[above=6] at (T1bra) {$T_{1_{[x, y]}}^\text{bra}$};
			\draw[thick, fill = gray] (T1bra) circle (\tensorSize);

            \node[above right=5] at (C2) {$C_{2_{[x, y]}}$};
			\draw[thick, fill = gray] (C2) circle (\tensorSize);

            \node[right=6] at (T2ket) {$T_{2_{[x, y]}}^\text{ket}$};
			\draw[thick, fill = gray] (T2ket) circle (\tensorSize);







            \node[left=6] at (T4ket) {$T_{4_{[x, y]}}^\text{ket}$};
			\draw[thick, fill = gray] (T4ket) circle (\tensorSize);

            \node at (text) {$\rho_{\text{yellow}}^T =$};
		\end{scope}

        \begin{scope}[shift = {(0.0, 0.0*\tensorSep)}]
            \coordinate (ket) at ({-0.5*\tensorSep}, {+0.5*\tensorSep});
            \coordinate (bra) at ({+0.5*\tensorSep}, {-0.5*\tensorSep});
  
			\coordinate (C1) at ({-1.5*\tensorSep}, {+1.5*\tensorSep});
            \coordinate (T1ket) at ({-0.5*\tensorSep}, {+1.5*\tensorSep});
            \coordinate (T1bra) at ({+0.5*\tensorSep}, {+1.5*\tensorSep});
            \coordinate (C2) at ({+1.5*\tensorSep}, {+1.5*\tensorSep});
            \coordinate (T2ket) at ({+1.5*\tensorSep}, {+0.5*\tensorSep});
            \coordinate (T2bra) at ({+1.5*\tensorSep}, {-0.5*\tensorSep});
            \coordinate (C3) at ({+1.5*\tensorSep}, {-1.5*\tensorSep});
            \coordinate (T3bra) at ({+0.5*\tensorSep}, {-1.5*\tensorSep});
            \coordinate (T3ket) at ({-0.5*\tensorSep}, {-1.5*\tensorSep});
            \coordinate (C4) at ({-1.5*\tensorSep}, {-1.5*\tensorSep});
            \coordinate (T4bra) at ({-1.5*\tensorSep}, {-0.5*\tensorSep});
            \coordinate (T4ket) at ({-1.5*\tensorSep}, {+0.5*\tensorSep});

            \coordinate (text) at ({-3.5*\tensorSep}, {-1.0*\tensorSep});

            \draw[rounded corners, gray] ({-3*\tensorSep}, {+0.05*\tensorSep}) --  ({3*\tensorSep}, {+0.05*\tensorSep}) -- ({3*\tensorSep}, {-2.25*\tensorSep}) -- ({-3*\tensorSep}, {-2.25*\tensorSep}) -- cycle;

            \draw[thick, decorate, decoration={coil, aspect=0}] ($(bra) + (-0.3333333333333333*\tensorSep, 0.5*\tensorSep)$) to (bra);


            \draw[thick] (bra) to (T2bra);
            \draw[thick] (bra) to (T3bra);
            \draw[thick] (bra) to (T4bra);

            \draw[line width=0.5mm,] (T2bra) to (C3);
            \draw[line width=0.5mm,] (C3) to (T3bra);
            \draw[line width=0.5mm, dashed, dash pattern=on 0.5pt off 0.5pt] (T3bra) to (T3ket);
            \draw[line width=0.5mm,] (T3ket) to (C4);
            \draw[line width=0.5mm,] (C4) to (T4bra);

            \draw[thick] (bra) to ($(bra) + (0, 0.5*\tensorSep)$);
            \draw[white, line width=3pt] (T3ket) to ($(T3ket) + (0, 1.5*\tensorSep)$);
            \draw[thick] (T3ket) to ($(T3ket) + (0, 1.5*\tensorSep)$);
            \draw[line width=0.5mm, dashed, dash pattern=on 0.5pt off 0.5pt] (T2bra) to ($(T2bra) + (0, 0.5*\tensorSep)$);
            \draw[line width=0.5mm, dashed, dash pattern=on 0.5pt off 0.5pt] (T4bra) to ($(T4bra) + (0, 0.5*\tensorSep)$);

            \node[above right=1] at (bra) {$\psi^{*}_{x,y}$};
            \draw[thick, fill=googleB] (bra) circle (\tensorSize);






            \node[right=6] at (T2bra) {$T_{2_{[x, y]}}^\text{bra}$};
			\draw[thick, fill = gray] (T2bra) circle (\tensorSize);

            \node[below right=5] at (C3) {$C_{3_{[x, y]}}$};
			\draw[thick, fill = gray] (C3) circle (\tensorSize);

            \node[below=6] at (T3bra) {$T_{3_{[x, y]}}^\text{bra}$};
			\draw[thick, fill = gray] (T3bra) circle (\tensorSize);

            \node[below=6] at (T3ket) {$T_{3_{[x, y]}}^\text{ket}$};
			\draw[thick, fill = gray] (T3ket) circle (\tensorSize);

            \node[below left=5] at (C4) {$C_{4_{[x, y]}}$};
			\draw[thick, fill = gray] (C4) circle (\tensorSize);

            \node[left=6] at (T4bra) {$T_{4_{[x, y]}}^\text{bra}$};
			\draw[thick, fill = gray] (T4bra) circle (\tensorSize);


            \node at (text) {$\rho_{\text{yellow}}^B =$};
		\end{scope}
    \end{tikzpicture}
    \end{minipage}
    \begin{minipage}{.9\columnwidth}
      \begin{minipage}{\textwidth}
        \begin{align*}
            \mathcal{N}_{\text{yellow}} &=~ %
            \begin{tikzpicture}[baseline=-0.65ex]%
		\def\tensorSep{1.1}%
	    \def\tensorSize{0.2}%
		\begin{scope}[shift={(0, 0.75*\tensorSep)}]%
            \coordinate (rhoTop) at ({0*\tensorSep}, {0*\tensorSep});%
            \coordinate (rhoBottom) at ({0*\tensorSep}, {-1.5*\tensorSep});%
            \draw[line width=0.5mm, dashed, dash pattern=on 0.5pt off 0.5pt] ($(rhoTop) + (-0.8*\tensorSep, 0)$) -- ($(rhoTop) + (-0.8*\tensorSep, -0.75*\tensorSep)$);%
            \draw[thick] ($(rhoTop) + (-0.4*\tensorSep, 0)$) -- ($(rhoTop) + (-0.4*\tensorSep, -0.75*\tensorSep)$);%
            \draw[line width=0.5mm, dashed, dash pattern=on 0.5pt off 0.5pt] ($(rhoTop) + (0.8*\tensorSep, 0)$) -- ($(rhoTop) + (0.8*\tensorSep, -0.6*\tensorSep)$);%
            \draw[thick] ($(rhoTop) + (0.4*\tensorSep, 0)$) -- ($(rhoTop) + (0.4*\tensorSep, -0.6*\tensorSep)$);%
            \draw[thick, decorate, decoration={coil, aspect=0}] ($(rhoTop) + (-0.1*\tensorSep, -0.75*\tensorSep)$) -- ($(rhoTop) + (-0.1*\tensorSep, 0)$);%
            \draw[line width=0.5mm, dashed, dash pattern=on 0.5pt off 0.5pt] ($(rhoBottom) + (-0.8*\tensorSep, 0)$) -- ($(rhoBottom) + (-0.8*\tensorSep, 0.75*\tensorSep)$);%
            \draw[thick] ($(rhoBottom) + (-0.4*\tensorSep, 0)$) -- ($(rhoBottom) + (-0.4*\tensorSep, 0.75*\tensorSep)$);%
            \draw[line width=0.5mm, dashed, dash pattern=on 0.5pt off 0.5pt] ($(rhoBottom) + (0.8*\tensorSep, 0)$) -- ($(rhoBottom) + (0.8*\tensorSep, 0.6*\tensorSep)$);%
            \draw[thick] ($(rhoBottom) + (0.4*\tensorSep, 0)$) -- ($(rhoBottom) + (0.4*\tensorSep, 0.6*\tensorSep)$);%
            \draw[thick, decorate, decoration={coil, aspect=0}] ($(rhoBottom) + (-0.1*\tensorSep, 0.75*\tensorSep)$) -- ($(rhoBottom) + (-0.1*\tensorSep, 0)$);%
            \draw[draw,fill=gray!30] ($(rhoTop) - (1.0*\tensorSep, 0.25*\tensorSep)$) rectangle ++(2.0*\tensorSep, 0.5*\tensorSep);%
            \node at (rhoTop) {$\rho_{\text{yellow}}^T$};%
            \draw[draw,fill=gray!30] ($(rhoBottom) - (1.0*\tensorSep, 0.25*\tensorSep)$) rectangle ++(2.0*\tensorSep, 0.5*\tensorSep);%
            \node at (rhoBottom) {$\rho_{\text{yellow}}^B$};%
		\end{scope}%
      \end{tikzpicture}%
            ~=~ %
            \begin{tikzpicture}[baseline=-0.65ex]%
\def\tensorSep{1.1}%
\def\tensorSize{0.2}%
\begin{scope}[shift={(0, 0.75*\tensorSep)}]%
    \coordinate (Vh) at ({0*\tensorSep}, {0*\tensorSep});%
    \coordinate (S) at ({-0.6*\tensorSep}, {-0.75*\tensorSep});%
    \coordinate (U) at ({0*\tensorSep}, {-1.5*\tensorSep});%
    \draw[line width=0.5mm, dashed, dash pattern=on 0.5pt off 0.5pt] ($(Vh) + (-0.8*\tensorSep, 0)$) -- ($(Vh) + (-0.8*\tensorSep, -0.75*\tensorSep)$);%
    \draw[thick] ($(Vh) + (-0.4*\tensorSep, 0)$) -- ($(Vh) + (-0.4*\tensorSep, -0.75*\tensorSep)$);%
    
    \draw[thick, decorate, decoration={coil, aspect=0}] ($(Vh) + (-0.1*\tensorSep, 0)$) -- ($(Vh) + (-0.1*\tensorSep, -0.6*\tensorSep)$);%
    
    \draw[line width=0.5mm, dashed, dash pattern=on 0.5pt off 0.5pt] ($(Vh) + (0.8*\tensorSep, 0)$) -- ($(Vh) + (0.8*\tensorSep, -0.6*\tensorSep)$);%
    \draw[thick] ($(Vh) + (0.4*\tensorSep, 0)$) -- ($(Vh) + (0.4*\tensorSep, -0.6*\tensorSep)$);%
    \draw[line width=0.5mm, dashed, dash pattern=on 0.5pt off 0.5pt] ($(U) + (-0.8*\tensorSep, 0)$) -- ($(U) + (-0.8*\tensorSep, 0.75*\tensorSep)$);%
    \draw[thick] ($(U) + (-0.4*\tensorSep, 0)$) -- ($(U) + (-0.4*\tensorSep, 0.75*\tensorSep)$);%

    \draw[thick, decorate, decoration={coil, aspect=0}] ($(U) + (-0.1*\tensorSep, 0)$) -- ($(U) + (-0.1*\tensorSep, 0.64*\tensorSep)$);%
    
    \draw[line width=0.5mm, dashed, dash pattern=on 0.5pt off 0.5pt] ($(U) + (0.8*\tensorSep, 0)$) -- ($(U) + (0.8*\tensorSep, 0.6*\tensorSep)$);%
    \draw[thick] ($(U) + (0.4*\tensorSep, 0)$) -- ($(U) + (0.4*\tensorSep, 0.6*\tensorSep)$);%
    \draw[draw,fill=gray!30] ($(U) - (1.0*\tensorSep, 0.25*\tensorSep)$) rectangle ++(2.0*\tensorSep, 0.5*\tensorSep);%
    \node at (U) {$U_{\text{yellow}}$};%
    \draw[draw,fill=gray!30] ($(S) - (0.4*\tensorSep, 0.25*\tensorSep)$) rectangle ++(1.0*\tensorSep, 0.5*\tensorSep);%
    \node at ($(S) + (0.1*\tensorSep, 0)$) {$S_{\text{yellow}}$};%
    \draw[draw,fill=gray!30] ($(Vh) - (1.0*\tensorSep, 0.25*\tensorSep)$) rectangle ++(2.0*\tensorSep, 0.5*\tensorSep);%
    \node at (Vh) {$V_{\text{yellow}}^\dagger$};%
\end{scope}%
\end{tikzpicture}%
        \end{align*}
      \end{minipage}
      \begin{minipage}{\textwidth}
        \begin{align*}
            \mathbb{1} \approx~ %
            \begin{tikzpicture}[baseline=-0.65ex]%
		\def\tensorSep{1.1}%
	    \def\tensorSize{0.2}%
		\begin{scope}[shift={(0, 0.75*\tensorSep)}]%
            \coordinate (projTop) at ({0*\tensorSep}, {0*\tensorSep});%
            \coordinate (projBottom) at ({0*\tensorSep}, {-1.5*\tensorSep});%
            \draw[line width=0.5mm, dashed, dash pattern=on 0.5pt off 0.5pt] (projTop) to (projBottom);
            \draw[line width=0.5mm, dashed, dash pattern=on 0.5pt off 0.5pt] ($(projTop) + (-0.5*\tensorSep, 0)$) -- ($(projTop) + (-0.5*\tensorSep, 0.5*\tensorSep)$);%
            \draw[thick] ($(projTop) + (0.5*\tensorSep, 0)$) -- ($(projTop) + (0.5*\tensorSep, 0.5*\tensorSep)$);%
            \draw[thick, decorate, decoration={coil, aspect=0}] ($(projTop) + (-0.0*\tensorSep, 0)$) -- ($(projTop) + (-0.0*\tensorSep, 0.5*\tensorSep)$);%
            \draw[line width=0.5mm, dashed, dash pattern=on 0.5pt off 0.5pt] ($(projBottom) + (-0.5*\tensorSep, 0)$) -- ($(projBottom) + (-0.5*\tensorSep, -0.5*\tensorSep)$);%
            \draw[thick] ($(projBottom) + (0.5*\tensorSep, 0)$) -- ($(projBottom) + (0.5*\tensorSep, -0.5*\tensorSep)$);%
            \draw[thick, decorate, decoration={coil, aspect=0}] ($(projBottom) + (-0.0*\tensorSep, 0)$) -- ($(projBottom) + (-0.0*\tensorSep, -0.5*\tensorSep)$);%
            \node[isosceles triangle, isosceles triangle apex angle=100, draw, shape border rotate=-90, fill=googleY] at (projTop) {$P_T$};%
            \node[isosceles triangle, isosceles triangle apex angle=100, minimum size=1, draw, shape border rotate=90, fill=googleY] at (projBottom) {$P_B$};%
		\end{scope}%
      \end{tikzpicture}%
            ~=~~ %
            \begin{tikzpicture}[baseline=-0.65ex]%
		\def\tensorSep{1.1}%
	    \def\tensorSize{0.2}%
		\begin{scope}%
            \coordinate (Uh) at ({0.2*\tensorSep}, {1.25*\tensorSep});%
            \coordinate (Stop) at ({0.6*\tensorSep}, {0.5*\tensorSep});%
            \coordinate (Sbottom) at ({0.6*\tensorSep}, {-0.5*\tensorSep});%
            \coordinate (V) at ({0.2*\tensorSep}, {-1.25*\tensorSep});%
            \coordinate (rhoBottom) at ({-1*\tensorSep}, {0.5*\tensorSep});%
            \coordinate (rhoTop) at ({-1*\tensorSep}, {-0.5*\tensorSep});%
            \draw[thick] ($(rhoBottom) + (0.55*\tensorSep, 0)$) -- ($(rhoBottom) + (0.55*\tensorSep, 0.75*\tensorSep)$);%
            \draw[line width=0.5mm, dashed, dash pattern=on 0.5pt off 0.5pt] ($(rhoBottom) + (0.75*\tensorSep, 0)$) -- ($(rhoBottom) + (0.75*\tensorSep, 0.75*\tensorSep)$);%
            \draw[thick, decorate, decoration={coil, aspect=0}] ($(rhoBottom) + (0.1*\tensorSep, 0.75*\tensorSep)$) -- ($(rhoBottom) + (0.1*\tensorSep, 0)$);%
            
            \draw[line width=0.5mm, dashed, dash pattern=on 0.5pt off 0.5pt] ($(rhoBottom) + (-0.65*\tensorSep, 0)$) -- ($(rhoBottom) + (-0.65*\tensorSep, 0.75*\tensorSep)$);%
            
            \draw[thick] ($(rhoBottom) + (-0.25*\tensorSep, 0)$) -- ($(rhoBottom) + (-0.25*\tensorSep, 0.75*\tensorSep)$);%
            \draw[thick] ($(rhoTop) + (0.55*\tensorSep, 0)$) -- ($(rhoTop) + (0.55*\tensorSep, -0.75*\tensorSep)$);%
            
            \draw[line width=0.5mm, dashed, dash pattern=on 0.5pt off 0.5pt] ($(rhoTop) + (0.75*\tensorSep, 0)$) -- ($(rhoTop) + (0.75*\tensorSep, -0.75*\tensorSep)$);%
            
            \draw[thick, decorate, decoration={coil, aspect=0}] ($(rhoTop) + (0.1*\tensorSep, -0.75*\tensorSep)$) -- ($(rhoTop) + (0.1*\tensorSep, 0)$);%
            
            \draw[line width=0.5mm, dashed, dash pattern=on 0.5pt off 0.5pt] ($(rhoTop) + (-0.65*\tensorSep, 0)$) -- ($(rhoTop) + (-0.65*\tensorSep, -0.75*\tensorSep)$);%
            
            \draw[thick] ($(rhoTop) + (-0.25*\tensorSep, 0)$) -- ($(rhoTop) + (-0.25*\tensorSep, -0.75*\tensorSep)$);%
            \draw[line width=0.5mm, dashed, dash pattern=on 0.5pt off 0.5pt] ($(Stop) + (0, 0.1*\tensorSep)$) to ($(Sbottom) + (0, -0.1*\tensorSep)$);%
            
            \draw[line width=0.5mm, dashed, dash pattern=on 0.5pt off 0.5pt, out=90, in=0] ($(Stop) + (0, 0.25*\tensorSep)$) to ($(Uh) + (0.15*\tensorSep, 0)$);%
            \draw[line width=0.5mm, dashed, dash pattern=on 0.5pt off 0.5pt, out=270, in=0] ($(Sbottom) + (0, -0.25*\tensorSep)$) to ($(V) + (0.15*\tensorSep, 0)$);%
            %
            \draw[draw,fill=gray!30] ($(V) - (0.8*\tensorSep, 0.25*\tensorSep)$) -- ($(V) + (-0.8*\tensorSep, 0.25*\tensorSep)$) -- ($(V) + (-0.1*\tensorSep, 0.25*\tensorSep)$) -- ($(V) + (0.16*\tensorSep, 0*\tensorSep)$) -- ($(V) + (-0.1*\tensorSep, -0.25*\tensorSep)$) -- cycle;
            \node at ($(V) + (-0.35*\tensorSep, 0)$) {$\widetilde{V}_{\text{yellow}}$};%
            \draw[draw,fill=gray!30] ($(Stop) - (0.4*\tensorSep, 0.25*\tensorSep)$) rectangle ++(0.8*\tensorSep, 0.5*\tensorSep);%
            \node at (Stop) {$\widetilde{S}_{\text{yellow}}^{-1/2}$};%
            \draw[draw,fill=gray!30] ($(Sbottom) - (0.4*\tensorSep, 0.25*\tensorSep)$) rectangle ++(0.8*\tensorSep, 0.5*\tensorSep);%
            \node at (Sbottom) {$\widetilde{S}_{\text{yellow}}^{-1/2}$};%
            %
            \draw[draw,fill=gray!30] ($(Uh) - (0.8*\tensorSep, 0.25*\tensorSep)$) -- ($(Uh) + (-0.8*\tensorSep, 0.25*\tensorSep)$) -- ($(Uh) + (-0.1*\tensorSep, 0.25*\tensorSep)$) -- ($(Uh) + (0.16*\tensorSep, 0*\tensorSep)$) -- ($(Uh) + (-0.1*\tensorSep, -0.25*\tensorSep)$) -- cycle;
            \node at ($(Uh) + (-0.35*\tensorSep, 0)$) {$\widetilde{U}_{\text{yellow}}^\dagger$};%
            \draw[draw,fill=gray!30] ($(rhoBottom) - (0.85*\tensorSep, 0.25*\tensorSep)$) rectangle ++(1.7*\tensorSep, 0.5*\tensorSep);%
            \node at (rhoBottom) {$\rho_{\text{yellow}}^B$};%
            \draw[draw,fill=gray!30] ($(rhoTop) - (0.85*\tensorSep, 0.25*\tensorSep)$) rectangle ++(1.7*\tensorSep, 0.5*\tensorSep);%
            \node at (rhoTop) {$\rho_{\text{yellow}}^T$};%
		\end{scope}%
      \end{tikzpicture}%
        \end{align*}
      \end{minipage}
    \end{minipage}
    \caption{Alternative construction of the yellow projectors. The initial networks for $\rho^B_{\text{yellow}}$ and $\rho^T_{\text{yellow}}$ are the same as in Fig.~\ref{fig:constructionYellowProjectors}, however, 
    the physical index is traced over and therefore contained in the renormalization step implemented by these alternative projectors.}
    \label{fig:alternativeConstructionYellowProjectors}
\end{figure*}

\begin{figure*}[t]
    \centering
    \begin{minipage}{1.1\columnwidth}
        \input{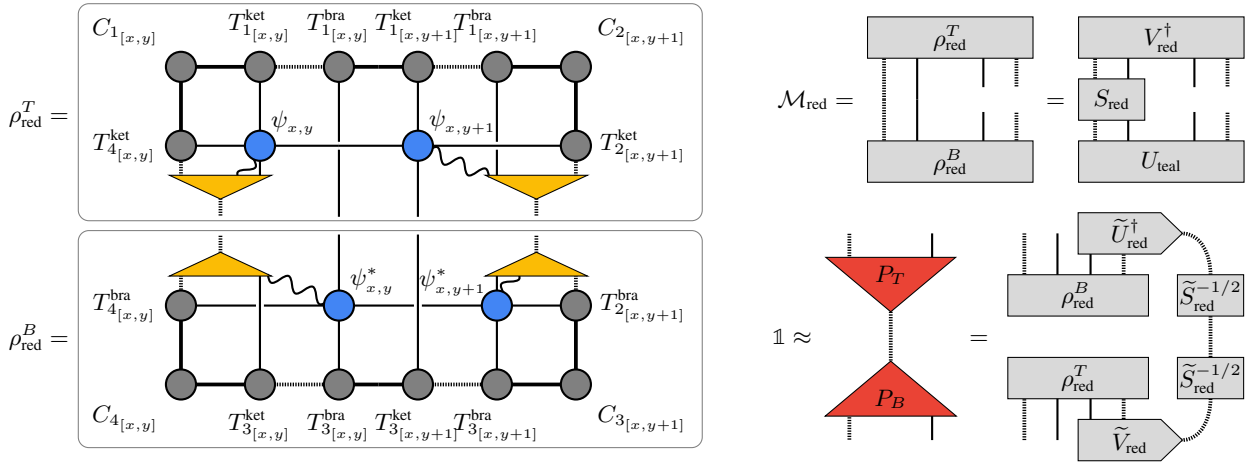}
    \end{minipage}
    \begin{minipage}{0.9\columnwidth}
      \begin{minipage}{\textwidth}
        \begin{align*}
          \mathcal{M}_{\text{red}} &=~ %
             \begin{tikzpicture}[baseline=-0.65ex]%
    \def\tensorSep{1.1}%
    \def\tensorSize{0.2}%
    \begin{scope}[shift={(0, 0.75*\tensorSep)}]%
        \coordinate (rhoTop) at ({0*\tensorSep}, {0*\tensorSep});%
        \coordinate (rhoBottom) at ({0*\tensorSep}, {-1.5*\tensorSep});%
        \draw[line width=0.5mm, dashed, dash pattern=on 0.5pt off 0.5pt] ($(rhoTop) + (-0.8*\tensorSep, 0)$) -- ($(rhoTop) + (-0.8*\tensorSep, -0.75*\tensorSep)$);%
        
        \draw[thick] ($(rhoTop) + (-0.4*\tensorSep, 0)$) -- ($(rhoTop) + (-0.4*\tensorSep, -0.75*\tensorSep)$);%
        

        \draw[line width=0.5mm, dashed, dash pattern=on 0.5pt off 0.5pt] ($(rhoTop) + (0.8*\tensorSep, 0)$) -- ($(rhoTop) + (0.8*\tensorSep, -0.6*\tensorSep)$);%
        \draw[thick] ($(rhoTop) + (0.4*\tensorSep, 0)$) -- ($(rhoTop) + (0.4*\tensorSep, -0.6*\tensorSep)$);%
        
        %
        
        \draw[line width=0.5mm, dashed, dash pattern=on 0.5pt off 0.5pt] ($(rhoBottom) + (-0.8*\tensorSep, 0)$) -- ($(rhoBottom) + (-0.8*\tensorSep, 0.75*\tensorSep)$);%
        \draw[thick] ($(rhoBottom) + (-0.4*\tensorSep, 0)$) -- ($(rhoBottom) + (-0.4*\tensorSep, 0.75*\tensorSep)$);%
        \draw[line width=0.5mm, dashed, dash pattern=on 0.5pt off 0.5pt] ($(rhoBottom) + (0.8*\tensorSep, 0)$) -- ($(rhoBottom) + (0.8*\tensorSep, 0.6*\tensorSep)$);%
        \draw[thick] ($(rhoBottom) + (0.4*\tensorSep, 0)$) -- ($(rhoBottom) + (0.4*\tensorSep, 0.6*\tensorSep)$);%
        \draw[draw,fill=gray!30] ($(rhoTop) - (1.0*\tensorSep, 0.25*\tensorSep)$) rectangle ++(2.0*\tensorSep, 0.5*\tensorSep);%
        \node at (rhoTop) {$\rho_{\text{red}}^T$};%
        \draw[draw,fill=gray!30] ($(rhoBottom) - (1.0*\tensorSep, 0.25*\tensorSep)$) rectangle ++(2.0*\tensorSep, 0.5*\tensorSep);%
        \node at (rhoBottom) {$\rho_{\text{red}}^B$};%
    \end{scope}%
  \end{tikzpicture}%
            ~=~ 
            \begin{tikzpicture}[baseline=-0.65ex]%
\def\tensorSep{1.1}%
\def\tensorSize{0.2}%
\begin{scope}[shift={(0, 0.75*\tensorSep)}]%
    \coordinate (Vh) at ({0*\tensorSep}, {0*\tensorSep});%
    \coordinate (S) at ({-0.6*\tensorSep}, {-0.75*\tensorSep});%
    \coordinate (U) at ({0*\tensorSep}, {-1.5*\tensorSep});%
    \draw[line width=0.5mm, dashed, dash pattern=on 0.5pt off 0.5pt] ($(Vh) + (-0.8*\tensorSep, 0)$) -- ($(Vh) + (-0.8*\tensorSep, -0.75*\tensorSep)$);%
    \draw[thick] ($(Vh) + (-0.4*\tensorSep, 0)$) -- ($(Vh) + (-0.4*\tensorSep, -0.75*\tensorSep)$);%
    
    \draw[line width=0.5mm, dashed, dash pattern=on 0.5pt off 0.5pt] ($(Vh) + (0.8*\tensorSep, 0)$) -- ($(Vh) + (0.8*\tensorSep, -0.6*\tensorSep)$);%
    \draw[thick] ($(Vh) + (0.4*\tensorSep, 0)$) -- ($(Vh) + (0.4*\tensorSep, -0.6*\tensorSep)$);%
    \draw[line width=0.5mm, dashed, dash pattern=on 0.5pt off 0.5pt] ($(U) + (-0.8*\tensorSep, 0)$) -- ($(U) + (-0.8*\tensorSep, 0.75*\tensorSep)$);%
    \draw[thick] ($(U) + (-0.4*\tensorSep, 0)$) -- ($(U) + (-0.4*\tensorSep, 0.75*\tensorSep)$);%


    %

    \draw[line width=0.5mm, dashed, dash pattern=on 0.5pt off 0.5pt] ($(U) + (0.8*\tensorSep, 0)$) -- ($(U) + (0.8*\tensorSep, 0.6*\tensorSep)$);%
    \draw[thick] ($(U) + (0.4*\tensorSep, 0)$) -- ($(U) + (0.4*\tensorSep, 0.6*\tensorSep)$);%
    \draw[draw,fill=gray!30] ($(U) - (1.0*\tensorSep, 0.25*\tensorSep)$) rectangle ++(2.0*\tensorSep, 0.5*\tensorSep);%
    \node at (U) {$U_{\text{teal}}$};%
    \draw[draw,fill=gray!30] ($(S) - (0.4*\tensorSep, 0.25*\tensorSep)$) rectangle ++(0.8*\tensorSep, 0.5*\tensorSep);%
    \node at (S) {$S_{\text{red}}$};%
    \draw[draw,fill=gray!30] ($(Vh) - (1.0*\tensorSep, 0.25*\tensorSep)$) rectangle ++(2.0*\tensorSep, 0.5*\tensorSep);%
    \node at (Vh) {$V_{\text{red}}^\dagger$};%
\end{scope}%
\end{tikzpicture}%
        \end{align*}
      \end{minipage}
      \begin{minipage}{\textwidth}
        \begin{align*}
            \mathbb{1} \approx~ %
            \begin{tikzpicture}[baseline=-0.65ex]%
		\def\tensorSep{1.1}%
	    \def\tensorSize{0.2}%
		\begin{scope}[shift={(0, 0.75*\tensorSep)}]%
            \coordinate (projTop) at ({0*\tensorSep}, {0*\tensorSep});%
            \coordinate (projBottom) at ({0*\tensorSep}, {-1.5*\tensorSep});%
            \draw[line width=0.5mm, dashed, dash pattern=on 0.5pt off 0.5pt] (projTop) to (projBottom);
            \draw[line width=0.5mm, dashed, dash pattern=on 0.5pt off 0.5pt] ($(projTop) + (-0.5*\tensorSep, 0)$) -- ($(projTop) + (-0.5*\tensorSep, 0.5*\tensorSep)$);%
            %
            %
            \draw[thick] ($(projTop) + (0.5*\tensorSep, 0)$) -- ($(projTop) + (0.5*\tensorSep, 0.5*\tensorSep)$);%
            \draw[line width=0.5mm, dashed, dash pattern=on 0.5pt off 0.5pt] ($(projBottom) + (-0.5*\tensorSep, 0)$) -- ($(projBottom) + (-0.5*\tensorSep, -0.5*\tensorSep)$);%
            %
            %
            \draw[thick] ($(projBottom) + (0.5*\tensorSep, 0)$) -- ($(projBottom) + (0.5*\tensorSep, -0.5*\tensorSep)$);%
            \node[isosceles triangle, isosceles triangle apex angle=100, draw, shape border rotate=-90, fill=googleR] at (projTop) {$P_T$};%
            \node[isosceles triangle, isosceles triangle apex angle=100, minimum size=1, draw, shape border rotate=90, fill=googleR] at (projBottom) {$P_B$};%
		\end{scope}%
      \end{tikzpicture}%
            ~=~~ %
                  \begin{tikzpicture}[baseline=-0.65ex]%
		\def\tensorSep{1.1}%
	    \def\tensorSize{0.2}%
		\begin{scope}%
            \coordinate (Uh) at ({0*\tensorSep}, {1.25*\tensorSep});%
            \coordinate (Stop) at ({0.6*\tensorSep}, {0.5*\tensorSep});%
            \coordinate (Sbottom) at ({0.6*\tensorSep}, {-0.5*\tensorSep});%
            \coordinate (V) at ({0*\tensorSep}, {-1.25*\tensorSep});%
            \coordinate (rhoBottom) at ({-1*\tensorSep}, {0.5*\tensorSep});%
            \coordinate (rhoTop) at ({-1*\tensorSep}, {-0.5*\tensorSep});%
            \draw[line width=0.5mm, dashed, dash pattern=on 0.5pt off 0.5pt] ($(rhoBottom) + (0.55*\tensorSep, 0)$) -- ($(rhoBottom) + (0.55*\tensorSep, 0.75*\tensorSep)$);%
            \draw[thick] ($(rhoBottom) + (0.15*\tensorSep, 0)$) -- ($(rhoBottom) + (0.15*\tensorSep, 0.75*\tensorSep)$);%
            \draw[line width=0.5mm, dashed, dash pattern=on 0.5pt off 0.5pt] ($(rhoBottom) + (-0.65*\tensorSep, 0)$) -- ($(rhoBottom) + (-0.65*\tensorSep, 0.75*\tensorSep)$);%
            \draw[thick] ($(rhoBottom) + (-0.25*\tensorSep, 0)$) -- ($(rhoBottom) + (-0.25*\tensorSep, 0.75*\tensorSep)$);%
            %

            \draw[line width=0.5mm, dashed, dash pattern=on 0.5pt off 0.5pt] ($(rhoTop) + (0.55*\tensorSep, 0)$) -- ($(rhoTop) + (0.55*\tensorSep, -0.75*\tensorSep)$);%
            \draw[thick] ($(rhoTop) + (0.15*\tensorSep, 0)$) -- ($(rhoTop) + (0.15*\tensorSep, -0.75*\tensorSep)$);%

            \draw[line width=0.5mm, dashed, dash pattern=on 0.5pt off 0.5pt] ($(rhoTop) + (-0.65*\tensorSep, 0)$) -- ($(rhoTop) + (-0.65*\tensorSep, -0.75*\tensorSep)$);%
            \draw[thick] ($(rhoTop) + (-0.25*\tensorSep, 0)$) -- ($(rhoTop) + (-0.25*\tensorSep, -0.75*\tensorSep)$);%
            %
            \draw[line width=0.5mm, dashed, dash pattern=on 0.5pt off 0.5pt] ($(Stop) + (0, 0.1*\tensorSep)$) to ($(Sbottom) + (0, -0.1*\tensorSep)$);%
            \draw[line width=0.5mm, dashed, dash pattern=on 0.5pt off 0.5pt, out=90, in=0] ($(Stop) + (0, 0.25*\tensorSep)$) to ($(Uh) + (0.25*\tensorSep, 0)$);%
            \draw[line width=0.5mm, dashed, dash pattern=on 0.5pt off 0.5pt, out=270, in=0] ($(Sbottom) + (0, -0.25*\tensorSep)$) to ($(V) + (0.25*\tensorSep, 0)$);%
            %
            \draw[draw,fill=gray!30] ($(V) - (1.0*\tensorSep, 0.25*\tensorSep)$) -- ($(V) + (-1.0*\tensorSep, 0.25*\tensorSep)$) -- ($(V) + (0*\tensorSep, 0.25*\tensorSep)$) -- ($(V) + (0.26*\tensorSep, 0*\tensorSep)$) -- ($(V) + (0*\tensorSep, -0.25*\tensorSep)$) -- cycle;
            \node at ($(V) + (-0.375*\tensorSep, 0)$) {$\widetilde{V}_{\text{red}}$};%
            \draw[draw,fill=gray!30] ($(Stop) - (0.4*\tensorSep, 0.25*\tensorSep)$) rectangle ++(0.8*\tensorSep, 0.5*\tensorSep);%
            \node at (Stop) {$\widetilde{S}_{\text{red}}^{-1/2}$};%
            \draw[draw,fill=gray!30] ($(Sbottom) - (0.4*\tensorSep, 0.25*\tensorSep)$) rectangle ++(0.8*\tensorSep, 0.5*\tensorSep);%
            \node at (Sbottom) {$\widetilde{S}_{\text{red}}^{-1/2}$};%
            %
            \draw[draw,fill=gray!30] ($(Uh) - (1.0*\tensorSep, 0.25*\tensorSep)$) -- ($(Uh) + (-1.0*\tensorSep, 0.25*\tensorSep)$) -- ($(Uh) + (0*\tensorSep, 0.25*\tensorSep)$) -- ($(Uh) + (0.26*\tensorSep, 0*\tensorSep)$) -- ($(Uh) + (0*\tensorSep, -0.25*\tensorSep)$) -- cycle;%
            \node at ($(Uh) + (-0.375*\tensorSep, 0)$) {$\widetilde{U}_{\text{red}}^\dagger$};%
            \draw[draw,fill=gray!30] ($(rhoBottom) - (0.85*\tensorSep, 0.25*\tensorSep)$) rectangle ++(1.7*\tensorSep, 0.5*\tensorSep);%
            \node at (rhoBottom) {$\rho_{\text{red}}^B$};%
            \draw[draw,fill=gray!30] ($(rhoTop) - (0.85*\tensorSep, 0.25*\tensorSep)$) rectangle ++(1.7*\tensorSep, 0.5*\tensorSep);%
            \node at (rhoTop) {$\rho_{\text{red}}^T$};%
		\end{scope}%
      \end{tikzpicture}%
        \end{align*}
      \end{minipage}
    \end{minipage}
    \caption{Alternative construction of the red projectors. The initial networks for $\rho^B_{\text{red}}$ and $\rho^T_{\text{red}}$ differ from the ones in Fig.~\ref{fig:constructionRedProjectors}, as the physical index has already been incorporated in the proceeding 
    yellow projectors.}
    \label{fig:alternativeConstructionRedProjectors}
\end{figure*}

The leading complexity of the split-CTMRG algorithm can be reduced even further by moving to an analog to of the \textit{half} projectors~\cite{Naumann2024} for the construction of the teal projectors, as this construction is the only one containing operations which scale as \begin{align}
    \mathcal{O}(\chi_E^3\chi_B^4) \xrightarrow{\chi_E \,   \approx \, \chi_B^2} \mathcal{O}(\chi_B^{10}).
\end{align} 
Half projectors provide computationally more efficient alternative to the presented full projectors, while possibly being less accurate. 
They can be useful even in variational optimization, as long  as sufficiently high environment bond dimensions are accessible, or in order to preconverge the simulations before moving to the more accurate projectors for final convergence.
The corresponding adapted networks for $\rho_{\text{teal}}^B$ and $\rho_{\text{teal}}^T$ are shown in Fig.~\ref{fig:constructionTealHalfProjectors}. 
The smaller initial size of the networks already reduces their contraction cost to
\begin{equation}
    \mathcal{O}(\chi_I^3\chi_B^3 p ) \xrightarrow{\chi_I \,   \approx \, \chi_B^2} \mathcal{O}(\chi_B^{9} p),
\label{eq:app:leading}
\end{equation}
so that the new leading cost now becomes the singular value decomposition of the $\mathcal{M}_{\text{yellow}}$ and $\mathcal{M}_{\text{red}}$ matrices, that is given by
\begin{equation}
     \mathcal{O}(\chi_I^3\chi_B^3 p^3 ) \xrightarrow{\chi_I \,   \approx \, \chi_B^2} \mathcal{O}(\chi_B^{9} p^3).
\label{eq:app:complexity_svd_Mred_yellow}
\end{equation}
However, this computational cost can be further reduced by switching the order of absorption of the physical legs. 
Specifically, incorporating the physical leg directly in the yellow projectors instead of the red projectors, as done in our initial proposal. 
In this scenario, the dependence of the scaling on the physical Hilbert space dimension $p$ in Eq.~\eqref{eq:app:complexity_svd_Mred_yellow} is eliminated. 
Thus, the leading cost again becomes the one in Eq.~\eqref{eq:app:leading}. 
The alternative constructions for the yellow and red projectors are presented in Fig.~\ref{fig:alternativeConstructionYellowProjectors} and Fig.~\ref{fig:alternativeConstructionRedProjectors}, respectively.

\bibliography{references}

\end{document}